\pdfoutput=1
 
\newcommand*{\ATLASLATEXPATH}{}
\documentclass[PAPER, atlasdraft=true, texlive=2016, cernpreprint, UKenglish]{\ATLASLATEXPATH atlasdoc}
 
\usepackage{\ATLASLATEXPATH atlaspackage}
\usepackage{afterpage}
\usepackage{\ATLASLATEXPATH atlasbiblatex}
 
\usepackage{\ATLASLATEXPATH atlasphysics}
 
\graphicspath{{logos/}{figures/}}
 
\usepackage{ANA-EGAM-2018-01-PAPER-defs}
 
\usepackage{tabularx,ragged2e}

\AtlasTitle{Electron and photon performance measurements with the ATLAS detector using the 2015--2017
LHC proton--proton collision data}
 
\AtlasAbstract{
This paper describes the reconstruction of electrons and photons with the ATLAS detector, employed for measurements and searches exploiting the complete LHC Run 2 dataset. An improved energy
clustering algorithm is introduced, and its implications for the measurement and identification of prompt electrons and photons are discussed in detail. Corrections and calibrations that affect performance,
including energy calibration, identification and isolation efficiencies, and the measurement of the charge of reconstructed electron candidates are determined using up to 81~fb$^{-1}$ of proton--proton collision data collected at $\sqrt{s}=13$~TeV between 2015 and 2017.
}
 
\author{The ATLAS Collaboration}
 
\AtlasRefCode{EGAM-2018-01}

\AtlasJournal{JINST}
 
\PreprintIdNumber{CERN-EP-2019-145}
 
\AtlasJournalRef{JINST 14 (2019) P12006}
\AtlasDOI{10.1088/1748-0221/14/12/P12006}

\AtlasCoverSupportingNote{Super-cluster reconstruction}{https://cds.cern.ch/record/2298955}
\AtlasCoverSupportingNote{Electron and photon energy calibration}{https://cds.cern.ch/record/2651890}
\AtlasCoverSupportingNote{Electron ID}{https://cds.cern.ch/record/2652163}
\AtlasCoverSupportingNote{Photon ID}{https://cds.cern.ch/record/2647979}
\AtlasCoverSupportingNote{Electron and photon isolation}{https://cds.cern.ch/record/2646265}
 
\AtlasCoverCommentsDeadline{10 July 2019}
 
\AtlasCoverAnalysisTeam{Maria Josefina Alconada Verzini, Francisco Alonso, Christos Anastopoulos, Nansi Andari, Ludovica Aperio Bella, Jean-Francois Arguin, Hicham Atmani, Cyril Becot, Maarten Boonekamp, Frued Braren, Kurt Brendlinger, Chav Chhiv Chau, Olympia Dartsi, Jean-Baptiste de Vivie de R\'egie, David Delgove, Lucia Di Ciaccio, David Di Valentino, Otilia Anamaria Ducu, Johannes Erdmann, Marc Escalier, Saskia Falke, Yi Fang, Marcello Fanti, Louis Fayard, Alexandra Fell, Lucas Macrorie Flores, Gregor Gessner, Dag Ingemar Gillberg, Dominique Godin, Thibault Guillemin, Sarah Heim, Rachel Hyneman, Lydia Iconomidou-Fayard, Oleh Kivernyk, Matthew Henry Klein, Thomas Koffas, Kevin Kroeninger, Joe Kroll, Antoine Laudrain, Yanwen Liu, Kristin Lohwasser, Stefano Manzoni, Julien Maurer, Jovan Mitrevski, Kazuya Mochizuki, Emmanuel Monnier, Miha Muskinja, Roger Felipe Naranjo Garcia, Isabel Nitsche, Ioannis Nomidis, Jose Ocariz, Andreas Petridis, Pavel Podberezko, Thomas Dennis Powell, Pascal Pralavorio, Nadezda Proklova, Joseph Reichert, Amartya Rej, Elias Michael Ruettinger, Abhishek Sharma, Nima Sherafati, Philip Sommer, Kerstin Tackmann, Grigore Tarna, Savannah Jennifer Thais, Evelyn Thomson, Artur Trofymov, Guillaume Unal, Hanlin Xu, Zhiqing Zhang}

\AtlasCoverEdBoardMember{Olivier Arnaez~(chair)}
\AtlasCoverEdBoardMember{Fares Djama}
\AtlasCoverEdBoardMember{Kun Liu}
 
\AtlasCoverEgroupEditors{atlas-EGAM-2018-01-editors@cern.ch}
 
\AtlasCoverEgroupEdBoard{atlas-EGAM-2018-01-editorial-board@cern.ch}

\LEcontact{Norman Mccubbin, norman.mccubbin@stfc.ac.uk}
\hypersetup{pdftitle={ATLAS document},pdfauthor={The ATLAS Collaboration}}
 
\begin{document}
 
\maketitle
\tableofcontents
\newpage
\section{Introduction}
With an integrated luminosity of about 147~\ifb, the proton--proton ($pp$) collision dataset collected by the ATLAS detector between 2015 and 2018 at a centre-of-mass energy of $\rts=13$~\tev\ will allow significant advances in the exploration of the electroweak scale. Optimal performance in the measurement of electrons and photons plays a fundamental role in searches for new particles, in the measurement of Standard Model cross-sections, and in the precise measurement of the properties of fundamental particles such as the Higgs and $W$ bosons and the top quark.
 
The ATLAS Collaboration
published three papers describing the performance of the reconstruction, identification and energy measurement of electrons and photons with 36~\ifb\ of $pp$ collision
data collected in 2015 and 2016~\cite{PERF-2017-02,PERF-2017-01,PERF-2017-03}. New algorithms for electron and photon reconstruction were introduced in 2017. The present paper describes the performance of these algorithms, and extends the
analysis to the dataset collected between 2015 and 2017, which corresponds to an integrated luminosity of about 81~\ifb. The discussion is limited to electrons and photons reconstructed in the central calorimeters, covering the pseudorapidity range $|\eta|<2.5$.
 
The transition from the reconstruction of electrons and photons based on fixed-size clusters of calorimeter cells towards a dynamical, topological cell clustering algorithm~\cite{PERF-2014-07} represents the most important modification. The algorithms used for the identification of the candidates and the estimation of their energy
have been updated accordingly. The performance of these changes is discussed in detail. In addition, methods allowing an improved rejection of misreconstructed or non-isolated candidates are presented, and are of particular importance for measurements of processes with low cross-sections or high backgrounds, such as the associated production of a Higgs boson with a top-quark pair, or vector-boson scattering at high energy.
 
After a summary of the experimental apparatus and the samples used for this analysis in Sections~\ref{sec:detector} and \ref{sec:samples}, Section~\ref{sec:reconstruction} describes the new reconstruction of clusters of energy deposits in the electromagnetic (EM) calorimeter, the
estimation of their energy, and the use of information from the inner tracking detector to distinguish between electrons and photons. Section~\ref{sec:calib} summarizes the energy calibration corrections and the
associated systematic uncertainties. Sections~\ref{sec:eID} and~\ref{sec:pID} present the re-optimized electron and photon identification algorithms. Section~\ref{sec:iso} discusses the
discrimination between prompt electrons and photons and backgrounds from hadron decays. Finally, studies dedicated to the electron and positron charge identification are reported
in Section~\ref{sec:QMisID}.
 
\section{ATLAS detector}
\label{sec:detector}
The ATLAS experiment \cite{PERF-2007-01,ATLAS-TDR-2010-19,PIX-2018-001} is a general-purpose particle
physics detector with a forward--backward symmetric cylindrical
geometry and almost 4$\pi$ coverage in solid angle.\footnote{ATLAS uses a right-handed coordinate system with its origin at the nominal interaction point (IP) in the centre of the detector and the $z$-axis along the beam pipe. The $x$-axis points from the IP to the centre of the LHC ring, and the $y$-axis points upward. Cylindrical coordinates $(r,\phi)$ are used in the transverse plane, $\phi$ being the azimuthal angle around the $z$-axis. The pseudorapidity is defined in terms of the polar angle $\theta$ as $\eta=-\ln\tan(\theta/2)$. The angular distance $\Delta R$ is defined as $\Delta R \equiv \sqrt{(\Delta\eta)^2+(\Delta\phi)^2}$. The transverse energy is $\et = E/\cosh(\eta)$.}
The inner tracking detector (ID) covers the pseudorapidity range $|\eta|<2.5$ and
consists of a silicon pixel detector, a silicon microstrip detector
(SCT), and a transition  radiation tracker (TRT) in the range
$|\eta|<2.0$. The TRT provides electron identification capability through the detection of transition radiation photons.
It consists of small-radius drift tubes (`straws') interleaved with a polymer material creating transition radiation for particles with a large Lorentz factor. This radiation is absorbed by the
Xe-based gas mixture filling the straws, discriminating electrons from hadrons over a wide energy range. Due to gas leaks, some TRT modules are filled with an Ar-based gas mixture.
The ID is surrounded by a superconducting  solenoid producing a 2~T magnetic field and provides accurate reconstruction of
tracks from the primary $pp$ collision region. It also
identifies  tracks from secondary vertices, permitting an efficient
reconstruction of photon conversions in the ID up to a radius of about
800~mm.
 
The EM calorimeter is a lead/liquid-argon (LAr) sampling calorimeter with an accordion geometry. It is divided into a
barrel section (EMB) covering the pseudorapidity region
$|\eta|<1.475$,\footnote{The EMB is split into two half-barrel modules,
which cover the positive and negative $\eta$ regions.} and two endcap
sections (EMEC) covering $1.375<|\eta|<3.2$.
The barrel and endcap calorimeters are immersed in three LAr-filled cryostats, and are segmented into three layers for $|\eta|<2.5$. The first layer, covering $|\eta|<1.4$ and $1.5<|\eta|<2.4$, has a thickness
of about 4.4 radiation lengths ($X_{0}$) and is finely segmented in the $\eta$ direction, typically $0.003\times 0.1$ in $\Delta  \eta \times \Delta \phi$ in the EMB, to provide an event-by-event
discrimination between single-photon showers and overlapping showers from the decays of neutral hadrons.
The second layer (L2), which collects most of the energy  deposited in the calorimeter by photon and electron showers, has a thickness of about $17 X_{0}$  and a granularity of $0.025\times 0.025$ in
$\Delta \eta \times \Delta \phi$. A third layer, which has a granularity of $0.05\times 0.025$ in $\Delta \eta \times \Delta \phi$ and a depth of about 2$X_{0}$, is used to correct for leakage beyond
the EM calorimeter for high-energy showers.
In front of the accordion calorimeter, a thin presampler layer (PS), covering the pseudorapidity interval $|\eta|<1.8$, is used to correct for energy loss upstream of the
calorimeter. The PS consists of an active LAr layer with a thickness of 1.1~cm (0.5~cm) in the barrel (endcap) and has a granularity of $\Delta \eta \times \Delta \phi = 0.025 \times 0.1$.
The transition region between the EMB and the EMEC, $1.37<|\eta|<1.52$, has a large amount of material in front of the first active calorimeter layer ranging from 5 to almost $10 X_{0}$. This section
is instrumented with scintillators located between the barrel and endcap cryostats, and extending up to $|\eta|=1.6$.
 
The hadronic calorimeter, surrounding the EM calorimeter, consists of an iron/scintillator tile calorimeter in the range
$|\eta|<1.7$ and two copper/LAr calorimeters  spanning $1.5<|\eta|<3.2$. The acceptance is extended by two copper/LAr and tungsten/LAr
forward calorimeters extending up to $|\eta| =$ 4.9, and hosted in the same cryostats as the EMEC.
Electron reconstruction in the forward calorimeters is not discussed in this paper.

The muon spectrometer, located beyond the calorimeters, consists of
three large air-core superconducting toroid systems with eight coils each, with precision
tracking chambers providing accurate muon tracking for $|\eta|<2.7$
and fast-triggering detectors up to $|\eta|=2.4$.
 
A two-level trigger system~\cite{TRIG-2016-01} is used to select events. The first-level trigger is implemented
in hardware and uses a subset of the detector information to reduce the accepted rate to a maximum of about \SI{100}{\kHz}.
This is followed by a software-based trigger that reduces the accepted event rate to \SI{1}{\kHz} on average, depending
on the data-taking conditions.
 
 
\section{Collision data and simulation samples}
\label{sec:samples}
\subsection{Dataset}
\label{sec:data_set}
 
The analyses described in this paper use the full $pp$ collision dataset recorded by ATLAS between 2015 and 2017
with the LHC operating at a centre-of-mass energy of \rts = 13 \tev\ and a bunch spacing of 25~ns.
The dataset is divided into two subsamples according to the typical mean number of interactions per bunch crossing,
\muhat, with which it was recorded :
\begin{itemize}
\item The `low-$\mu$' sample was recorded in 2017 with $\muhat \sim 2$; after application of data-quality requirements,
the integrated luminosity amounts to 147 \ipb.
\item The `high-$\mu$' sample corresponds to an integrated luminosity of 80.5 \ifb; for this sample, \muhat\
was on average 13, 25 and 38 for 2015, 2016 and 2017 data, respectively. The corresponding integrated luminosities are 3.2 \ifb, 33.0 \ifb\ and 44.3 \ifb. In 2016, a small sample corresponding to
0.7 \ifb\ of data was recorded without magnetic field in the muon system; it is added to the `high-$\mu$' sample for
electron reconstruction and identification studies.
\end{itemize}
Two different LHC filling schemes were used in 2017. The nominal filling scheme, labelled 48b
in the following, corresponding to an integrated
luminosity of 17.9 \ifb\ and $\muhat \sim 32$, was built from `sub-trains' of
48 filled bunches followed by seven empty bunches.
Simulated event samples use this configuration,\footnote{The simulation used in conjunction with 2015 and 2016 data
has a similar bunch configuration, consisting of 72 filled bunches followed by eight empty bunches.} as it represents about 70\% of the collected data; the implications of this approximation for the energy calibration are discussed in Section~\ref{sec:calib}.
The second scheme, labelled 8b4e,
corresponding to an integrated luminosity of 26.4~\ifb\ and $\muhat \sim 42$, was made of sub-trains
of eight filled bunches followed by four empty bunches. To sustain
these conditions, a levelling of the instantaneous luminosity at $2\times10^{34}$ cm$^{-2}$s$^{-1}$ was
necessary at the beginning of the fill, resulting in a peak \muhat\ around 60.
The noise induced by pile-up, or multiple $pp$ interactions occurring in the same bunch crossing as the event of interest or in nearby crossings, is 10\% smaller than for the standard configuration for a given $\mu$.
The LHC filling scheme for the `low-$\mu$' data sample was 8b4e.
 
Several levels of object identification and isolation criteria are employed to select the event samples
used in the analyses described in this paper. Electrons are identified using a likelihood-based method
combining information from the EM calorimeter and the ID. Different identification working points, Loose, Medium
and Tight are defined~\cite{PERF-2017-01}. Similar levels are used at trigger level (online), with slightly different
inputs. A Very Loose working point is also defined for the online selection. Photons are selected
using a set of cuts on calorimeter variables~\cite{PERF-2017-02} in the pseudorapidity range $|\eta| < 2.37$,
with the transition region between the barrel and endcap calorimeters, $1.37 < |\eta| < 1.52$, excluded.
Two levels of identification, Loose and Tight,
are considered. A Loose identification is used at trigger level to select a sample of inclusive photons.

The measurements of the electromagnetic energy response and of the electron identification efficiency
use a large sample of \Zee\ events selected with single-electron and dielectron triggers. The dielectron
high-level triggers use a transverse energy (\et) threshold ranging from 12 \gev\ (2015) to 17 or 24 \gev\ (2016 and 2017)
and a Loose (2015) or Very Loose (2016 and 2017) identification criterion.
The single-electron high-level trigger has an \et threshold ranging from 24 \gev\ in 2015 and most of 2016 to
26 \gev\ at the end of 2016 and during 2017; it requires a Tight identification and loose tracking-based
isolation criteria. The offline selection for the energy calibration measurement requires two electrons
with Medium identification and loose isolation~\cite{PERF-2017-01} with $\et > 27$ \gev, resulting in $\sim$
36 million \Zee\ candidate events.
 
A sample of $J/\psi \to ee$ events with at least two electron candidates with $\et > 4.5\ \gev$ and $|\eta| < 2.47$ was collected for studies with low-\et\ electrons
using dedicated prescaled dielectron triggers with electron \et thresholds ranging from 4 to 14~\gev. Each of
these triggers requires Tight trigger identification and \et above a certain threshold for one trigger object,
while only demanding the electromagnetic cluster \et to be higher than some other (lower) threshold for
the second object.
 
Samples of $Z \to \ell\ell\gamma$ events, used to validate the photon energy scale and measure photon identification
and isolation efficiencies at low \et, were selected with the same triggers as for the \Zee\ sample for the electron channel
and single-muon or dimuon triggers in the muon channel. The dimuon (single-muon) trigger transverse momentum (\pt) threshold was 14 (26) \gev\
at the high-level trigger; a loose tracking-based isolation criterion was applied at the high-level trigger for the single-muon trigger.
The $\mu\mu\gamma$ ($ee\gamma$) samples, after requiring two muons (electrons) with Medium identification~\cite{PERF-2015-10},
$\pt > 15$ \gev\ (18 \gev) and one tightly identified and loosely isolated photon with $\et > 15$ \gev, contain $\sim$ 110000 ($\sim$~54000) events.
 
Single-photon triggers with Loose identification and large prescale factors are used
for measurements of the photon identification and isolation efficiencies. The lowest transverse energy threshold of
these triggers is 10~\gev.

\subsection{Simulation samples}
\label{sec:monte_carlo}
 
Large Monte Carlo (MC) samples of $Z \to \ell\ell$ events ($\ell = e, \mu$)
were simulated at next-to-leading order (NLO) in QCD using \textsc{Powheg}~\cite{powheg}
interfaced to the \textsc{Pythia8}~\cite{pythia8}
parton shower model. The CT10~\cite{ct10} parton distribution function (PDF)
set was used in the matrix element. The AZNLO set of tuned parameters~\cite{STDM-2012-23} was used, with PDF set CTEQ6L1~\cite{cteq6}, for the
modelling of non-perturbative effects.
\textsc{Photos++} 3.52~\cite{photospp} was used for QED emissions from electroweak vertices and charged leptons.
To model the background in photon identification and isolation measurements using radiative $Z$ decays, samples of $Z \to \ell\ell$ events
with up to two additional partons at NLO in QCD and four additional partons at leading order (LO) in QCD were simulated
with \textsc{Sherpa}~\cite{sherpa} version 2.2.1, using the NNPDF30NNLO~\cite{nnpdf30} PDF in conjunction with the dedicated parton
shower tuning developed by the \textsc{Sherpa} authors.
 
Both non-prompt (originating from $b$-hadron decays) and prompt (not originating from $b$-hadron decays) $J/\psi \to ee$ samples were generated
using \textsc{Pythia8}. The A14 set of tuned parameters~\cite{ATL-PHYS-PUB-2014-021} was used together with the CTEQ6L1 PDF set.
 
Samples of $Z \to \ell\ell\gamma$ events with transverse energy of the photon above 10 \gev\ were generated
with \textsc{Sherpa} version 2.1.1 using QCD leading-order matrix elements with up to three additional
partons in the final state. The CT10 PDF set was used.
 
Samples of inclusive photon production were generated using \textsc{Pythia8}. The signal includes
LO photon-plus-jet events from the hard subprocesses $qg \to q\gamma$ and $\qqbar \to g\gamma$, and photon production from quark fragmentation
in LO QCD dijet events. The fragmentation component was modelled by QED radiation arising from
calculations of all 2 $\to$ 2 QCD processes involving light partons (gluons and up, down and strange quarks).
 
A large sample of backgrounds to prompt photon and electron production was generated with \textsc{Pythia8}, including all tree-level
2 $\to$ 2 QCD processes as well as top-quark pair and weak vector-boson production, filtered at particle level to mimic a first-level EM
trigger requirement. For this sample and the inclusive-photon samples, the A14 set of tuned parameters was used together
with the NNPDF23LO PDF set~\cite{nnpdf23}.
 
The \textsc{Pythia8} sample production used the \textsc{EvtGen} 1.2.0 program~\cite{evtgen} to model $b$- and $c$-hadron decays.
 
The generated events were processed through the full ATLAS detector simulation~\cite{SOFT-2010-01} based on \textsc{Geant4}~\cite{geant4}.
The MC events were simulated with additional interactions in the same or neighbouring bunch crossings to match the pile-up conditions
during LHC operations.
The overlaid $pp$ collisions were generated with the soft QCD processes of \textsc{Pythia8}
using the A3 set of
tuned parameters~\cite{ATL-PHYS-PUB-2016-017} and the NNPDF23LO PDF. Although this set of tuned parameters improves the modelling
of minimum-bias data relative to the set used previously (A2~\cite{ATL-PHYS-PUB-2012-003}), it overestimates
by roughly 3\% the hadronic activity as measured using charged-particle tracks. Simulated events
were weighted to reproduce the distribution of the average number of interactions per bunch crossing in data, scaled
down by a factor 1.03.
 
Many analyses
rely on MC samples generated with the ATLAS fast simulation, which uses a parameterized
response of the calorimeters~\cite{SOFT-2010-01}.
Dedicated corrections to the reconstructed energy and identification efficiencies of electrons and photons were determined
for these samples to match the performance observed in the samples using the full simulation of the ATLAS detector.
 
The response of the new reconstruction algorithm was optimized using samples of 40 million single-electron and single-photon events simulated without pile-up.
Their transverse energy distribution covers the range from 1 GeV to 3 TeV. Smaller samples with a flat \muhat\ spectrum between 0 and 60 were also simulated to assess the performance as a function of \muhat.

Studies presented throughout this paper using MC simulation select electrons originating from \Zee\ or $J/\psi\rightarrow ee$ decays using generator-level information.
The matching of reconstructed and generated electron is based on the ID track~\cite{PERF-2015-08} which can be reconstructed from the primary electron or from secondary particles produced in a material interaction of the primary electron or of final state radiation emitted collinearly. Similarly, reconstructed and generator-level photons are matched based on their distance in $\eta$--$\phi$ space.
 
\section{Electron and photon reconstruction}
\label{sec:reconstruction}
 
In replacement of the sliding-window algorithm previously exploited in ATLAS for the reconstruction of fixed-size clusters of calorimeter cells~\cite{ATL-LARG-PUB-2008-002,PERF-2017-01,PERF-2017-02}, the offline electron and photon reconstruction has been improved to use dynamic, variable-size clusters, called superclusters. While fixed-size clusters naturally provide a linear energy response and good stability as a function of pile-up, dynamic clusters change in size as needed to recover energy from bremsstrahlung photons or from electrons from photon conversions. The calibration techniques described in Ref.~\cite{PERF-2017-03} exploit this advantage of the dynamic clustering algorithm, while achieving similar linearity and stability as for fixed-size clusters.

An electron is defined as an object consisting of a cluster built from energy deposits in
the calorimeter (supercluster) and a matched track (or tracks). A converted photon is a cluster matched to a conversion
vertex (or vertices), and an unconverted photon is a
cluster matched to neither an electron track nor a conversion
vertex. About $20\%$ of photons at low $|\eta|$ convert in the ID, and
up to about $65\%$ convert at $|\eta| \approx 2.3$.
 
The reconstruction of electrons and photons with
$|\eta| < 2.5$ proceeds as shown in
Figure~\ref{fig:code_flow_cartoon}. The algorithm first prepares the
tracks and clusters it will use. It selects clusters of energy deposits measured in topologically connected EM and hadronic calorimeter cells~\cite{PERF-2014-07}, denoted topo-clusters, reconstructed
as described in Section~\ref{sec:topoclusters}. These clusters are matched to ID tracks, which are re-fitted accounting for bremsstrahlung. The algorithm also builds
conversion vertices and matches them to the selected topo-clusters.
The electron and photon supercluster-building steps then run
separately using the matched clusters as input. After applying initial
position corrections and energy calibrations to the resulting
superclusters, the supercluster-building algorithm matches tracks to
the electron superclusters and conversion vertices to the photon
superclusters. The electron and photon objects to be used for analyses
are then built, their energies are calibrated, and discriminating
variables used to separate electrons or photons from background are
added.
\begin{figure}
\begin{center}
\includegraphics[width=\columnwidth]{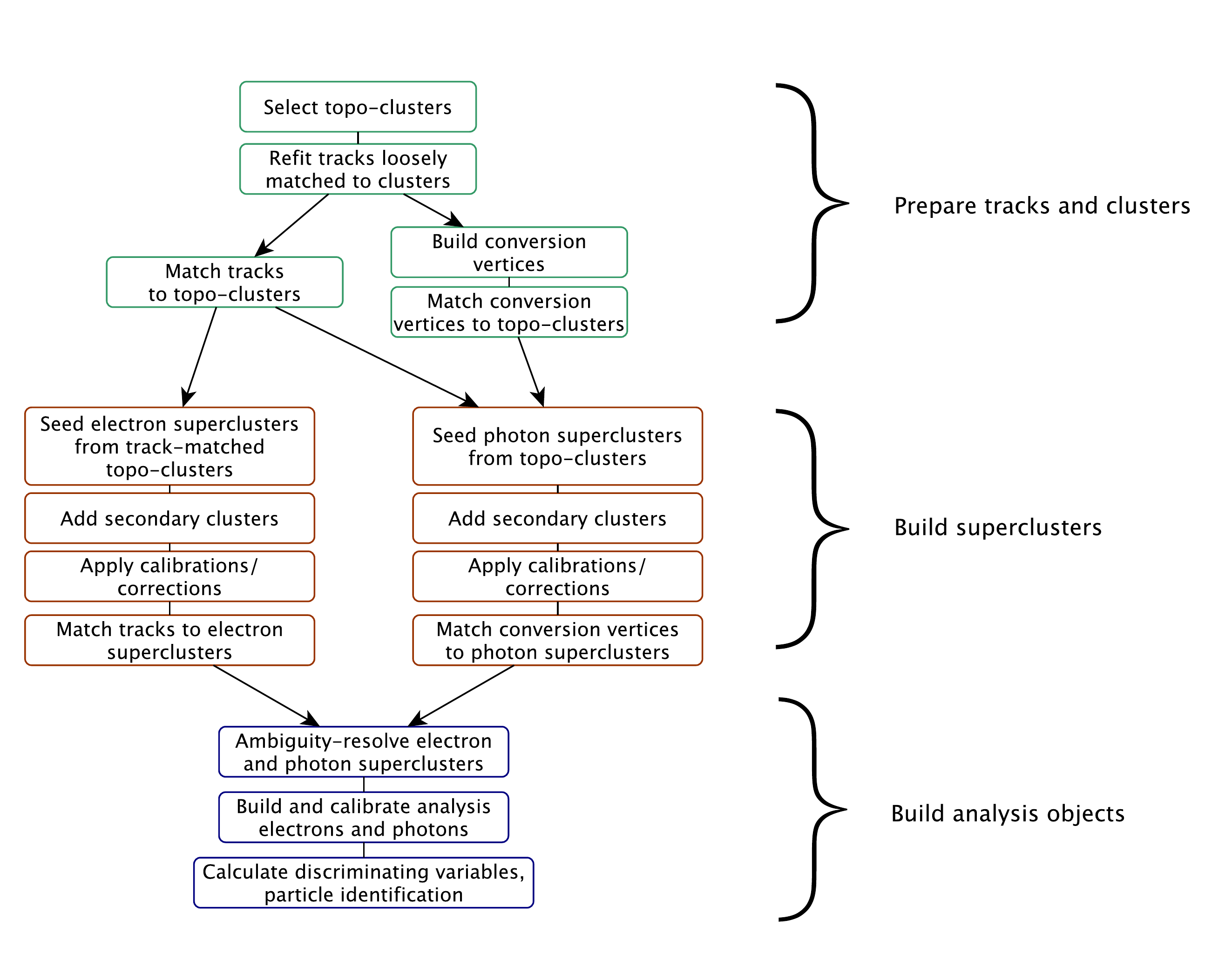}
\end{center}
\caption{Algorithm flow diagram for the electron and photon reconstruction.}
\label{fig:code_flow_cartoon}
\end{figure}
The steps are described in more detail below.
 
\subsection{Topo-cluster reconstruction}
\label{sec:topoclusters}
 
The topo-cluster reconstruction algorithm~\cite{ATL-LARG-PUB-2008-002,PERF-2014-07} begins by
forming proto-clusters in the EM and
hadronic calorimeters using a set of noise thresholds in which the cell initiating the cluster is required to have significance $\left|\cellsig\right| \geq 4$, where
\begin{equation}
\cellsig =  \frac{E_\text{cell}^\text{EM}}{\sigma_\text{noise,cell}^\text{EM}},
\nonumber
\end{equation}
$E_\text{cell}^\text{EM}$ is the cell energy at the EM
scale\footnote{The EM scale
is the basic signal scale accounting correctly for the energy deposited in the
calorimeter by electromagnetic showers.}
and
$\sigma_\text{noise,cell}^\text{EM}$ is the expected cell noise. The
expected cell noise includes the known electronic noise and an estimate of the pile-up noise corresponding to the average instantaneous luminosity expected for Run 2. In this initial stage, cells from the presampler and the first LAr EM calorimeter layer are
excluded from initiating proto-clusters, to suppress the
formation of noise clusters.  The proto-clusters then collect
neighbouring cells with significance $\left|\cellsig\right| \geq
2$. Each neighbour cell passing the threshold of
$\left|\cellsig\right| \geq 2$ becomes a seed cell in the next
iteration, collecting each of its neighbours in the proto-cluster. If
two proto-clusters contain the same cell with
$\left|\cellsig\right| \geq 2$ above the noise threshold, these
proto-clusters are merged.
A crown of nearest-neighbour cells
is added to the cluster independently on their energy.
In the presence of negative-energy cells induced by the calorimeter noise, the algorithm uses $\left|\cellsig\right|$ instead of $\cellsig$ to avoid biasing the cluster energy upwards, which would happen if only positive-energy
cells were used. This set of thresholds is commonly known as `4-2-0'
topo-cluster reconstruction. Proto-clusters with two or more local
maxima are split into separate clusters; a cell is considered a local maximum when it has $E_\text{cell}^\text{EM} > 500\,\MeV$, at least four
neighbours, and when none of the neighbours has a larger signal.
 
Electron and photon reconstruction starts from the topo-clusters but
only uses the energy from cells in the EM calorimeter, except in the
transition region of $1.37 < |\eta| < 1.63$, where the energy measured
in the presampler and the scintillator between the calorimeter cryostats is also added.
This is referred to as the EM energy of the cluster, and the EM fraction
($\emfrac$) is the ratio of the EM energy to the total cluster
energy. Only clusters with EM energy greater than $400\,\MeV$ are
considered.
The distribution of $\emfrac$ is shown in Figure~\ref{fig:emf} and the electron reconstruction
efficiency for various cuts on $\emfrac$ is shown in
Figure~\ref{fig:emfeff}, for electron clusters which
have been simulated with $\muhat = 0$, and for pile-up clusters. A
preselection requirement of $\emfrac > 0.5$ was chosen for the initial
topo-clusters, as it rejects $\sim 60\%$ of pile-up clusters without
affecting the efficiency for selecting true electron
topo-clusters.\footnote{In the transition region, some topo-clusters
are also selected as EM clusters, even if they fail the requirement on \emfrac,
when they satisfy $\et > 1\,\gev$,
in order to
increase the reconstruction efficiency in that region.} These
clusters are referred to as EM topo-clusters in the rest of this
paper.
\begin{figure}
\begin{center}
\subfloat[]{\includegraphics[width=0.49\columnwidth]{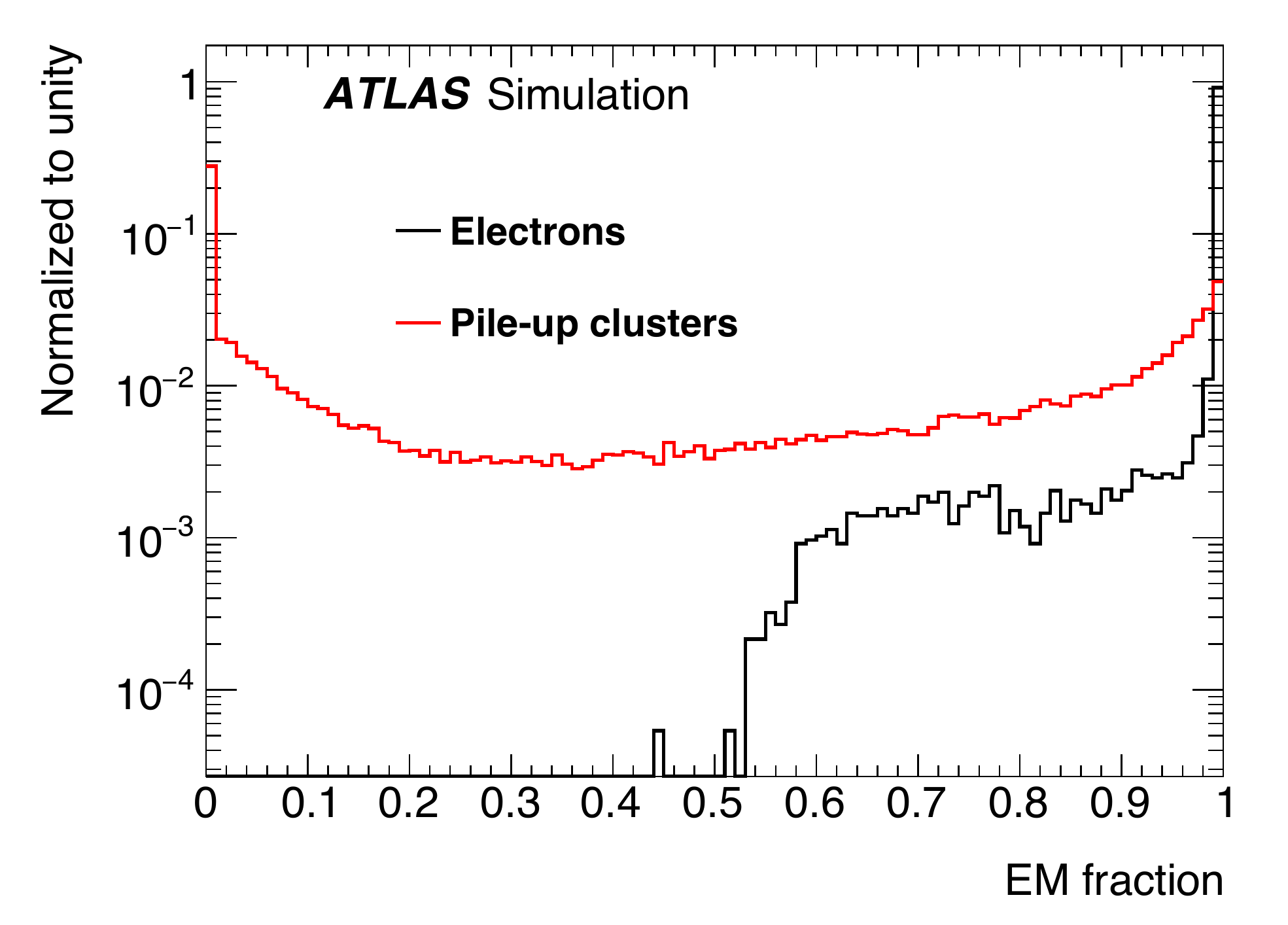}\label{fig:emf}}
\subfloat[]{\includegraphics[width=0.49\columnwidth]{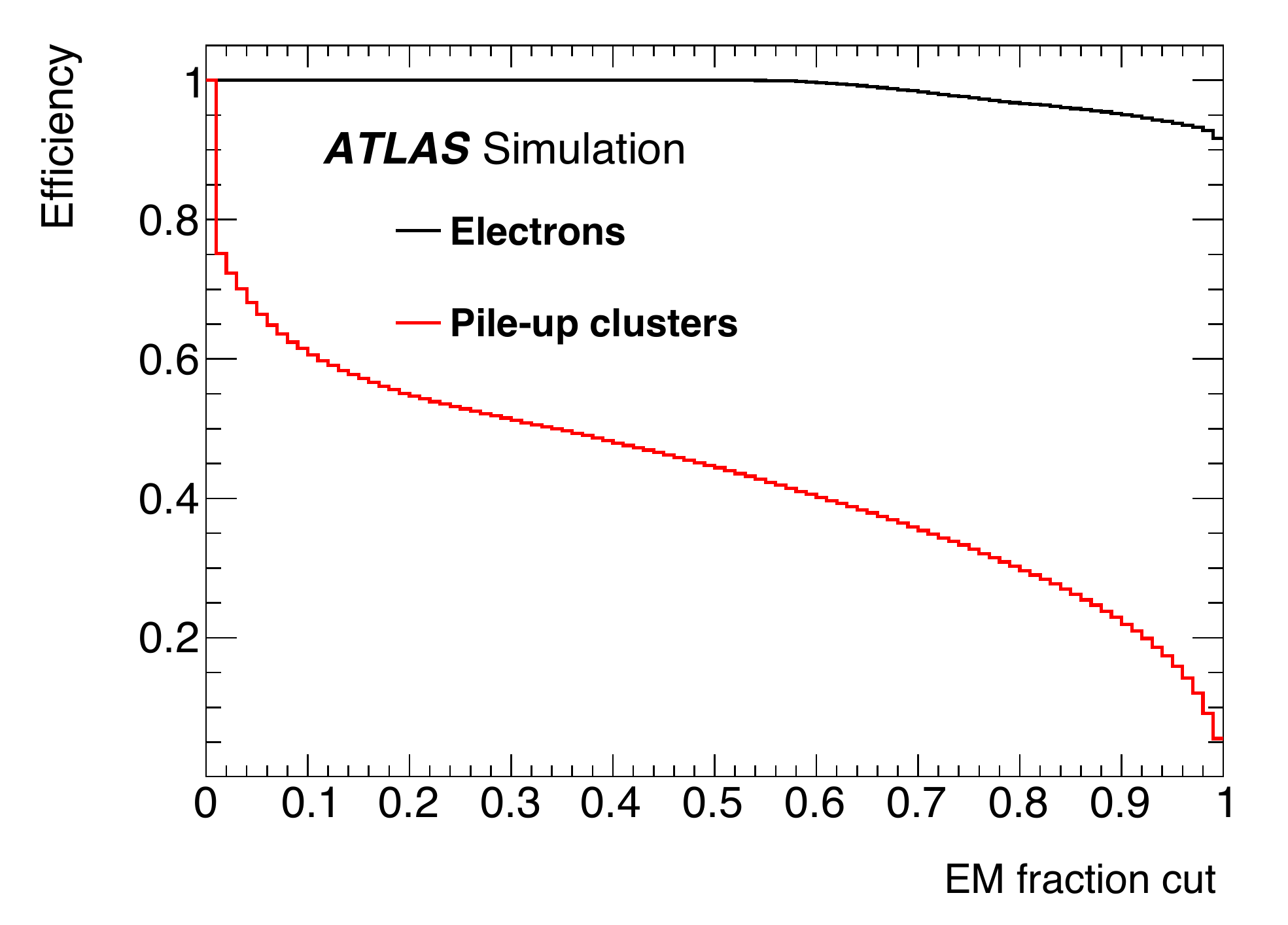}\label{fig:emfeff}}
\end{center}
\vspace{-0.4cm}
\caption{(a) Distribution of $\emfrac$ and (b) reconstruction efficiency
as a function of the \emfrac\ selection cut for simulated true
electron (black) and pile-up (red) clusters.
}
\end{figure}

\subsection{Track reconstruction, track--cluster matching, and photon conversion reconstruction}
\label{sec:tracking}
 
Track reconstruction for electrons is unchanged with respect to Ref.~\cite{PERF-2017-01,PERF-2017-02}.
A summary of the changes applied for photons is given below.
 
Standard track-pattern reconstruction~\cite{Cornelissen:1020106} is
first performed everywhere in the inner detector. However, fixed-size
clusters in
the calorimeter that have a longitudinal and lateral shower profile compatible with that of an EM shower
are used to create regions-of-interest
(ROIs). If the standard pattern
recognition fails for a silicon track seed (a set of silicon detector hits used to start a track) within an ROI, a
modified pattern recognition algorithm based on a Kalman filter formalism~\cite{Fruhwirth:1987fm} is used, allowing for up to
30\% energy loss at each material intersection. Track candidates are
then fitted with the global $\chi^2$
fitter~\cite{Cornelissen:2008zza}, allowing for additional energy loss when the standard track fit fails. Additionally, tracks with silicon hits loosely matched\footnote{The
match must be within $|\Delta\eta| < 0.05$ and
$-0.20 < q\cdot(\phi_\mathrm{track} -
\phi_\mathrm{clus}) < 0.05$ when using the track energy to
extrapolate from the last inner detector hit, or $|\Delta\eta| < 0.05$
and
$-0.10 < q\cdot(\phi_\mathrm{track} -
\phi_\mathrm{clus}) < 0.05$ when using the cluster energy to
extrapolate from the track perigee; $q$ refers to the reconstructed charge of the track.} to fixed-size clusters are
re-fitted using a Gaussian sum filter (GSF) algorithm
\cite{ATLAS-CONF-2012-047}, a non-linear generalization of the Kalman
filter, for improved track parameter estimation.
 
The loosely matched, re-fitted tracks are then matched to the EM
topo-clusters described above, extrapolating the track from the
perigee to the second layer of the calorimeter, and using either the
measured track momentum or rescaling the magnitude of the momentum to
match the cluster energy. The
momentum rescaling is performed to improve track--cluster matching for electron candidates with significant energy loss due to
bremsstrahlung radiation in the tracker. A track is considered matched if, with
either momentum magnitude, $|\Delta\eta| < 0.05$ and
$-0.10 < q\cdot(\phi_\mathrm{track} -
\phi_\mathrm{clus}) < 0.05$, where $q$ refers to the reconstructed charge of the track.  The requirement on
$q\cdot(\phi_\mathrm{track} - \phi_\mathrm{clus})$  is asymmetric
because tracks sometimes miss some energy from radiated photons that
clusters measure.
 
If multiple tracks are matched to a cluster, they are ranked as follows. Tracks with hits in the pixel detector are
preferred, then tracks with hits in the SCT but not in the pixel
detector.
Within each category, tracks with a better $\Delta R$
match to the cluster in
the second layer of the calorimeter are preferred, unless the differences are small (less than 0.01).
The extrapolation of the track through the calorimeter is done first with the track momentum rescaled
to the cluster energy and successively without rescaling.
If both the first and the second extrapolation result in small $\Delta R$ differences,
the track with more pixel hits is preferred, giving an extra weight to
a hit in the innermost layer. The highest-ranked track is used to define
the reconstructed electron properties.
 
The photon conversion reconstruction is largely unchanged from the method
described in Ref.~\cite{PERF-2017-02}. Tracks loosely matched to
fixed-size clusters serve as input to the reconstruction of the conversion vertex.
Both tracks with silicon hits (denoted Si tracks) and
tracks reconstructed only in the TRT (denoted TRT tracks) are used for
the conversion reconstruction. Two-track conversion vertices are
reconstructed from two opposite-charge tracks forming a vertex consistent with that of
a massless particle, while single-track vertices are essentially
tracks without hits in the innermost sensitive layers.
To increase the
converted-photon purity, the tracks used to build conversion vertices
must have a high probability to be electron tracks as
determined by the TRT~\cite{ATLAS-CONF-2011-128}. The requirement is loose
for Si tracks but tight for TRT tracks used to build double-track conversions, and even tighter
for tracks used to build single-track conversions.
 
Changes were made with respect to the reconstruction software described in Ref.~\cite{PERF-2017-02},
both to improve the reconstruction efficiency of double-track Si conversions (conversions reconstructed with two Si tracks),
and to reduce the fraction of unconverted photons mistakenly reconstructed as
single- or double-track TRT conversions (conversions reconstructed with one or two TRT tracks).
The efficiency for double-track Si conversions was improved by modifying the tracking ambiguity processor, which determines which track seeds are retained to reconstruct tracks. For double-track
conversion topologies, the two tracks are expected to be close to each other, parallel, and potentially to have shared hits, so that frequently only one track is reconstructed. The optimization in the ambiguity processor results in the recovery of the second track that was previously discarded.
Overall, these modifications result in a $2$--$4\%$ improvement in efficiency for
double-track Si conversions, with larger improvements of up to $9\%$ for photons with conversion radii larger than 200~mm.
In addition to reconstructing the second track of
what would otherwise have been single-track Si conversions, the overall
conversion reconstruction efficiency is improved by about $1\%$ by reducing
the fraction of low-radius converted photons that are only reconstructed as electrons.
 
To reduce the fraction of unconverted photons reconstructed as double- or
single-track TRT conversions, requirements on the TRT tracks were tightened.
The tracks are required to have at least $30\%$ precision hits, where a precision hit is defined as a hit with a track-to-wire
distance within 2.5 times its uncertainty~\cite{IDET-2015-01}.
In addition, the requirement on
the probability of a track to correspond to an electron, as determined by the TRT, was
tightened to 0.75 for tracks used in double-track TRT conversions and
to 0.85 for tracks used in single-track TRT conversions,
compared with the previous requirement of 0.7 for tracks used in both conversion types.
The fraction of unconverted photons erroneously reconstructed as converted photons is below
$5\%$ for events with $\muhat < 60$, improving by a factor of two compared to the previous algorithm.
 
The conversion vertices are then matched to the EM
topo-clusters.\footnote{If the conversion vertex has tracks with
silicon hits, a conversion vertex is considered matched if, after
extrapolation, the tracks match the cluster to within
$|\Delta\eta| < 0.05$ and
$|\Delta\phi| < 0.05$. If the conversion vertex is made of only TRT
tracks, then if the first track is in the TRT barrel, a match
requires $|\Delta\eta| < 0.35$ and $|\Delta\phi| < 0.02$, and if the
first track is in the TRT endcap, a match requires $|\Delta\eta| < 0.2$ and $|\Delta\phi| < 0.02$.}
If there are multiple conversion vertices matched to a cluster,
double-track conversions with two silicon tracks are preferred over
other double-track conversions, followed by single-track
conversions. Within each category, the vertex with the smallest
conversion radius is preferred.

\subsection{Supercluster reconstruction}
 
The reconstruction of electron and photon superclusters proceeds
independently, each in two stages: in the first stage, EM topo-clusters
are tested for use as seed cluster candidates, which form the basis
of superclusters; in the second stage, EM topo-clusters near the seed
candidates are identified as satellite cluster candidates, which may
emerge from bremsstrahlung radiation or topo-cluster
splitting. Satellite clusters are added to the seed candidates to form
the final superclusters if they satisfy the necessary selection
criteria.
 
The steps to build superclusters proceed as follows. The initial list
of EM topo-clusters is sorted according to descending \et,
calculated using the EM energy.\footnote{An exception to the \et\
ordering is made for clusters in the transition region that fail the
standard selection but pass a looser selection; these are added at
the end.} The clusters are tested one by one in the sort order for use as seed clusters. For a
cluster to become an electron supercluster seed, it is required to
have a minimum \et of 1 \gev\ and must be matched to a track with
at least four hits in the silicon tracking detectors. For photon
reconstruction, a cluster must have \et greater than 1.5 \gev\
to qualify as a supercluster seed, with no requirement made on any
track or conversion vertex matching. A cluster cannot be used as a
seed cluster if it has already been added as a satellite cluster to
another seed cluster.
 
If a cluster meets the seed cluster requirements, the algorithm attempts to find satellite
clusters, using the process summarized in
Figure~\ref{fig:supercluster_algo_cartoon_common}.
\begin{figure}
\begin{center}
\includegraphics[width=0.8\columnwidth]{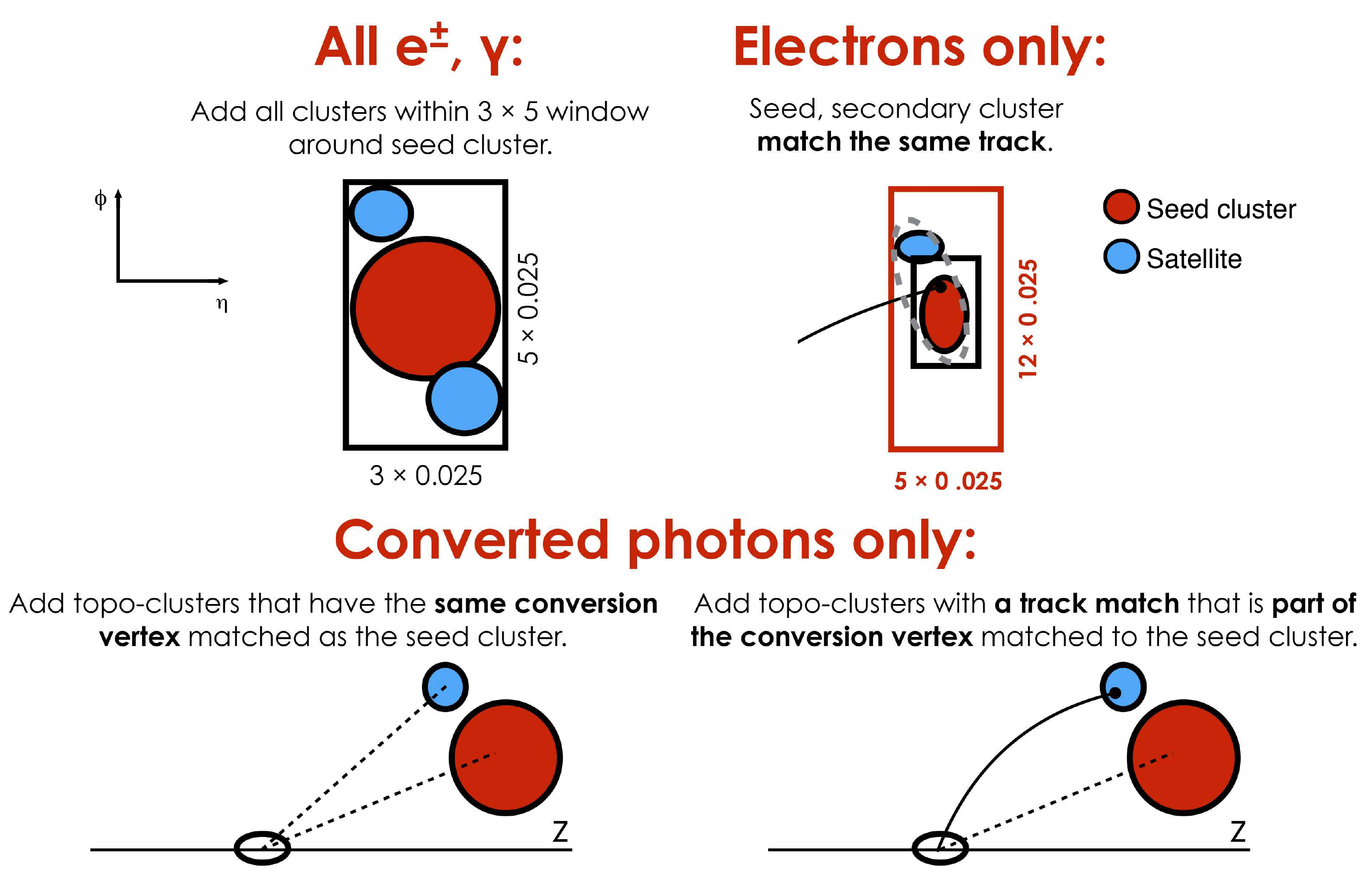}
\end{center}
\caption{Diagram of the superclustering algorithm for electrons and
photons. Seed clusters are shown in red, satellite clusters in blue.}
\label{fig:supercluster_algo_cartoon_common}
\end{figure}
For both electrons and photons, a cluster is considered a satellite if
it falls within a window of
$\Delta\eta \times \Delta\phi = 0.075 \times 0.125$ around the seed
cluster barycentre, as these cases tend to represent secondary EM
showers originating from the same initial electron or photon. For
electrons, a cluster is also considered a satellite if it is within a
window of $\Delta\eta \times \Delta\phi = 0.125 \times 0.300$ around the
seed cluster barycentre, and its `best-matched' track is also the
best-matched track for the seed cluster. For photons with conversion
vertices made up only of tracks containing silicon hits, a cluster is
added as a satellite if its best-matched (electron) track belongs to the conversion vertex matched to the seed cluster. These
steps rely on tracking information to discriminate distant radiative
photons or conversion electrons from pile-up noise or other unrelated
clusters.
 
The seed clusters with their associated satellite clusters are called
superclusters. The final step in the supercluster-building algorithm
is to assign calorimeter cells to a given supercluster.  Only
cells from the presampler and the first three LAr calorimeter layers are considered, except in the
transition region of $1.4 < |\eta| < 1.6$, where the energy measured
in the scintillator between the calorimeter cryostats is also added.
To limit the superclusters' sensitivity to pile-up noise, the size of
each constituent topo-cluster is restricted to a maximal width of
0.075 or 0.125 in the $\eta$ direction in the barrel or endcap
region, respectively. Because the magnetic field in the ID
is parallel to the beam-line, interactions between the
electron or photon and detector material generally cause the EM shower
to spread in the $\phi$ direction, so the restriction in $\eta$ still
generally allows the electron or photon energy to be captured. No
restriction is applied in the $\phi$-direction.

\subsection{Creation of electrons and photons for analysis}
 
After the electron and photon superclusters are built, an initial
energy calibration and position correction is applied to them, and
tracks are matched to electron superclusters and conversion vertices
to photon superclusters. The matching is performed the same way that
the matching to EM topo-clusters was performed, but using
the superclusters instead. Creating the
analysis-level electrons and photons follows. Because electron and
photon superclusters are built independently, a given seed cluster
can produce both an electron and a photon. In such cases, the
procedure presented in Figure~\ref{fig:ambiguity} is applied. The
purpose is that if a particular object can be easily identified only as
a photon (a cluster with no good track attached) or only as an electron
(a cluster with a good track attached and no good photon conversion vertex), then
only a photon or an electron object is created for analysis; otherwise, both an electron and a photon object
are created. Furthermore, these cases are marked explicitly as
ambiguous, allowing
the final classification of these objects to be determined based upon the specific requirements of each analysis.
\begin{figure}
\begin{center}
\includegraphics[width=\columnwidth]{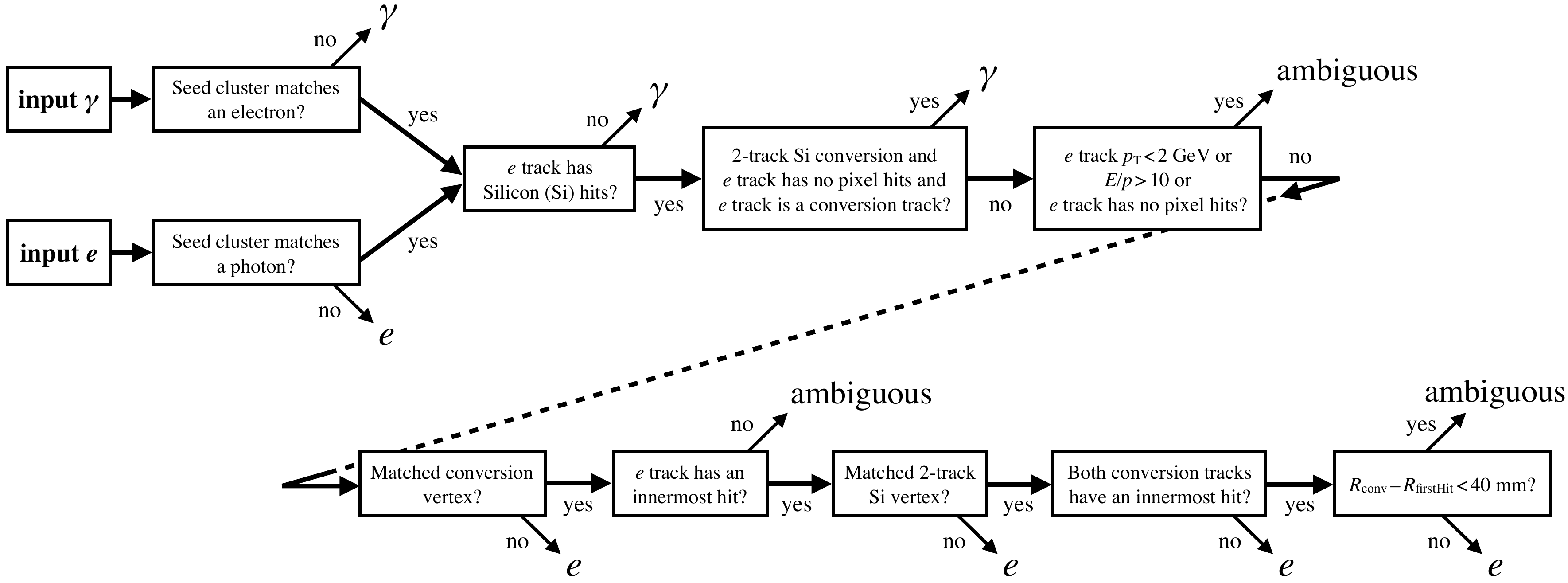}
\end{center}
\caption{Flowchart showing the logic of the ambiguity resolution for particles initially reconstructed both as electrons and photons. An `innermost hit' is a hit in the functioning pixel
nearest to the beam-line along the track trajectory, $E/p$ is
the ratio of the supercluster energy to the measured momentum of the matched track, $R_\mathrm{conv}$ is the radial position of the
conversion vertex, and $R_\mathrm{firstHit}$ is
the smallest radial position of a hit in the track or tracks that make
a conversion vertex.}
\label{fig:ambiguity}
\end{figure}
 
Because the energy calibration depends on matched tracks and
conversion vertices, and the initial supercluster calibration is
performed before the final track and conversion matching,
the energies of the electrons and photons are recalibrated,
following the procedure described in Ref.~\cite{PERF-2017-03}.
 
Subsequently, shower shape and other discriminating variables~\cite{PERF-2017-01, PERF-2017-02}
are calculated for electron and photon identification. A list is given
in Table~\ref{tab:IDcuts}, along with an indication if they are used
for electron or photon identification. The lateral shower shapes are based on the position of
the most energetic cell, so they are independent of the clustering used,
provided the same most energetic cell is included in the clusters.
More information about the
variables and the identification methods are given in
Sections~\ref{sec:eID} and \ref{sec:pID} for electrons and photons,
respectively.
\begin{table*}[tp]
\caption{Discriminating variables used for electron and photon
identification.  The usage column indicates if the variables are
used for the identification of electrons, photons, or both.
For variables calculated in the first EM layer, if the cluster has
more than one cell in the $\phi$ direction at a given $\eta$,
the two cells closest in $\phi$ to the cluster barycentre are merged
and the definitions below are given in terms of this merged cell.
The sign of \trackdO is conventionally chosen such that the coordinates of the perigee in the transverse plane
are $(x_0,y_0) = (-\trackdO \sin\phi, \trackdO \cos\phi)$, where $\phi$ is the azimuthal angle of the track momentum at the perigee. }
\def\arraystretch{1.3}
\centering
\footnotesize
\begin{tabular}{
l
>{\RaggedRight}p{0.55\textwidth}
lc}
\hline \hline
Category & Description & Name & Usage \\
\hline \hline
Hadronic leakage
& Ratio of $E_\mathrm{T}$ in the first layer of the hadronic calorimeter to $E_\mathrm{T}$ of the
EM cluster (used over the ranges $|\eta|<0.8$ and $|\eta|>1.37$)  & $\Rhadone$ & $e/\gamma$  \\
& Ratio of $E_\mathrm{T}$ in the hadronic calorimeter to $E_\mathrm{T}$ of the EM cluster
(used over the range $0.8<|\eta|<1.37$)  & $\Rhad$  & $e/\gamma$ \\
 
EM third layer
& Ratio of the energy in the third layer to the total energy in the EM calorimeter & \fIII & $e$ \\
 
EM second layer
& Ratio of the sum of the energies of
the cells contained in a $3\times7$ $\eta\times\phi$
rectangle (measured in cell units)  to the sum of
the cell energies in a $7\times7$ rectangle, both centred around the
most energetic cell & $\Reta$ & $e/\gamma$ \\
& Lateral shower width, $\sqrt{(\Sigma E_i \eta_i^2)/(\Sigma E_i)
-((\Sigma E_i\eta_i)/(\Sigma E_i))^2}$, where $E_i$ is the energy
and $\eta_i$ is the pseudorapidity of cell $i$ and the sum is calculated within a window of $3\times5$ cells
& $\wetatwo$ & $e/\gamma$ \\
& Ratio of the sum of the energies of
the cells contained in a $3\times3$ $\eta\times\phi$
rectangle (measured in cell units) to the sum of
the cell energies in a $3\times7$ rectangle, both centred around the
most energetic cell  & $\Rphi$ & $e/\gamma$ \\
 
EM first layer
& Total lateral shower width, $\sqrt{(\Sigma E_i
(i-i_\mathrm{max})^2)/(\Sigma E_i)}$, where $i$ runs over all
cells in a window of $\Delta\eta \approx 0.0625$ and
$i_{\textrm{max}}$ is the index of the highest-energy cell
& $\wtot$ & $e/\gamma$ \\
& Lateral shower width,
$\sqrt{(\Sigma E_i (i - i_{\textrm{max}})^2)/(\Sigma E_i)}$,
where $i$ runs over all cells in a window of 3 cells around the
highest-energy cell & $\wthree$ & $\gamma$ \\
& Energy fraction outside core of three central cells, within seven cells   & $\Fside$ & $\gamma$ \\
& Difference between the energy of the cell associated with the
second maximum, and the energy reconstructed
in the cell with the smallest value found between the first and
second maxima  & $\DeltaE$ & $\gamma$ \\
& Ratio of the energy difference between the maximum energy deposit and the energy deposit in a secondary maximum in the cluster to the sum of these energies   & $\Eratio$ & $e/\gamma$ \\
& Ratio of the energy measured in the first layer of the electromagnetic calorimeter to the total energy of the
EM cluster & $\fI$ & $e/\gamma$ \\
 
Track conditions
& Number of hits in the innermost pixel layer &   $n_\mathrm{innermost}$ & $e$ \\
& Number of hits in the pixel detector        &    $n_\mathrm{Pixel}$ & $e$ \\
& Total number of hits in the pixel and SCT detectors  &   $n_{\mathrm{Si}}$  & $e$ \\
& Transverse impact parameter relative to the beam-line &  \trackdO  & $e$ \\
& Significance of transverse impact parameter defined as
the ratio of \trackdO to its uncertainty &  |\dOSignificance|  & $e$  \\
&  Momentum lost by the track between the perigee and the last measurement point divided by
the momentum at perigee& \deltapoverp & $e$ \\
& Likelihood probability based on transition radiation in the TRT &   \TRTPID & $e$  \\
Track--cluster matching
& $\Delta\eta$ between the cluster position in the first layer
of the EM calorimeter and the extrapolated track &   \deltaeta & $e$  \\
& $\Delta\phi$ between the cluster position in the second layer
of the EM calorimeter and the momentum-rescaled track,
extrapolated from the perigee, times the charge $q$ & \deltaphires & $e$  \\
&  Ratio of the cluster energy to the measured track momentum  &       $E/p$   &  $e$ \\
 
\hline \hline
\end{tabular}
\label{tab:IDcuts}
\end{table*}

\subsection{Performance}
 
Figure~\ref{fig:truthElEff} shows the reconstruction
efficiencies for electrons. The reconstruction efficiency at high
\pt\ approaches the tracking efficiency, as expected. One interesting feature, however, is the difference between
the efficiency to reconstruct the cluster and track (green triangles) and
the efficiency to reconstruct an electron (purple inverted triangles) at lower
\pt. The reason for this is that tracks with silicon hits are
considered for matching to superclusters only if they have had a GSF
re-fit performed. The fixed-size clusters used for choosing the tracks on which
the GSF re-fit is performed introduce an \et\ threshold, which is the source of this
inefficiency.
To alleviate this feature, the EM topo-clusters as
defined in Section~\ref{sec:topoclusters} could be used to seed the GSF fit.
\begin{figure}
\begin{center}
\includegraphics[width=0.7\columnwidth]{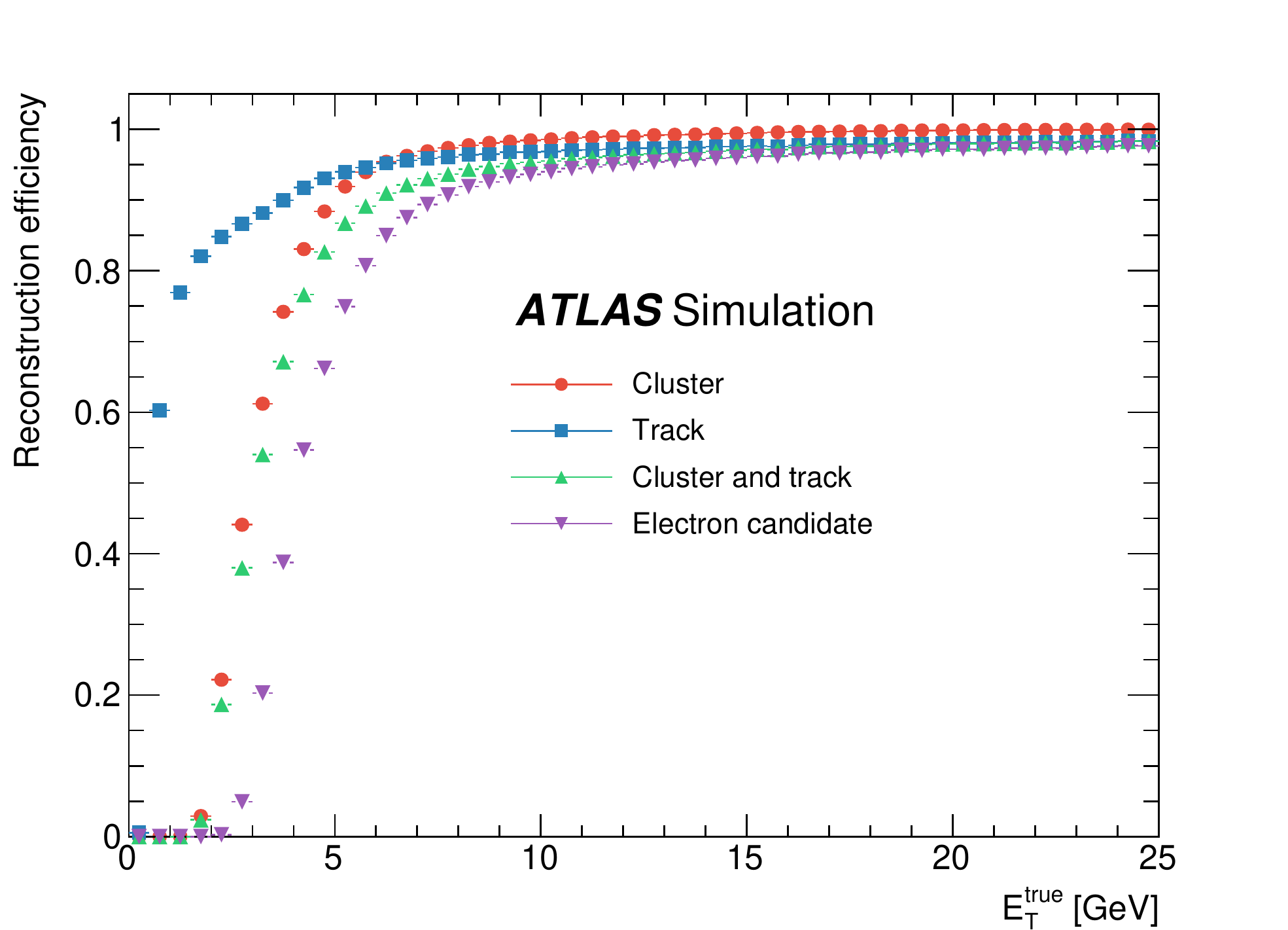}
\end{center}
\caption{The cluster, track, cluster and track, and electron
reconstruction efficiencies as a function of the generated electron \et.}
\label{fig:truthElEff}
\end{figure}
 
The top plot in Figure~\ref{fig:conv_mu_plots} shows the
reconstruction efficiency for converted photons as a function of the
true \et of the simulated photon for the previous version
of the reconstruction software, described in Ref.~\cite{PERF-2017-02},
and the current
version, described in Section~\ref{sec:tracking}, along with the
contributions of the different conversion types. For a photon to be
classified as a true converted photon, the true radius of the conversion
must be smaller than $800\,\mathrm{mm}$.  Only simulated photons with
transverse energy greater than 20 \gev\ are considered. The simulated
photons are distributed uniformly in $\feta$, with most of the photons
having a transverse momentum smaller than 200 \gev. The bottom left
plot of Figure~\ref{fig:conv_mu_plots} shows the reconstruction
efficiency for converted photons along with the contributions of the
different conversion types as a function of \muhat.  The improvement (see Section~\ref{sec:tracking})
in the reconstruction efficiency for double-track Si conversions and
the corresponding reduction of single-track Si conversions is clearly
visible in those two plots. A slight reduction in double- and single-track TRT
conversion efficiency is also visible, with the purpose of significantly reducing the probability for true
unconverted photons to be reconstructed as TRT conversions, as can be
seen in the bottom right plot of Figure~\ref{fig:conv_mu_plots}. The probability for true unconverted photons to be reconstructed as Si conversions is negligible in comparison.
\begin{figure}
\begin{center}
\includegraphics[width=0.49\columnwidth]{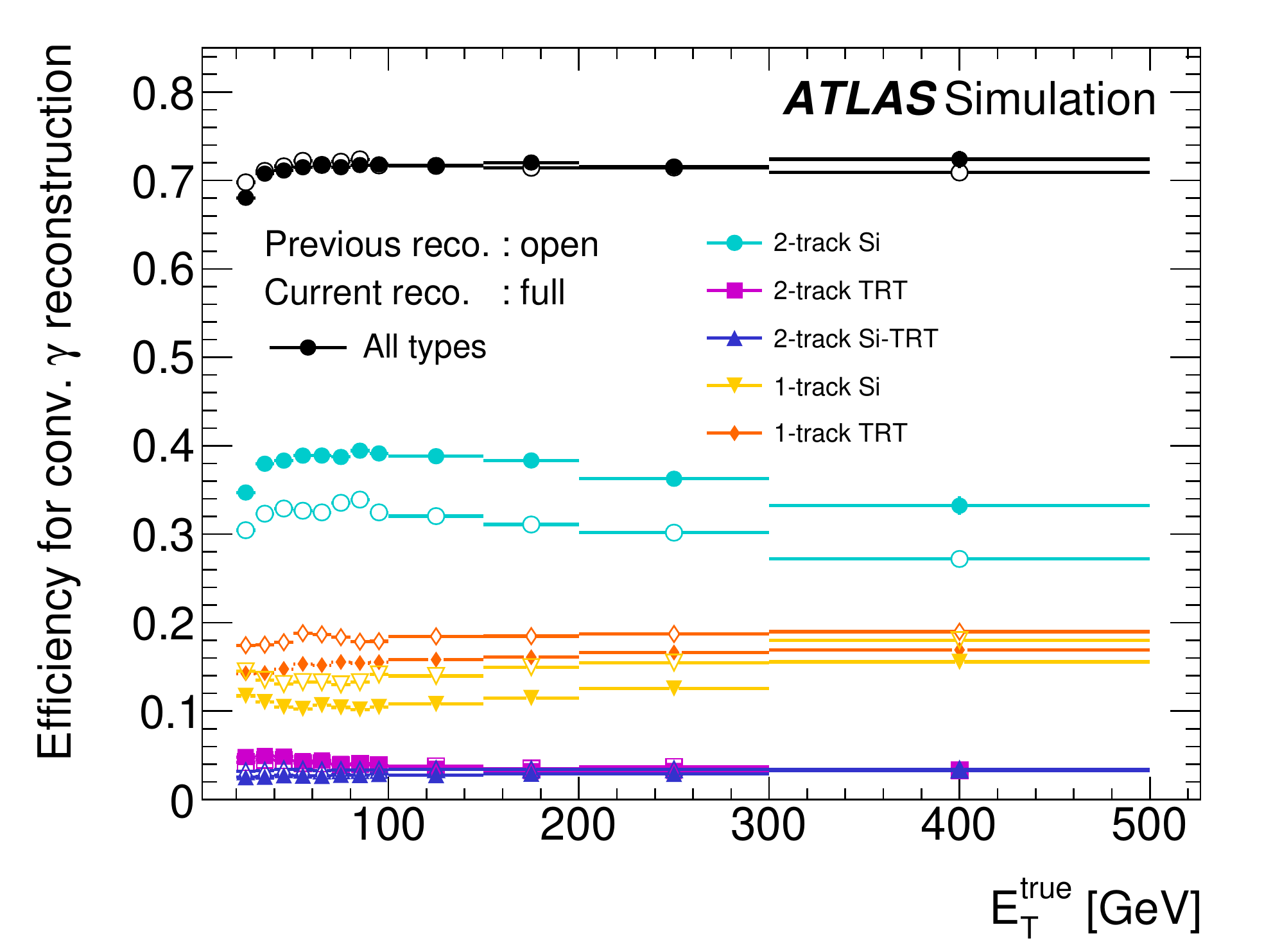}\\
\includegraphics[width=0.49\columnwidth]{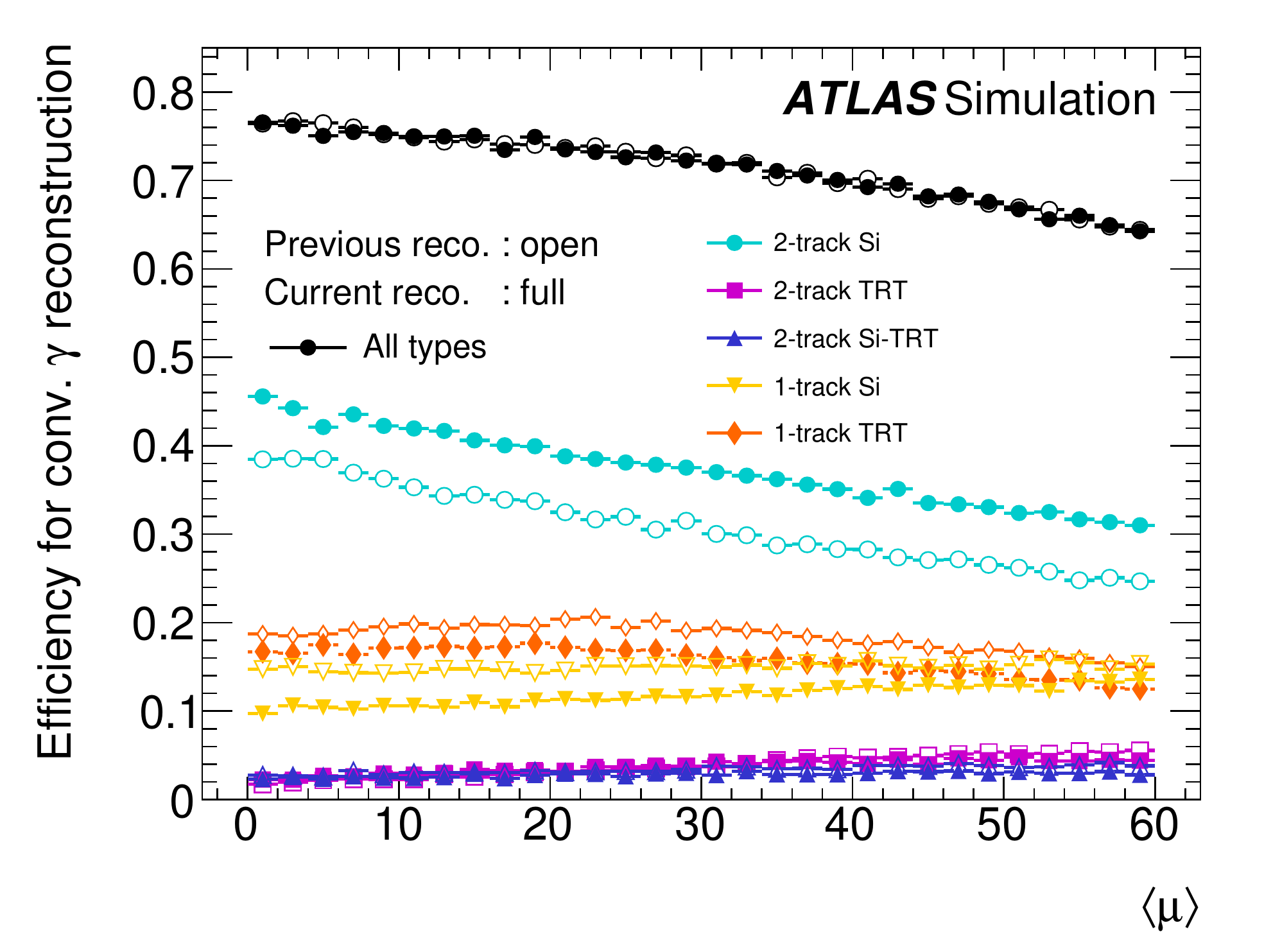}
\includegraphics[width=0.49\columnwidth]{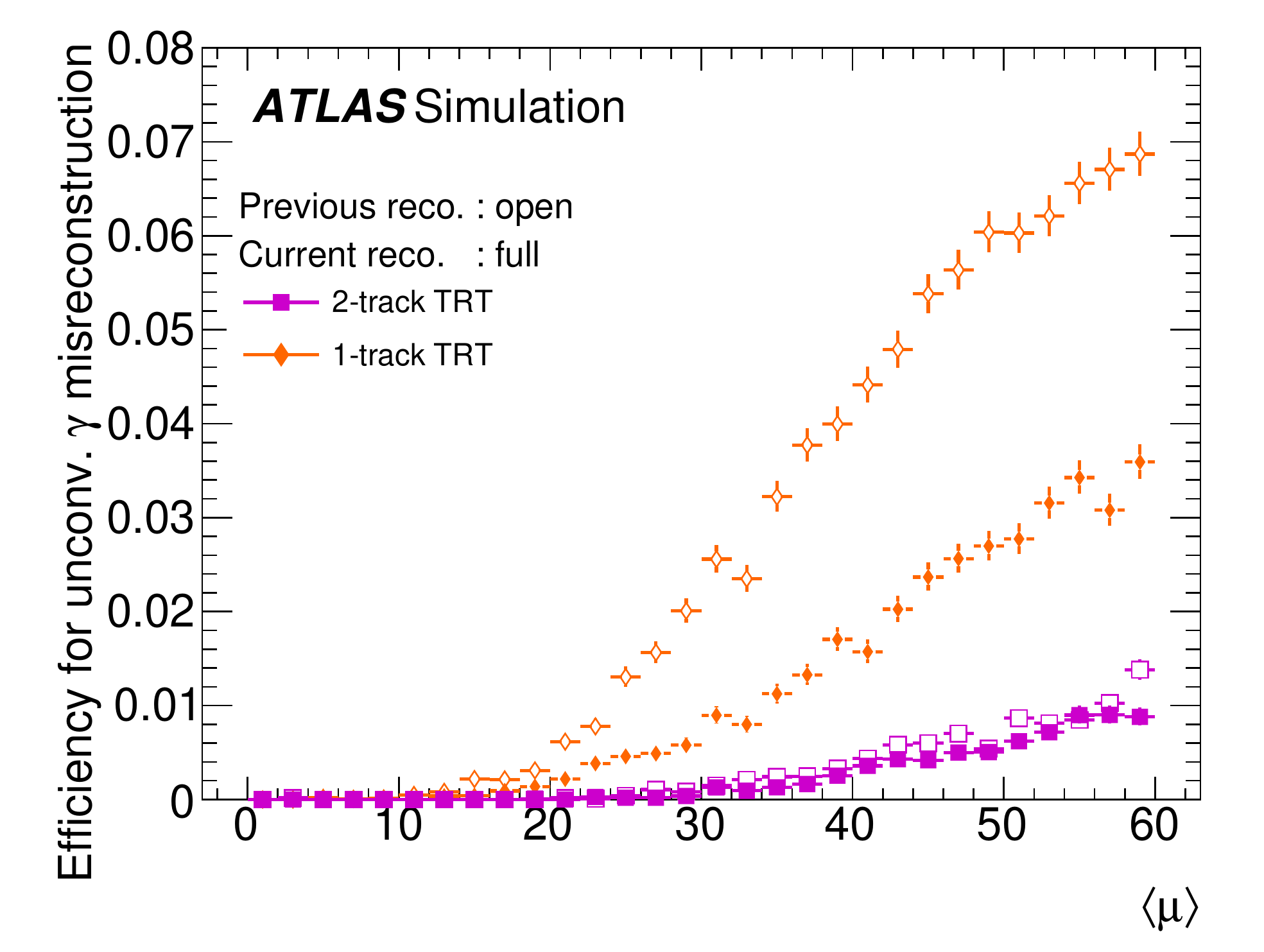}
\end{center}
\vspace{-0.5cm}
\caption{The top plot shows the converted photon reconstruction efficiency and contributions of the
different conversion types as a function of
$E_\text{T}^\text{true}$, averaged over \muhat\ for a uniform \muhat\ distribution between
0 and 60.
On the bottom, efficiency of the reconstruction of converted photons and contributions of the
different conversion types (left), and the probability of an unconverted photon to be
mistakenly reconstructed as a converted photon and contributions of the different
conversions types (right), both as a function of \muhat. }
\label{fig:conv_mu_plots}
\end{figure}
 
An important reason for using superclusters is the improved energy
resolution that superclusters provide by collecting more of the deposited
energy. The peaks of the energy response, $E_\text{calib} / \etrue$,
where $\etrue$ is the true energy of the simulated particle prior to
any detector simulation, and $E_\text{calib}$ is the calibrated
reconstructed energy,
do not deviate from one by more than $0.5\%$ for the different
particles. To quantify the width (resolution) of the energy response,
the \emph{effective interquartile range} is used, defined as
\begin{equation}
\text{IQE} = \frac{Q_3 - Q_1}{1.349},
\nonumber
\end{equation}
where $Q_1$ and $Q_3$ are the first and third quartiles of the distribution of $E_\text{calib} / \etrue$,
and the normalization factor is chosen such that the IQE of a Gaussian
distribution would equal its standard deviation.
 
Comparisons of the resolutions of the calibrated energy response of simulated single
electrons, converted photons, and unconverted photons, built using
fixed-size clusters and superclusters, are given in
Figure~\ref{fig:cal_iqe_results}. In particular,
Figure~\ref{fig:cal_iqe_results} shows the IQE of the two approaches in
different regions of $|\eta_\text{true}|$ and
$E_\text{T}^\text{true}$. The reconstructed electrons and photons in
these distributions are required to correspond to true primary
electrons and photons and to satisfy loose identification
requirements. After calibration, the supercluster algorithm shows a
significant improvement in resolution compared with the sliding-window
algorithm for electrons. In absence of pile-up, an improvement in resolution
of up to $20$--$30\%$ is found in some bins in the endcap region of
the detector, as well as in the central region for low-\et electrons.
\begin{figure}
\centering
\includegraphics[width=0.49\columnwidth]{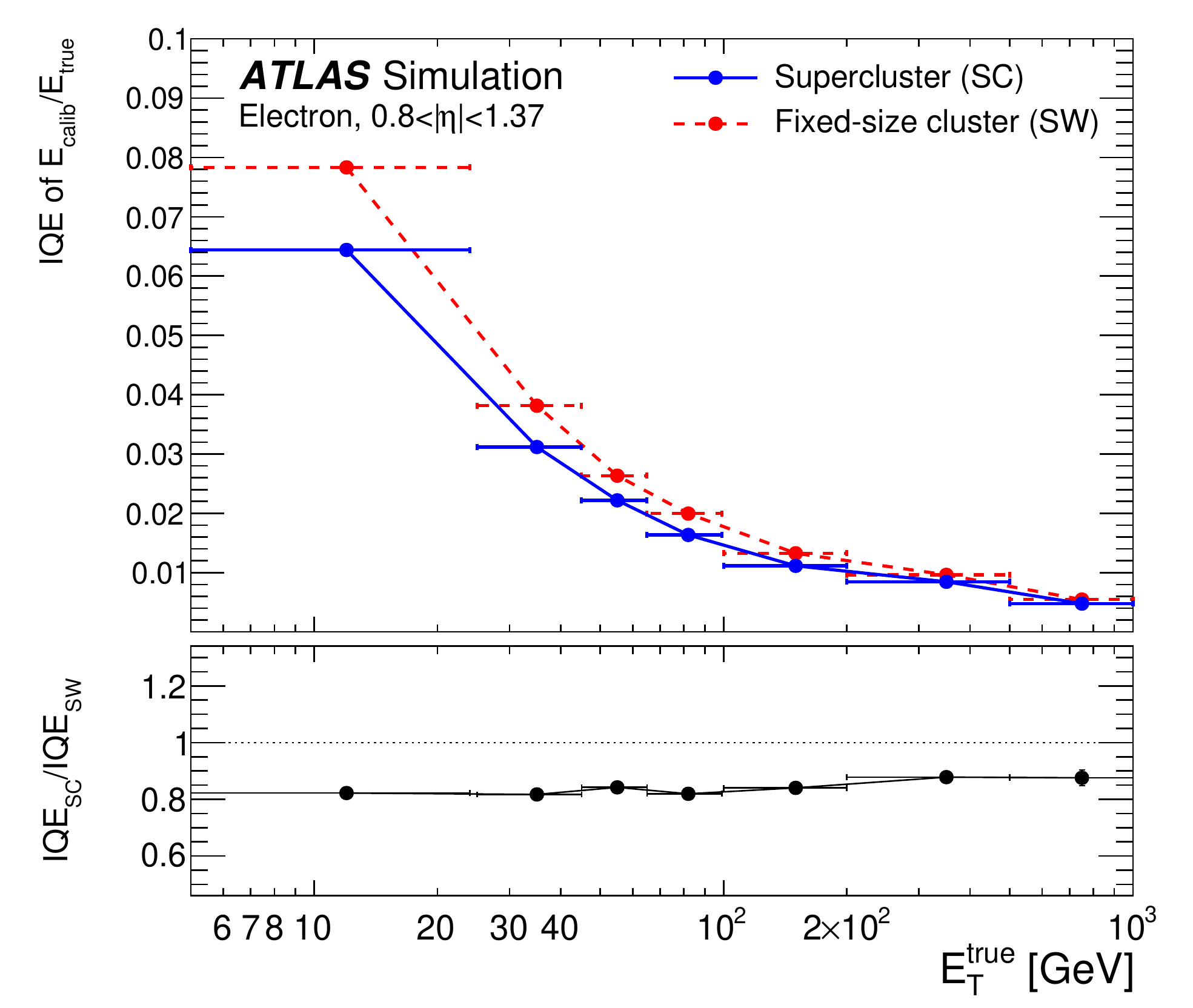}
\includegraphics[width=0.49\columnwidth]{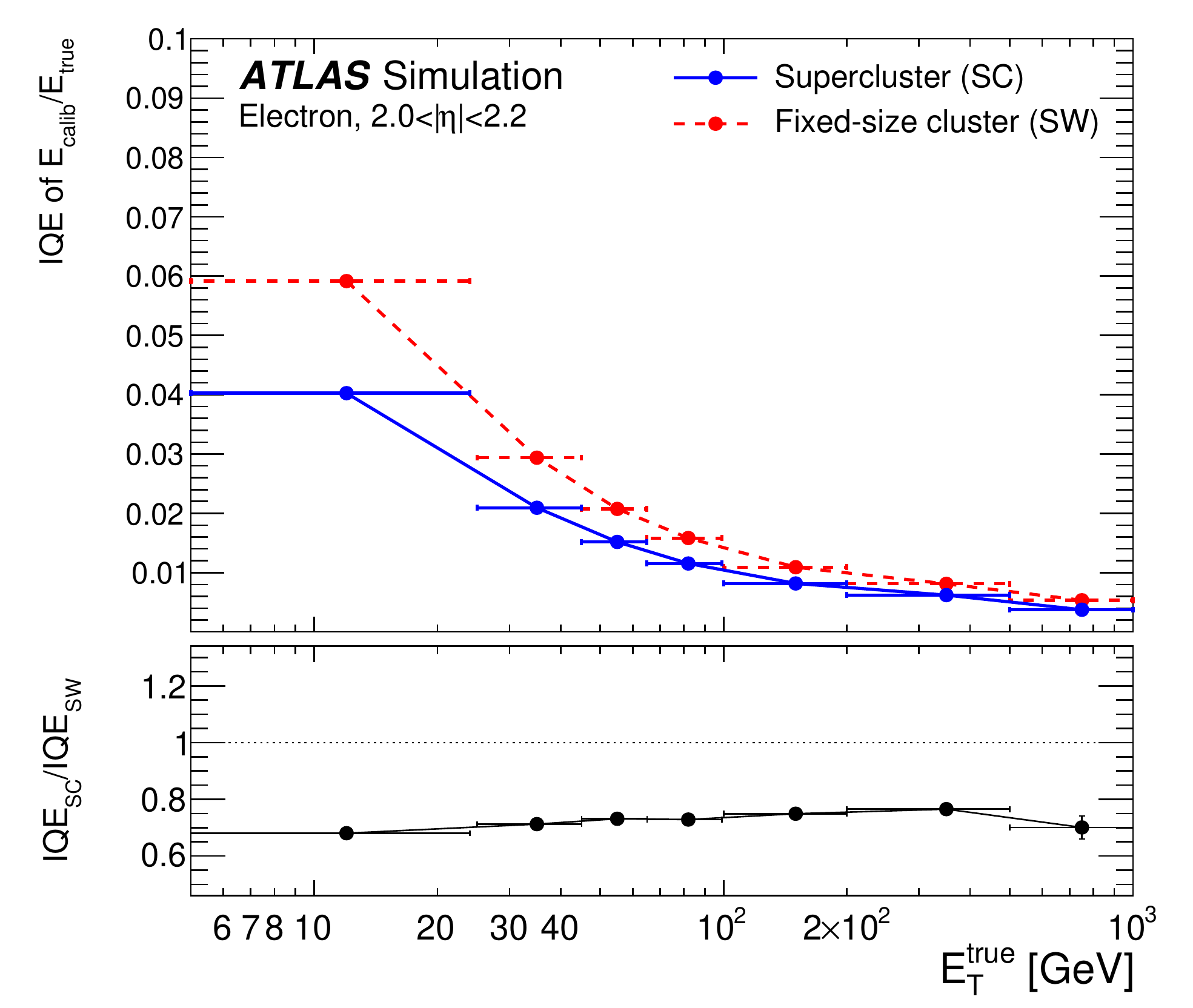}
\includegraphics[width=0.49\columnwidth]{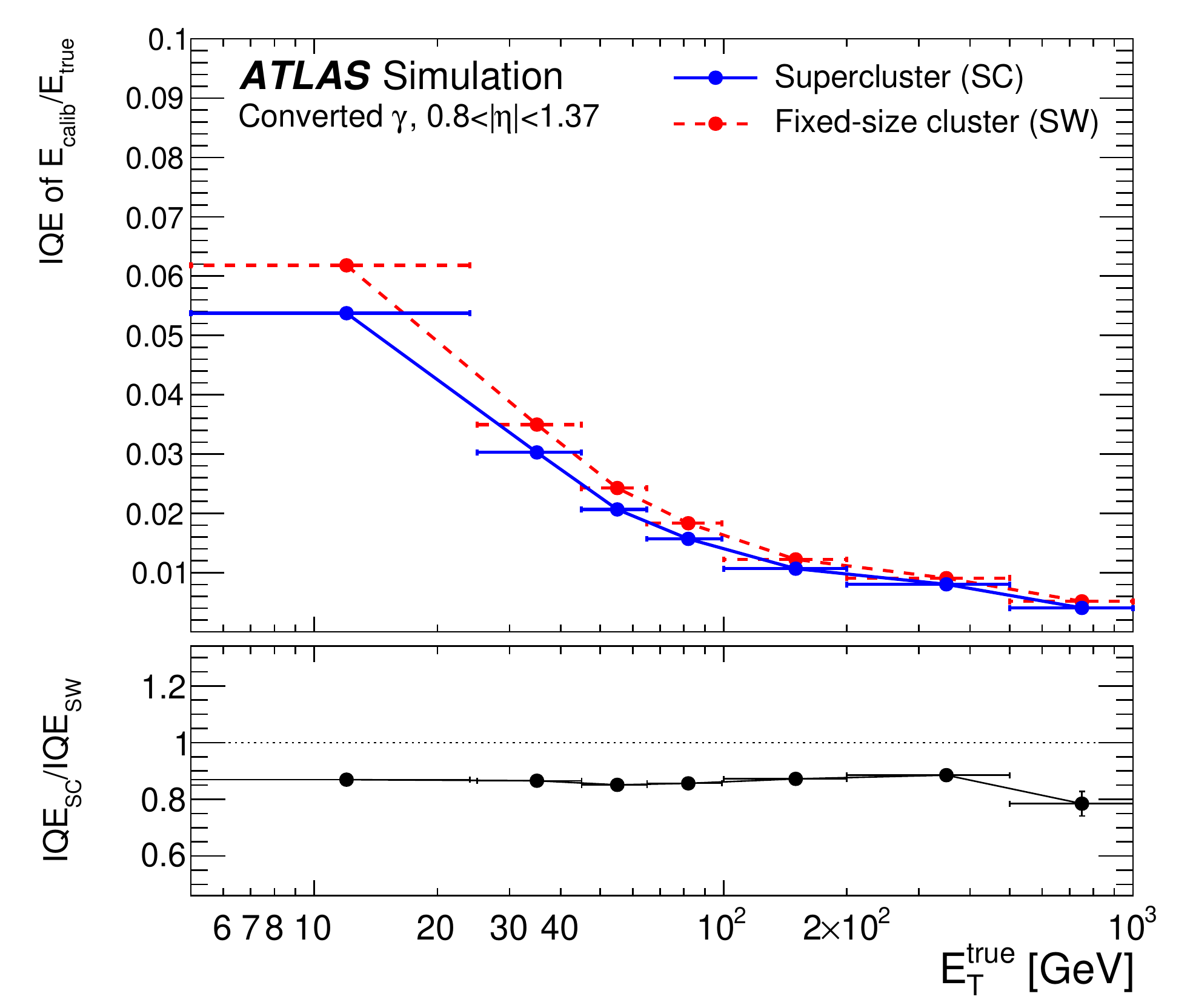}
\includegraphics[width=0.49\columnwidth]{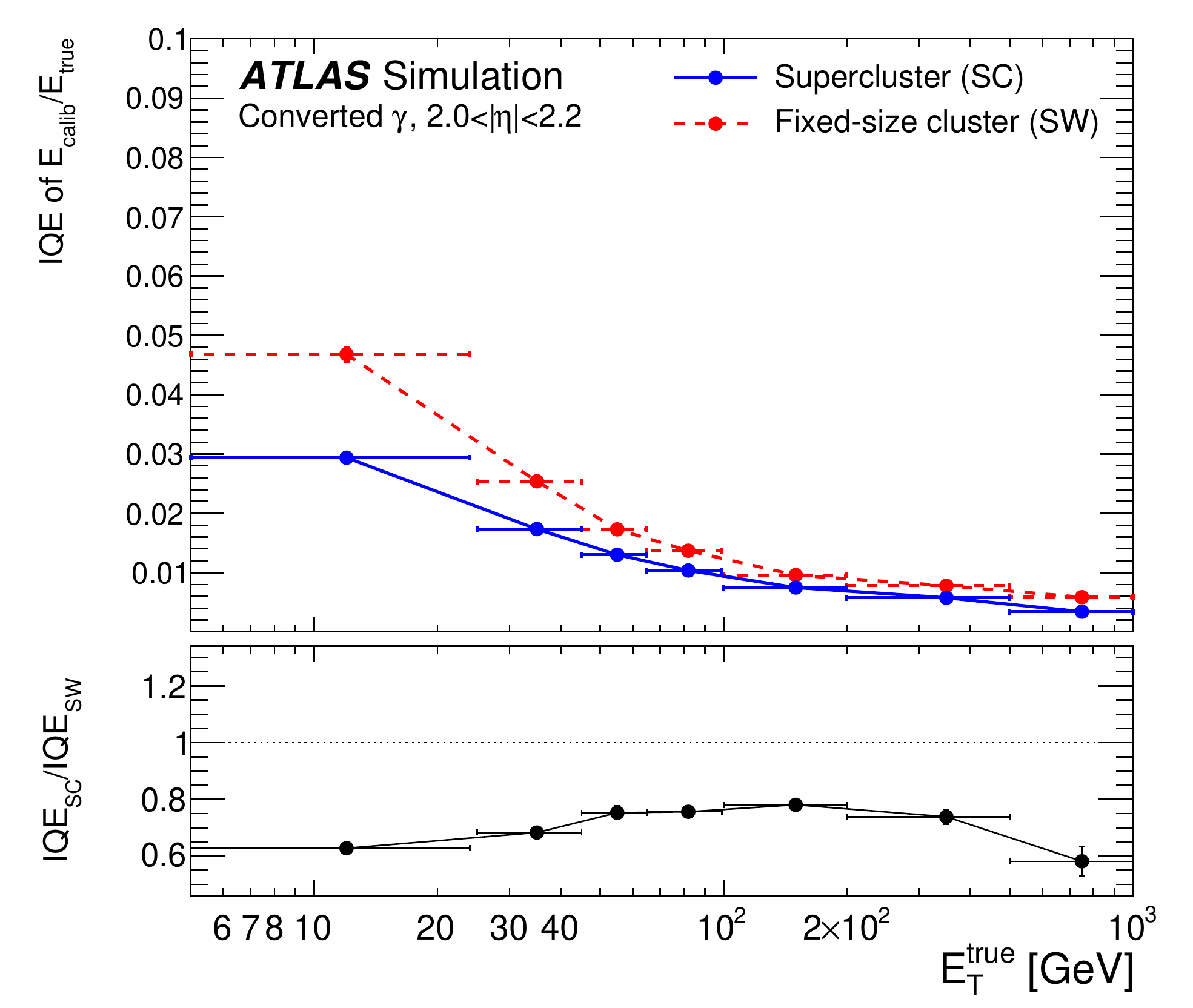}
\includegraphics[width=0.49\columnwidth]{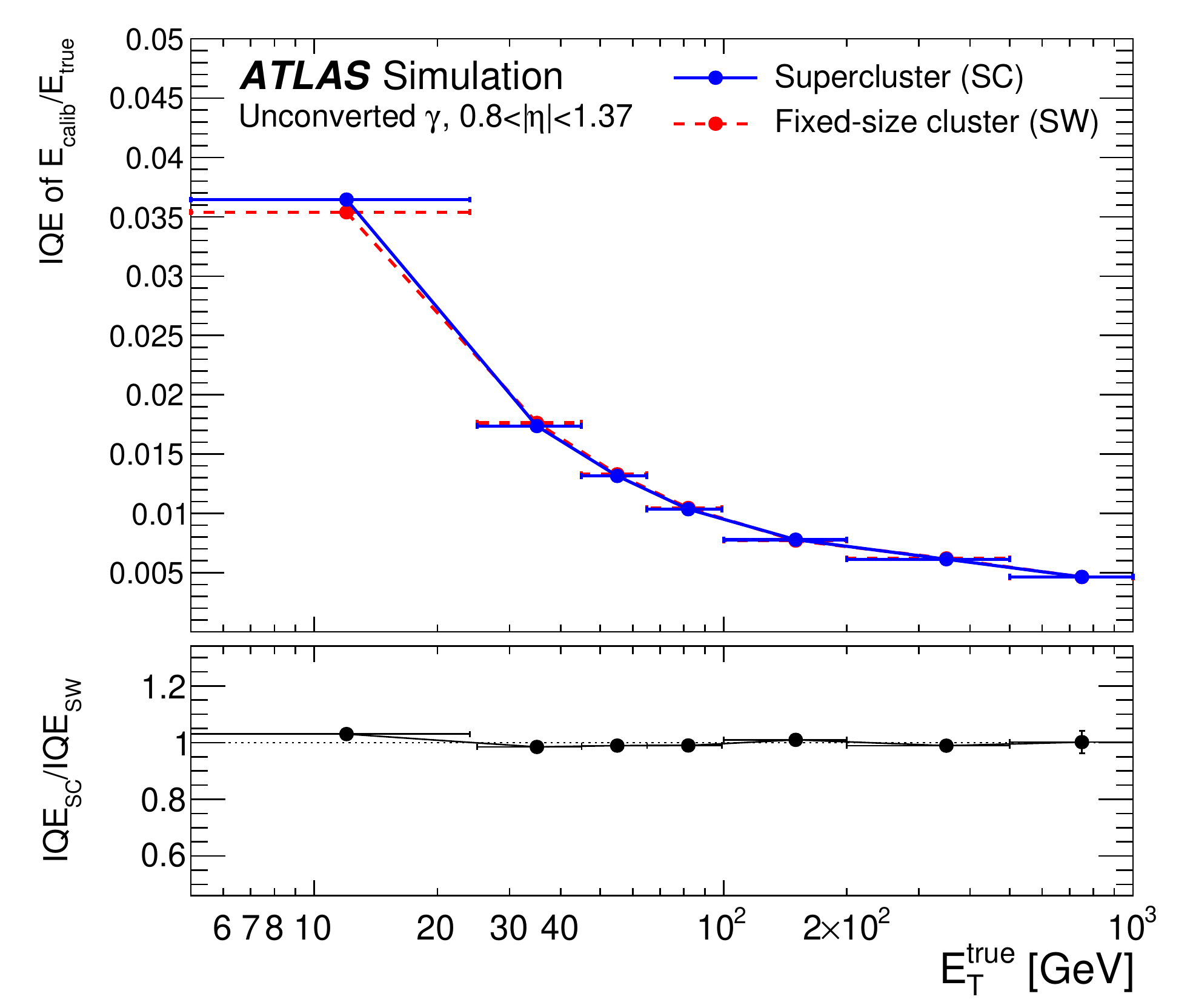}
\includegraphics[width=0.49\columnwidth]{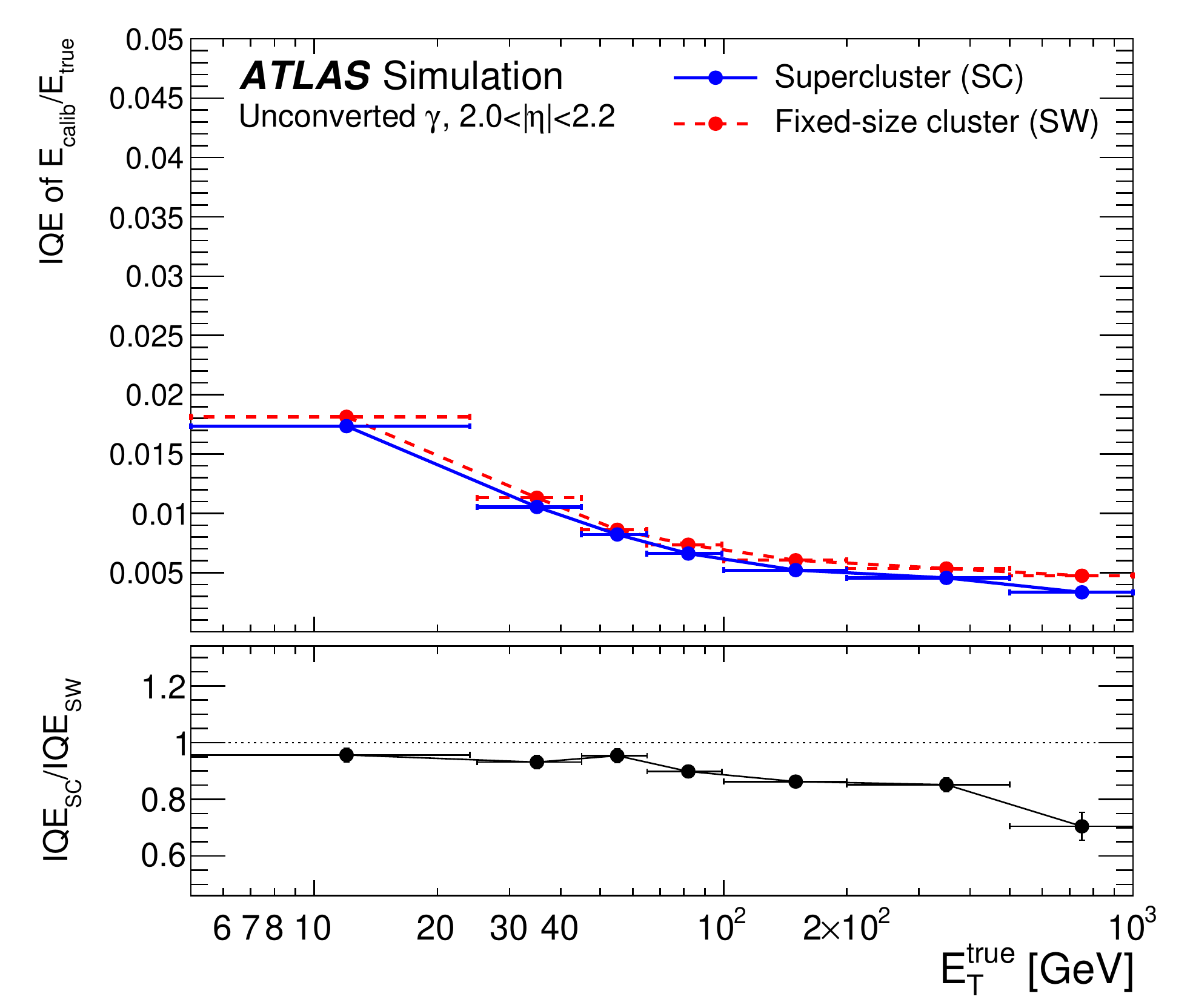}
\caption{Calibrated energy response resolution, expressed in terms of
IQE, for electrons (top), converted photons (middle), and
unconverted photons (bottom) simulated with $\muhat = 0$. Two representative pseudorapidity ranges are shown.
The response resolution for fixed-size clusters based on the sliding window method is shown in dashed red, while the
supercluster-based response resolution is shown in full blue. For all plots, the bottom panel shows the ratios between the IQE
obtained using the supercluster reconstruction and using the sliding window method. }
\label{fig:cal_iqe_results}
\end{figure}
Similarly, a large improvement in the resolution is seen for converted
photons, over $30\%$ in a few bins.  For unconverted photons, the
overall change in performance is small, due to the generally narrower
shower width. However, some improvement is observed for high \et bins in the endcap region.
In presence of pile-up, the improvement in resolution still reaches 15 to 20\%, depending on $\eta$ and \et.

An important consideration is the performance of the supercluster
reconstruction at different pile-up
levels. Figure~\ref{fig:pu_slices_iqe} shows the calibrated energy
response resolution at different \muhat\ levels for electrons,
converted photons, and unconverted photons, in two $|\eta|$ regions.
\begin{figure}[hp]
\centering
\includegraphics[width=0.45\columnwidth]{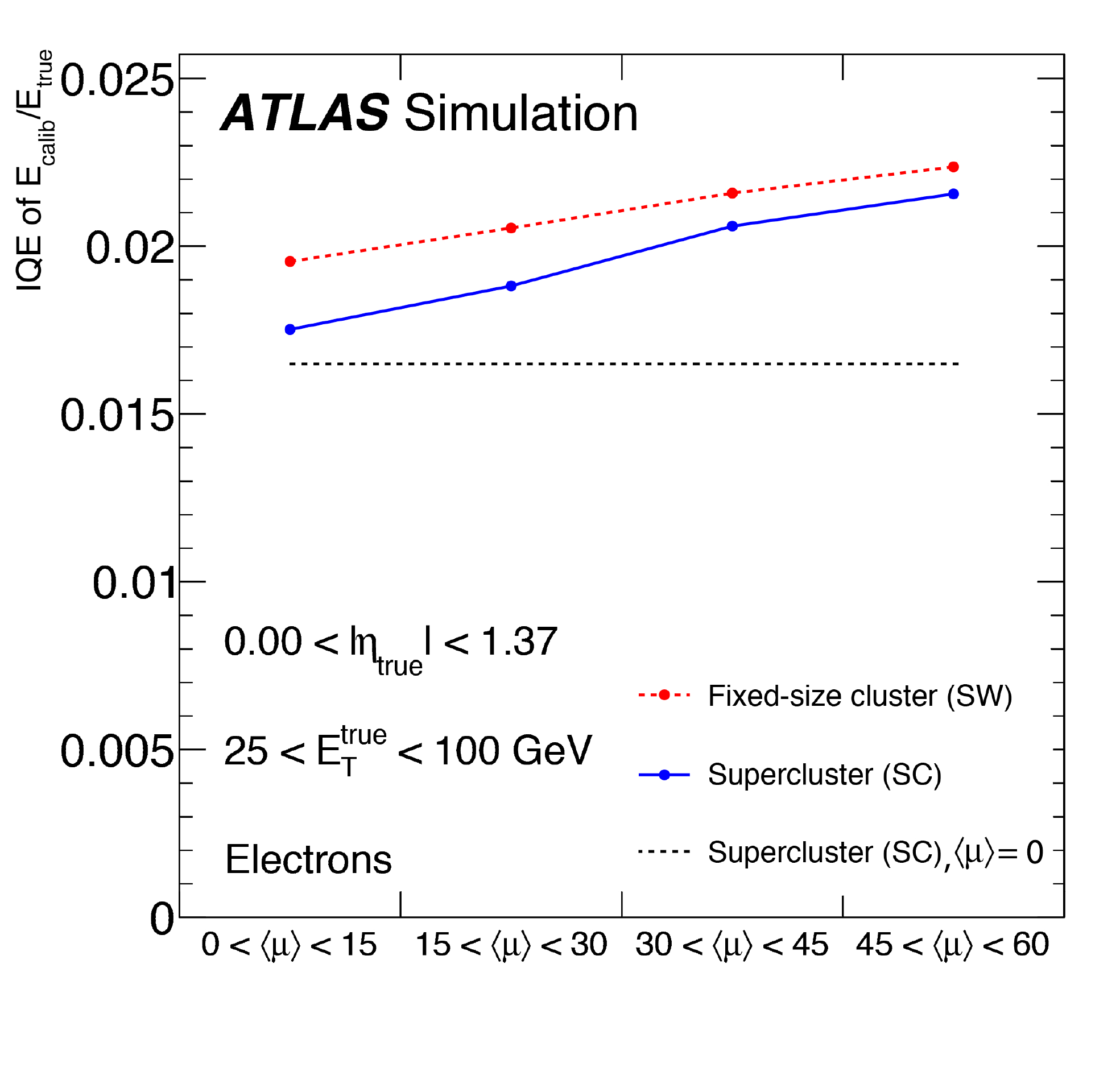}
\includegraphics[width=0.45\columnwidth]{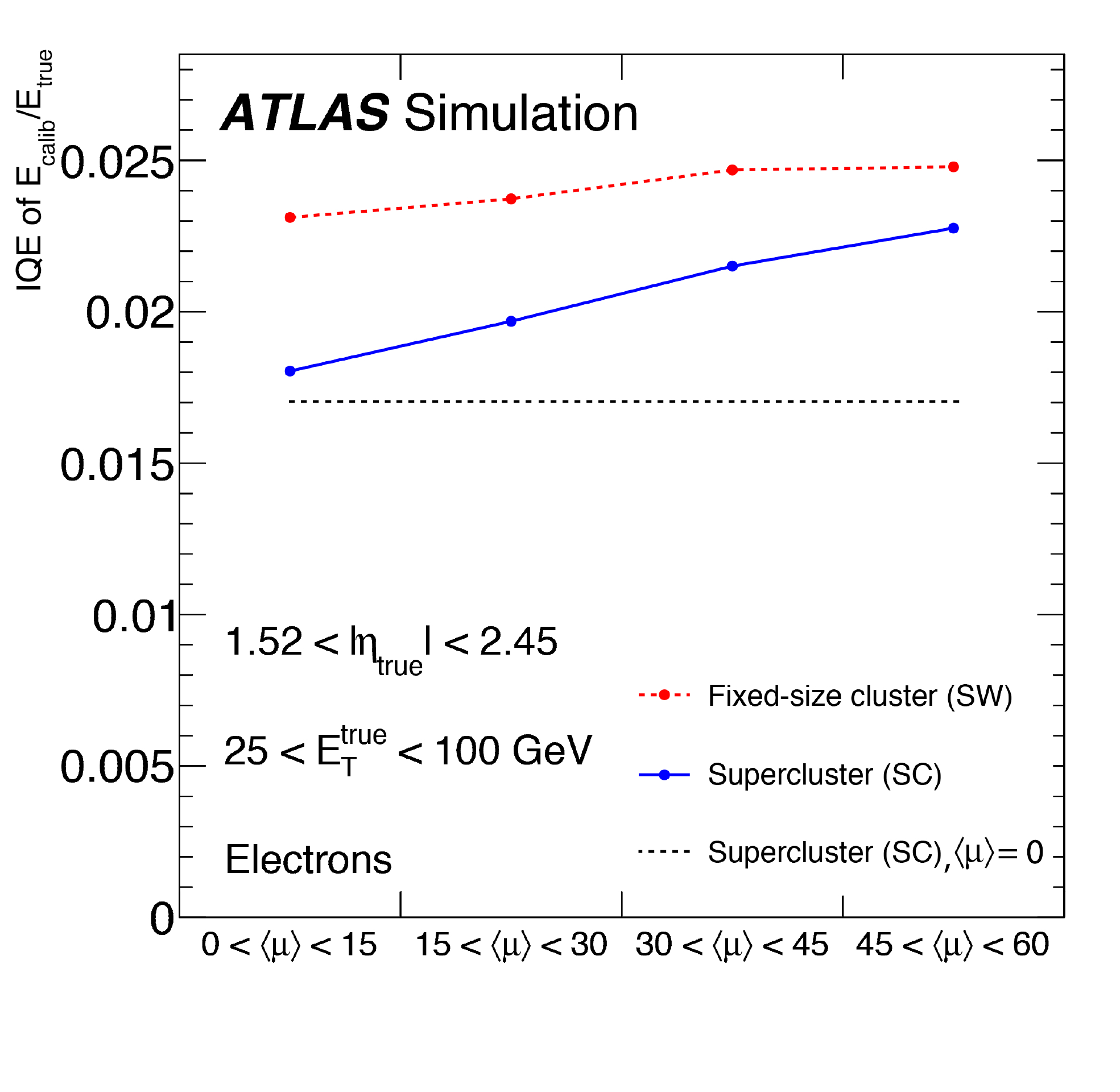}
\vspace{-0.2cm}
\includegraphics[width=0.45\columnwidth]{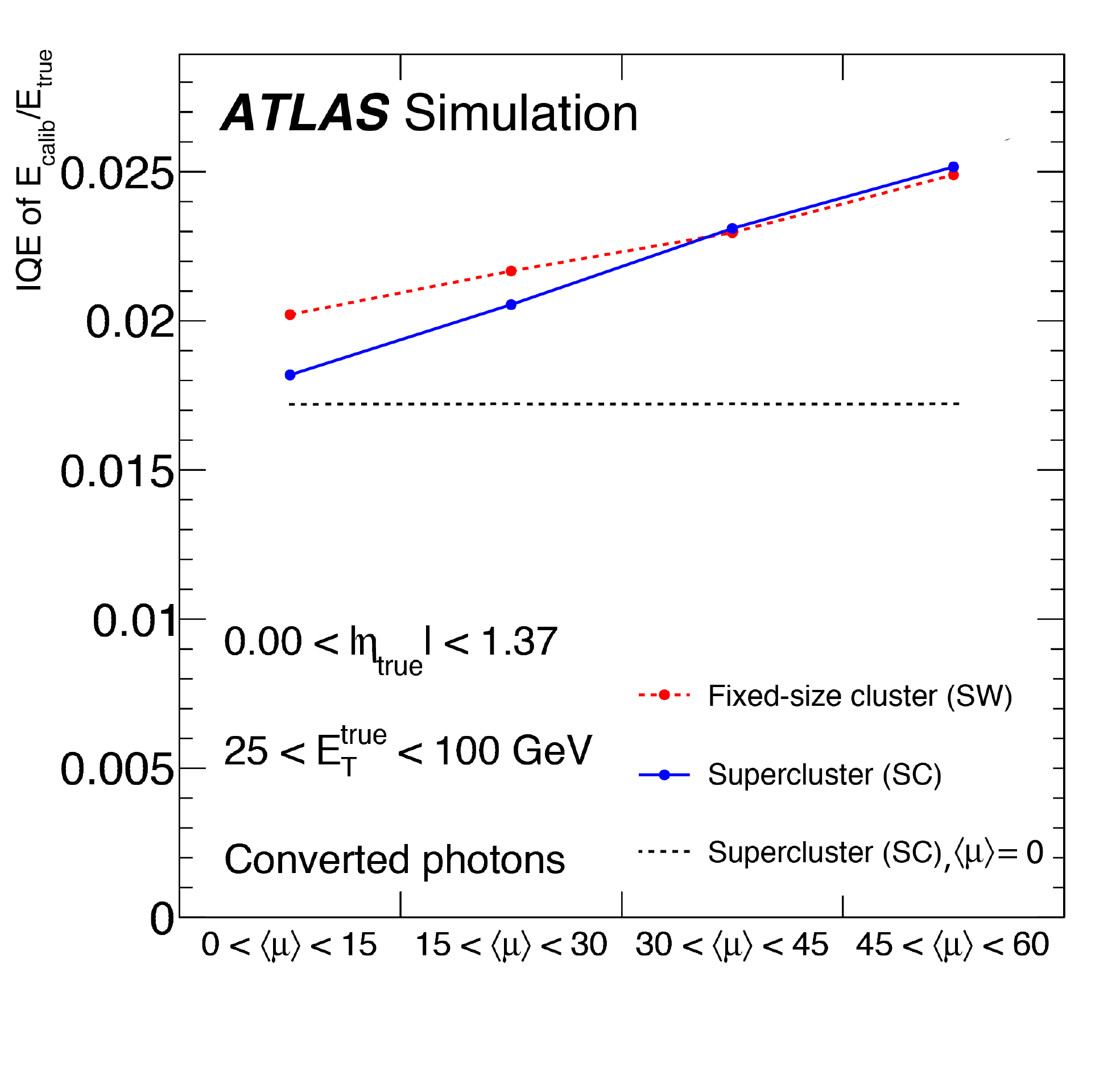}
\includegraphics[width=0.45\columnwidth]{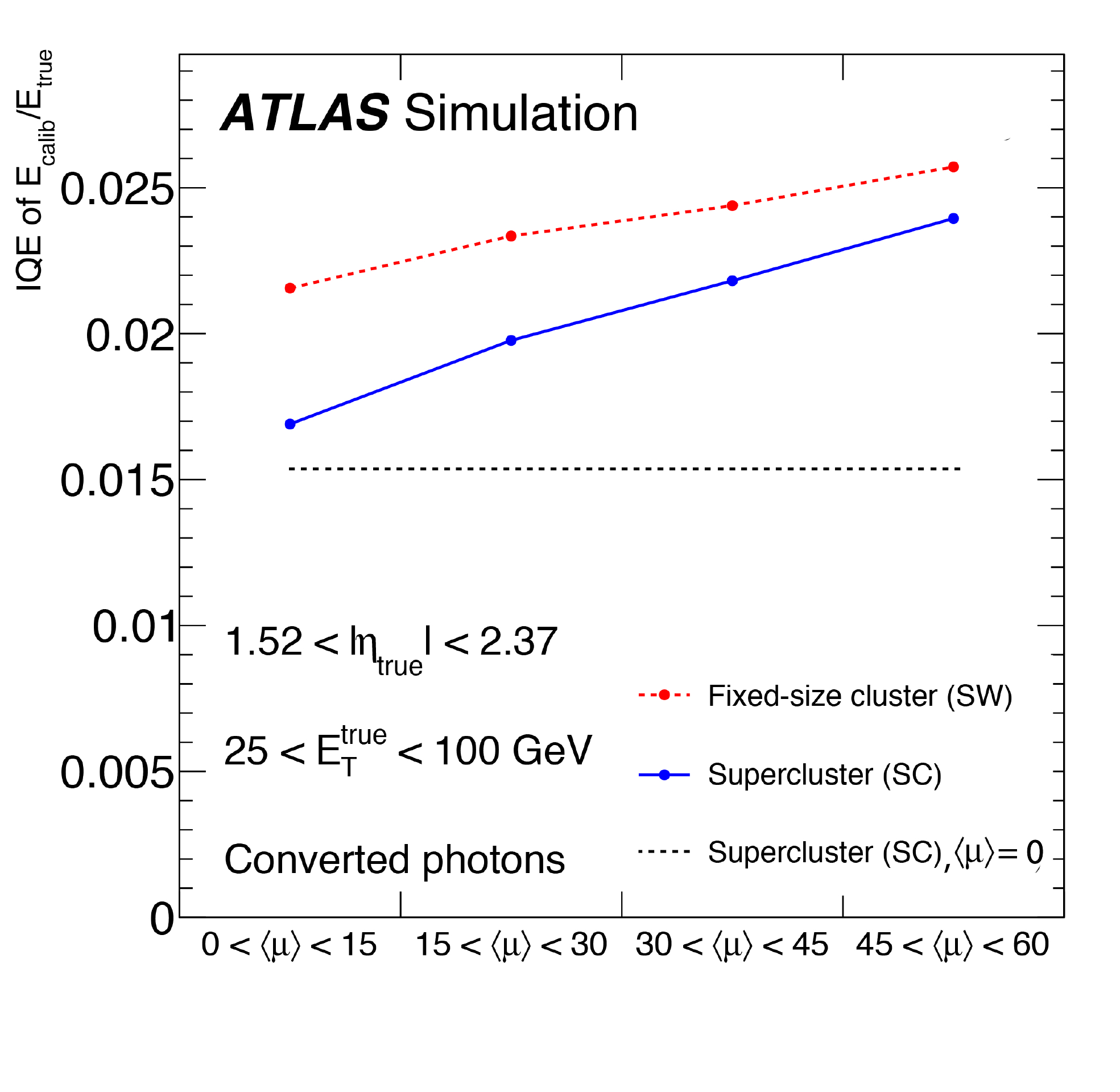}
\vspace{-0.2cm}
\includegraphics[width=0.45\columnwidth]{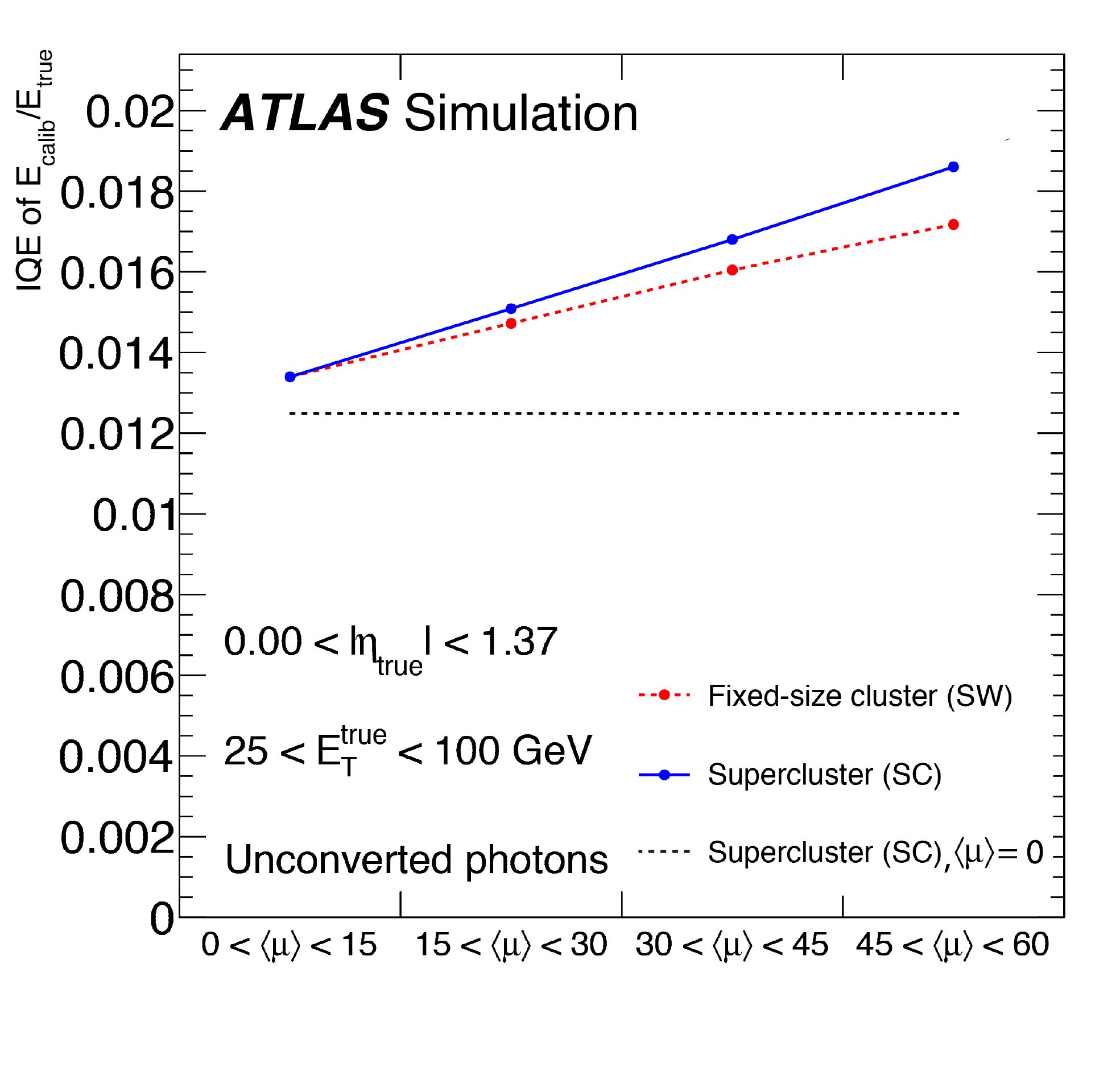}
\includegraphics[width=0.45\columnwidth]{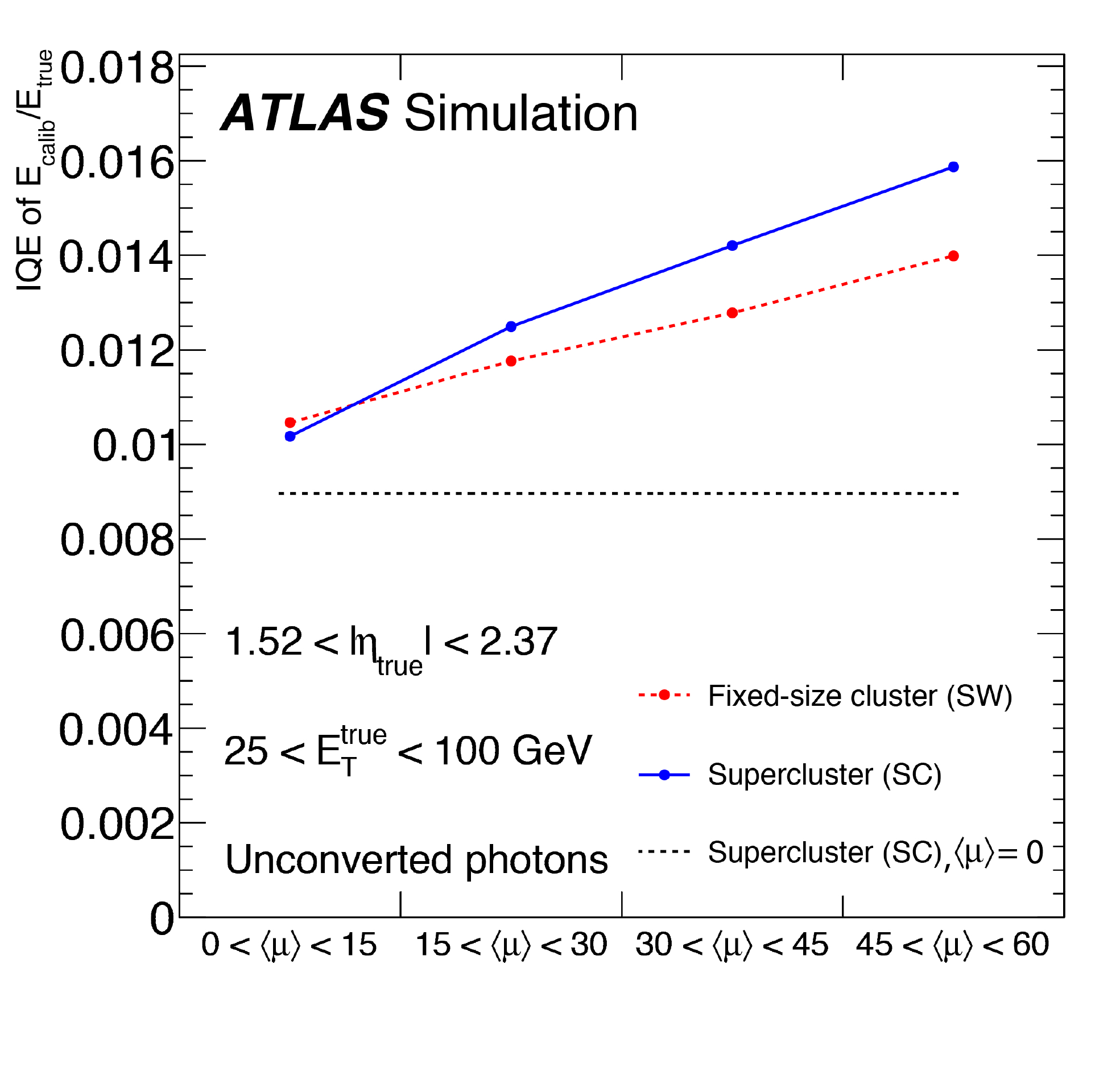}
\vspace{-0.2cm}
\caption{Calibrated energy response resolution, expressed in terms of
IQE, for simulated single electrons (top), converted photons (middle), and
unconverted photons (bottom) at different \muhat\ levels.  The plots on the left are for
the central calorimeter, while the plots on the right are for the
endcaps. The response for fixed-size clusters based on the
sliding-window algorithm is shown in dashed red, while the
supercluster-based response is shown in full blue. The
supercluster-based energy response resolution for $\muhat = 0$ is
also given as a black dashed line for comparison.
\label{fig:pu_slices_iqe}}
\end{figure}
The topo-cluster noise thresholds for the `high-$\mu$' data sample
were tuned for $\muhat \sim 40$. For electrons and converted photons,
the IQE of the supercluster reconstruction generally remains better,
although the supercluster-based response is more sensitive to pile-up, as
seen by its larger slope as a function of \muhat. Part of the reason
is that the topo-cluster noise thresholds remain fixed even though
\muhat\ changes. For unconverted photons, however, the supercluster
reconstruction shows worse IQE for $\muhat > 15$. This degradation
could be mitigated in particular by limiting the growth of the size of
the clusters.
 

\section{Electron and photon energy calibration}
\label{sec:calib}
The energy calibration of electrons and photons closely follows the procedure used in Ref.~\cite{PERF-2017-03}, updated for the new energy reconstruction described in
Section~\ref{sec:reconstruction}. The energy resolution of the electron or photon is optimized using a multivariate regression algorithm based on the properties of the shower development in the EM
calorimeter. The adjustment of the absolute energy scale using \Zee\ decays is updated, together with systematic uncertainties related to pile-up and material effects. The universality of
the energy scale is verified using radiative $Z$-boson decays.
 
\subsection{Energy scale and resolution measurements with \Zee\ decays}
\label{sec:calibScale}
 
The difference in energy scale between data and simulation is defined as $\alpha_i$, where $i$ corresponds to different regions in $\eta$. Similarly, the mismodelling of the energy resolution is parameterised as an $\eta$-dependent additional constant term, $c_i$. The corresponding energy scale correction is applied to the data, and the resolution correction is applied to the simulation as follows:
\begin{equation*}
E^\text{data,corr}=E^\text{data}/\left(1 + \alpha_i\right),  \qquad \left(\frac{\sigma_E}{E}\right)^\text{MC,corr} = \left(\frac{\sigma_E}{E}\right)^\text{MC}\oplus c_i\,,
\end{equation*}
where the symbol $\oplus$ denotes a sum in quadrature.
 
For samples of \Zee\ decays, with electrons reconstructed in $\eta$ regions $i$ and $j$, the effect of the energy scale correction on the dielectron invariant mass is given in first order by $m_{ij}^{\mathrm{data,corr}} = m_{ij}^{\mathrm{data}}/(1 + \alpha_{ij})$, with $ \alpha_{ij}= (\alpha_i+\alpha_j)/2$. Similarly, the difference in the simulated mass resolution is given by $({\sigma_m}/m)_{ij}^{\mathrm{MC,corr}} = ({\sigma_m}/m)_{ij}^{\mathrm{MC}} \oplus c_{ij}$, with $c_{ij} = (c_i \oplus c_j)/2$. The values of $\alpha_{ij}$ and $c_{ij}$ are determined by optimizing the agreement between the invariant mass distributions in data and simulation, separately for each $(i,j)$ category. The $\alpha_i$ and $c_i$ parameters are then extracted from a simultaneous fit of all categories.
 
Two methods are used for this comparison and the difference is taken as a systematic uncertainty. In the first method, the best estimates of $\alpha_{ij}$ and $c_{ij}$ are found by minimizing the $\chi^{2}$ of the difference
between data and simulation templates. The templates are created by shifting the mass scale in simulation by $\alpha_{ij}$ and by applying an extra resolution contribution of $c_{ij}$. In the second
method, used as a cross-check, a sum of three Gaussian functions is fitted to the data and simulated invariant mass distributions in each $(i,j)$ region; the $\alpha_i$ and $c_i$ are extracted from the differences, between data and simulation, of the means and widths of the fitted distributions.
 
Figures~\ref{fig:alpha_highmu} (a) and (b) show the results of $\alpha_i$ and $c_i$ derived in 68 and 24 $\eta$ intervals, respectively, separately for 2015, 2016 and 2017. The difference in $\alpha_i$ for the different years is mainly due to two effects: variations of the LAr temperature, and the increase of the instantaneous luminosity. The former effect induces a variation in the charge/energy collection, affecting the
energy response by about –2\%/K~\cite{deLaTaille:686091}. The latter implies an increased amount of deposited energy in the liquid-argon gap that creates a current in the high-voltage lines, reducing
the high voltage effectively applied to the gap and introducing a variation of the response of up to 0.1$\%$ in the endcap region. A prediction of the different effects that can impact
the results is presented in Ref.~\cite{PERF-2017-03}. Given the small size of the observed dependence, well within 0.3\%, dedicated energy scale corrections for each data taking year provide an adequate stability of the energy measurement.
 
For the constant term corrections $c_i$, a dependence on the pile-up level is observed through the different values obtained for 2015 to 2017 data;
this is addressed in Section~\ref{sec:calibSys}. A weighted average of the $c_i$ values for the different years is applied in the analyses of the complete dataset. The additional constant term of the energy resolution is typically less than 1$\%$ in most of the barrel and between 1$\%$ and 2$\%$ in the endcap.
 
\begin{figure}[htbp]
\centering
\subfloat[]{\includegraphics[width=0.49\textwidth]{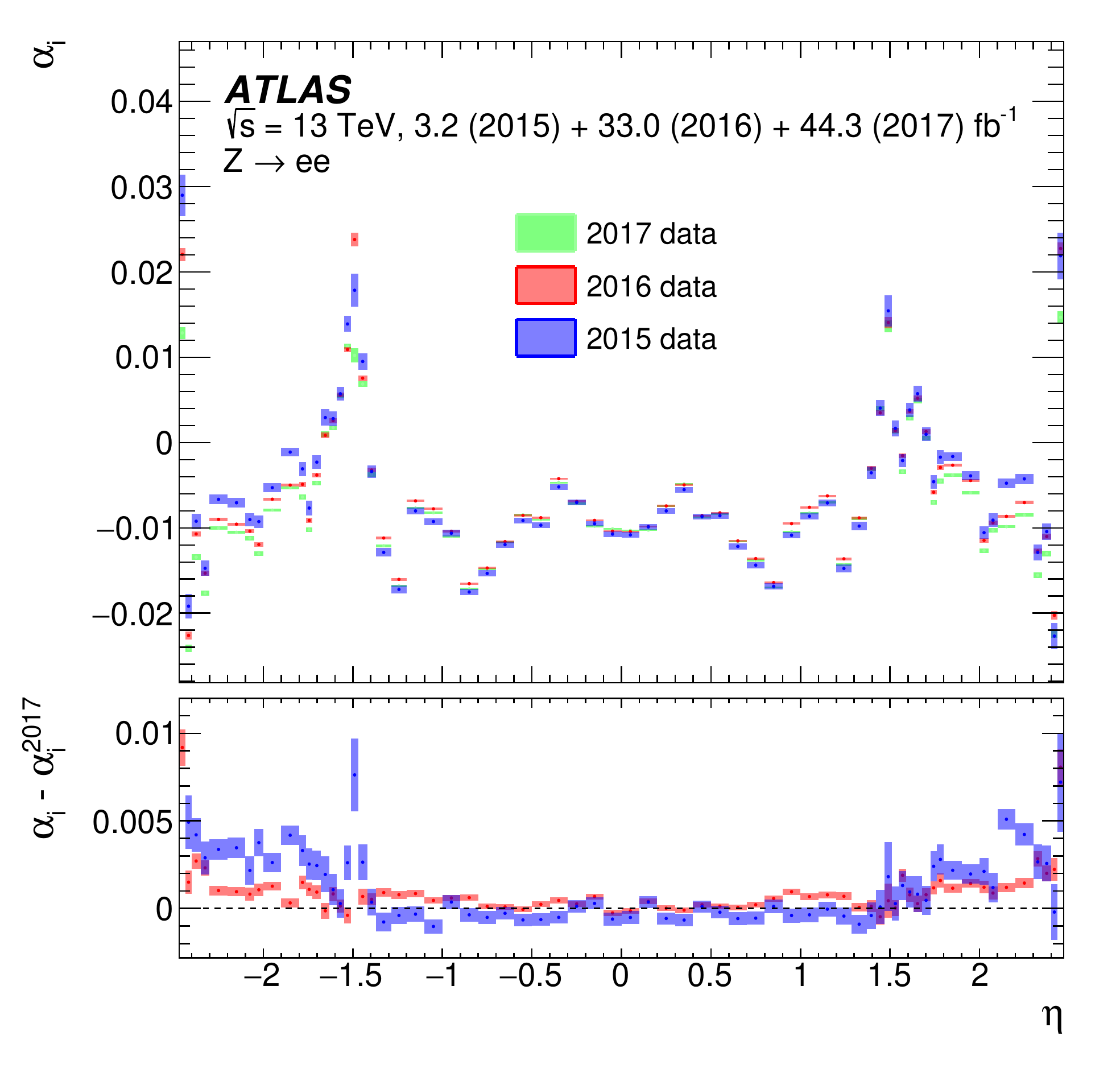}}
\subfloat[]{\includegraphics[width=0.49\textwidth]{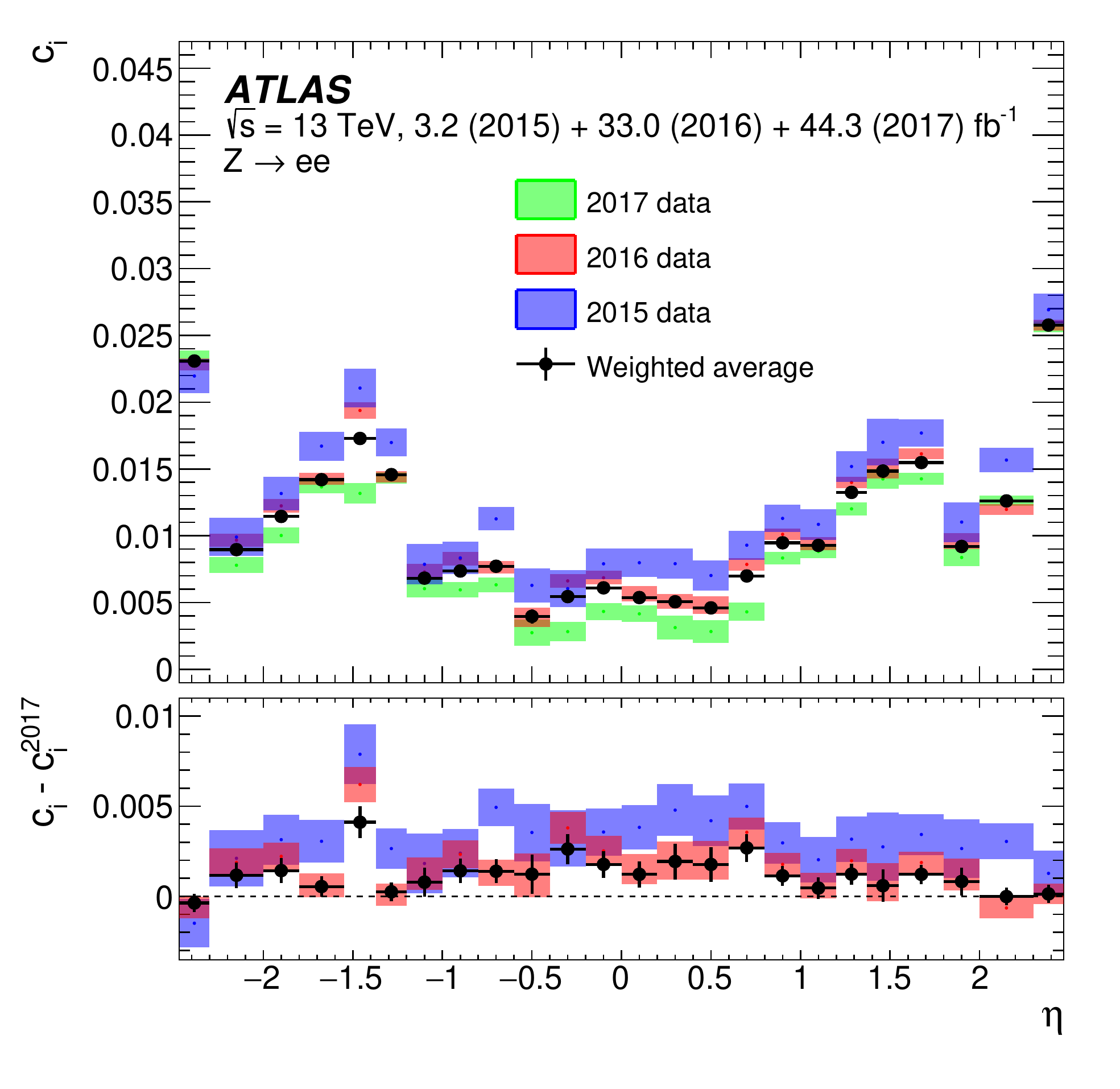}\label{fig:alpha_highmub}}
\caption{(a) Energy scale factors $\alpha_i$ and (b) additional constant term $c_i$, as a function of $\eta$. The shaded areas
correspond to the statistical uncertainties. The bottom panels show the differences between (a) $\alpha_i$ and (b) $c_i$ measured in
a given data-taking period and the measurements using 2017 data.}
\label{fig:alpha_highmu}
\end{figure}
 
Figure~\ref{fig:Meea} shows the invariant mass distribution for \Zee\ candidates for data and simulation after the energy scale correction has been applied to the data and the resolution
correction to the simulation. No background contamination is taken into account in this comparison, but it is expected to be at the level of 1$\%$ over the full shown mass range. The uncertainty band
corresponds to the propagation of the uncertainties in the $\alpha_i$ and $c_i$ factors, as discussed in Ref.~\cite{PERF-2017-03}. Within these uncertainties, the data and simulation are in fair agreement.
Figure~\ref{fig:Mee_vsMub} shows the stability of the reconstructed peak position of the dielectron mass distribution as a function of the average number of interactions per bunch crossing for the data collected in 2015, 2016 and 2017. The variation of the energy scale with \muhat\ is well below the 0.1$\%$ level in the data. The small increase of energy with \muhat\ observed in data is consistent with the MC expectation and is related to the new dynamical clustering used for the energy measurement, as introduced in Section~\ref{sec:reconstruction}.
 
\begin{figure}[htbp]
\subfloat[]{\includegraphics[width=0.48\textwidth]{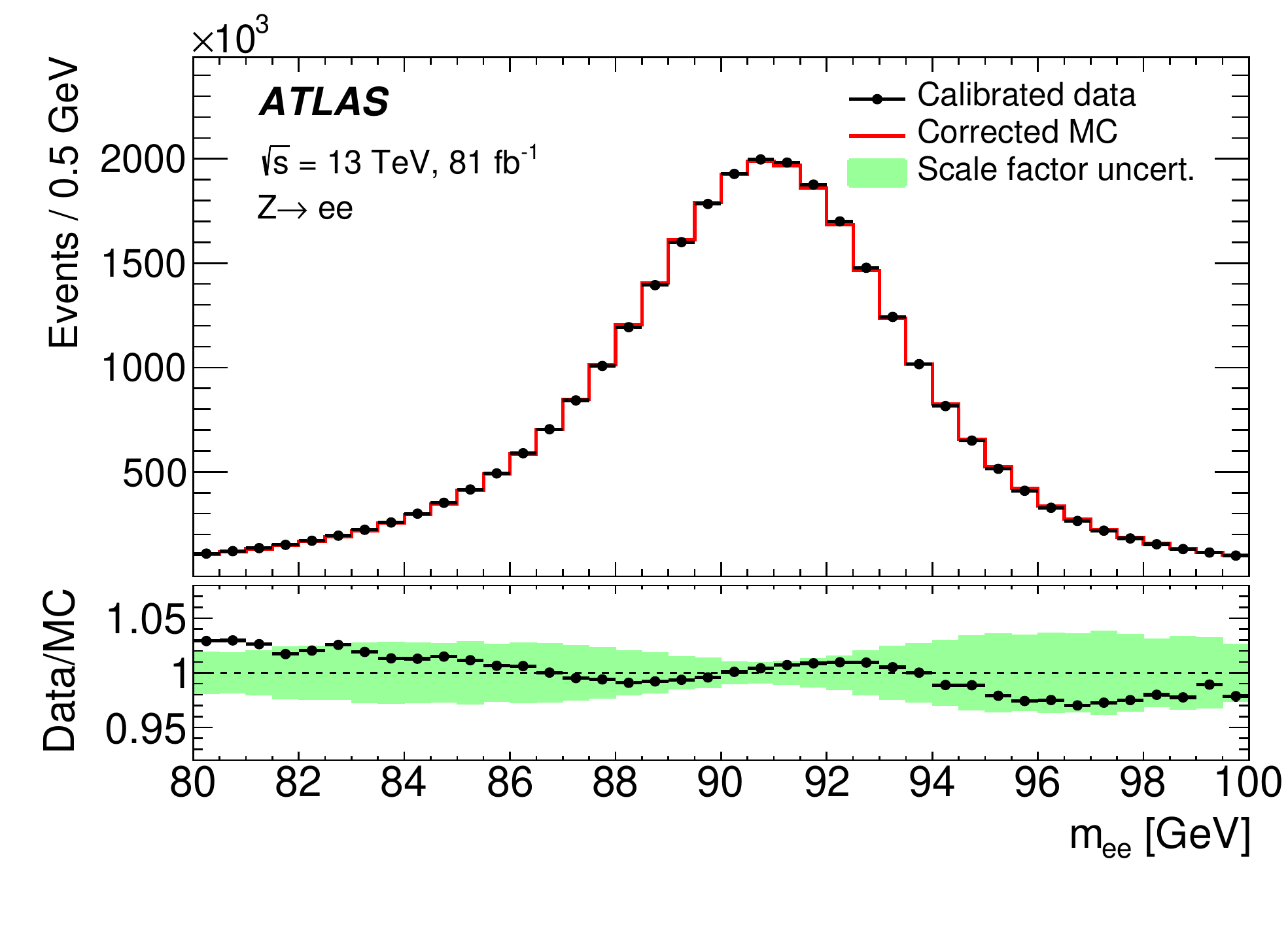}\label{fig:Meea}}
\subfloat[]{\includegraphics[width=0.48\textwidth]{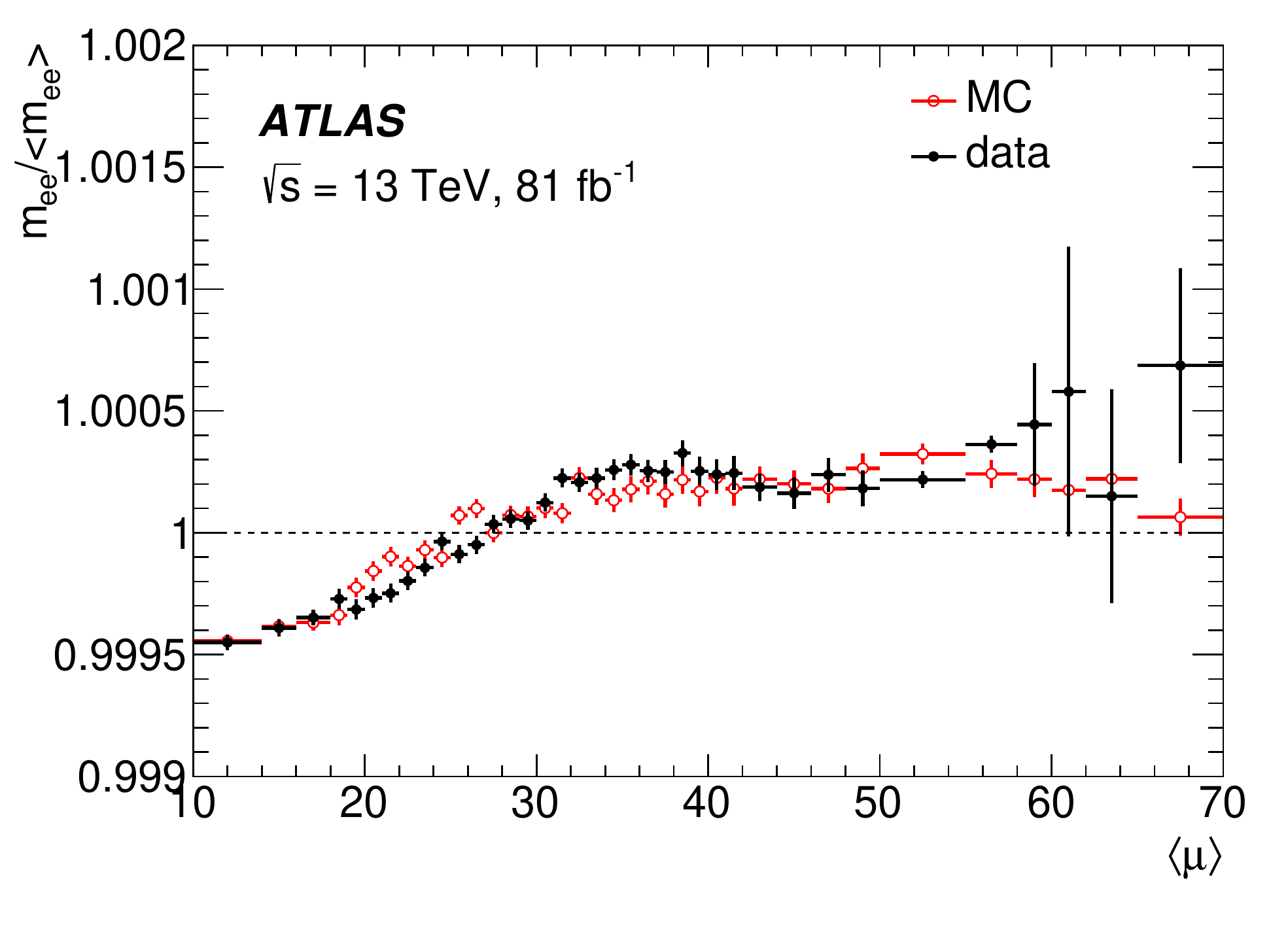}\label{fig:Mee_vsMub}}
 
\caption{(a) Comparison between data and simulation of the invariant mass distribution of the two electrons in the selected \Zee\ candidates, after the calibration and resolution corrections are applied. The total number of events in the simulation is normalized to the data. The uncertainty band of the bottom plot represents the impact of the uncertainties in the calibration and resolution correction factors. (b) Relative variation of the peak position of the reconstructed dielectron mass distribution in \Zee\ events as a function of the average number of interactions per bunch crossing. The error bars represent the statistical uncertainties.}
\end{figure}
 
\subsection{Systematic uncertainties}
\label{sec:calibSys}
 
Several systematic uncertainties impact the measurement of the energy of electrons or photons
in a way that depends on their transverse energy and pseudorapidity. These uncertainties were evaluated in Ref.~\cite{PERF-2017-03}. The amount of passive material located between the interaction
point and the EM calorimeter is measured using the ratio of the energies deposited by electrons from $Z$-boson decays in the first and second layer of the
EM calorimeter ($E_{1/2}$). The sensitivity of the calibrated energy to the detector material was re-evaluated
to reflect the changes in the reconstruction described above. The systematic uncertainty due to the material description of the innermost pixel detector layer and the services of the pixel detector
were updated with regards to Ref.~\cite{PERF-2017-03} using a more accurate description of these systems in the simulation~\cite{PERF-2015-07}.
 
The dependence of the constant term on the amount of pile-up, observed in Figure~\ref{fig:alpha_highmub}, is explained by the larger pile-up noise predicted by the simulation,
compared with that observed in the data. Figure~\ref{fig:fitnoise} shows an example of the evolution of the
second central moment of the cell energy deposit in data and simulation as a function of $\mu$ for the second layer and $1.0<|\eta|<1.1$ assuming $\phi$ symmetry. The contribution of the pile-up noise
varies linearly with $\sqrt{\mu}$, while the electronic noise remains constant. An average difference of 10$\%$ between the pile-up noise in data and simulation is observed. This mismodelling is
absorbed in the $c_i$ parameters for electrons of \et $\sim 40$ GeV, the average \et value for electrons from \Zee\ decays used to derive the energy corrections. The two methods used for the
extraction of the energy resolution corrections, described in Section~\ref{sec:calibScale}, are compared and the full difference is taken as an uncertainty in the energy resolution. This uncertainty
amounts to up to $0.2\%$ in the barrel and is due to the different sensitivities of the two methods to the pile-up. The impact of a 10$\%$ difference in pile-up noise at a different energy is
propagated to the energy resolution uncertainty relying on the predicted dependence of the pile-up noise effect as a function of the energy. For electrons and photons in the transverse energy range
30--60~\gev, the uncertainty in the energy resolution is of the order of $5\%$ to $10\%$. In order to mimic the pile-up noise estimation in the simulation, the pile-up rescaling factor, described in
Section~\ref{sec:samples}, is changed from 1.03 to 1.2 for the 48b filling scheme and to 1.3 for the 8b4e filling scheme. A systematic uncertainty in the energy scale is derived comparing the results
obtained with the two pile-up reweighting factors;
it is of the order of $2\times 10^{-4}$ in the barrel and of $5\times 10^{-4}$ in the endcap. The total systematic uncertainty in the energy scale amounts to $4\times 10^{-4}$ in the barrel and $2\times 10^{-3}$ in the endcap.
 
\begin{figure}
\centerline{
\includegraphics[width=0.7\textwidth]{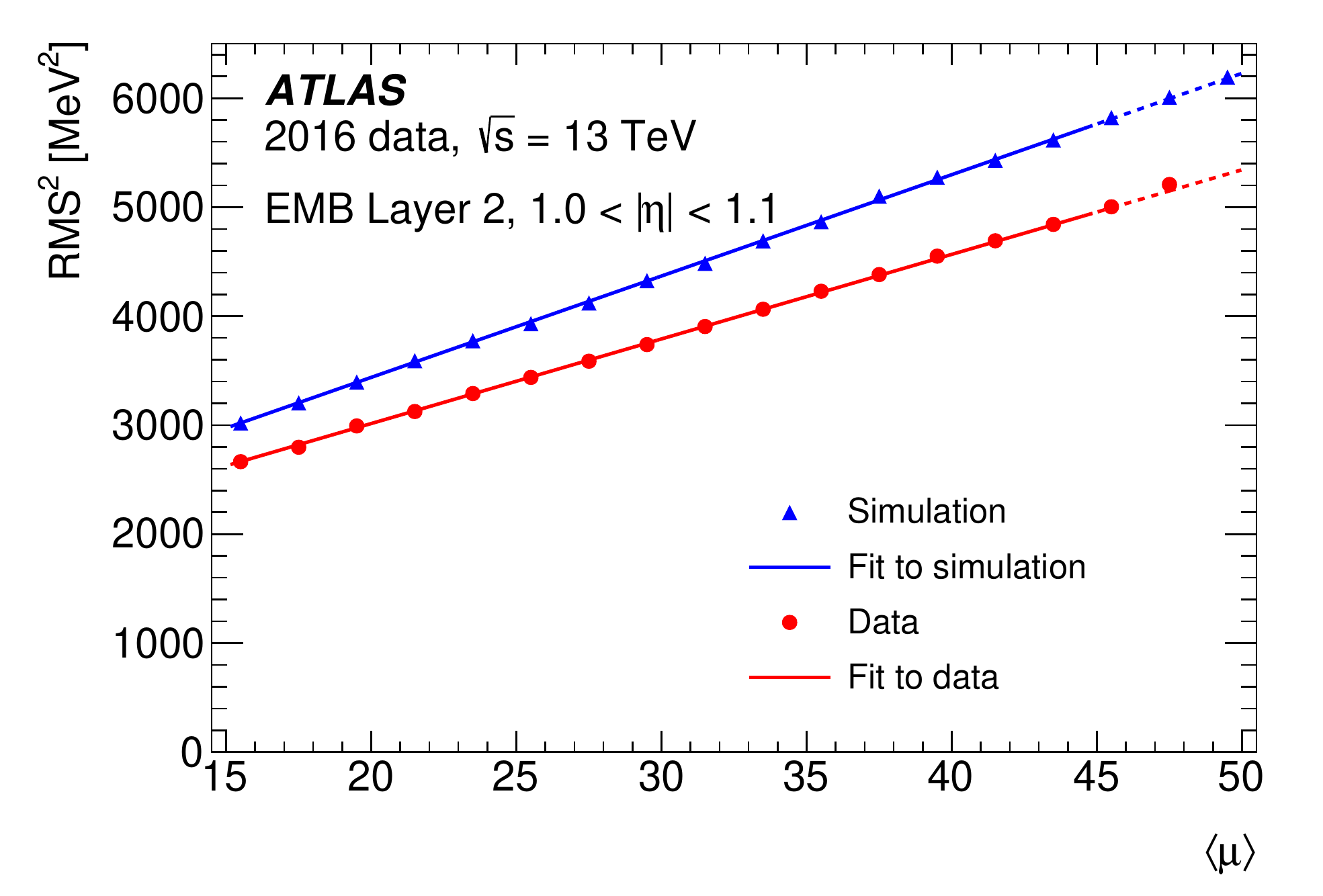}    }
\caption{Evolution of the squared noise as a function of \muhat\ in data (red points) and simulation (blue triangles), for one particular $\eta$ bin
in the second layer of the EM calorimeter. The lines show the result of linear fits to the points for $\muhat \in [15,45]$ and the dotted lines show the extrapolation to higher \muhat.}
\label{fig:fitnoise}
 
\end{figure}

\subsection{Validation of the photon energy scale with $Z \rightarrow \ell\ell\gamma$ decays}
 
The energy scale corrections extracted from \Zee\ decays, as described in Section~\ref{sec:calibScale}, are applied to correct the photon energy scale. A data-driven validation of the photon energy scale corrections is performed using radiative decays of the $Z$ boson, probing mainly the low-energy region. Residual energy scale factors for photons, $\Delta\alpha$, are derived by comparing the mass distribution of the $\ell\ell\gamma$ system in data and simulation after applying the $Z$-based energy scale corrections. The mass distribution of the $\ell\ell\gamma$ system in the simulation is modified by applying $\Delta\alpha$ to the photon energy and the value of $\Delta\alpha$ that minimizes the $\chi^{2}$ comparison between the data and the simulation is extracted. If the energy calibration is correct, $\Delta\alpha$ should be consistent with zero within the uncertainties described in Section~\ref{sec:calibSys}. An alternative method based on a binned extended maximum-likelihood fit with an analytic function to describe the mass distribution is used, and gives consistent results.
The electron and muon channels are analysed separately. In the electron channel, the electron energy scale uncertainty is accounted for in the determination of the residual photon energy scale. The
electron and muon results are found to agree, and are combined. Figure~\ref{fig:alphaZlly} shows the measured $\Delta\alpha$ as a function of \et and $|\eta|$, separately for
converted and unconverted photons. The dominant sources of uncertainty in the extrapolation to photons of the energy corrections derived in \Zee\ decays are related to the amount of passive material
in front of the EM calorimeter, and to the intercalibration of the calorimeter layers. The value of $\Delta\alpha$ is consistent with zero within about two standard deviations at most.
 
\begin{figure}[htbp]
\includegraphics[width=0.49\textwidth]{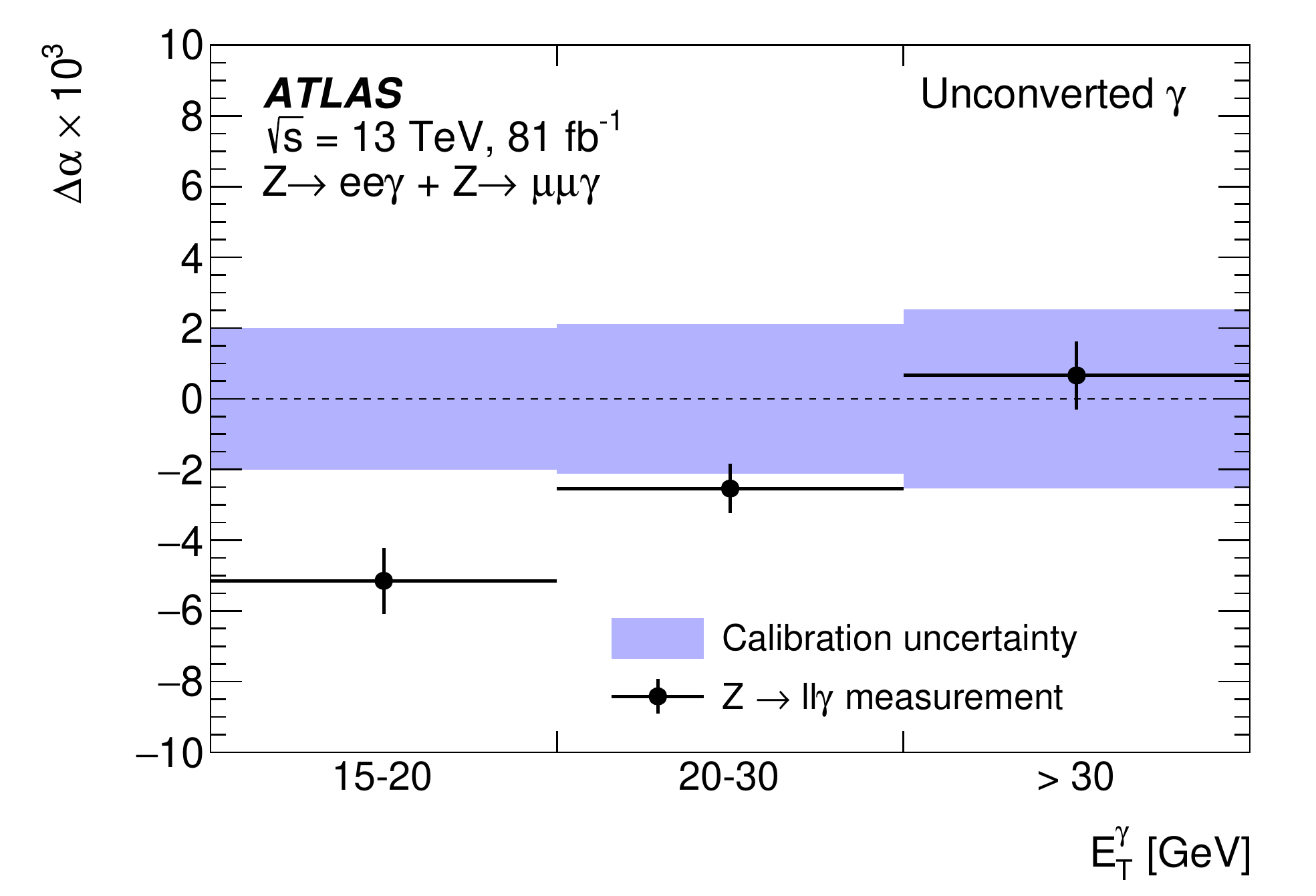}
\includegraphics[width=0.49\textwidth]{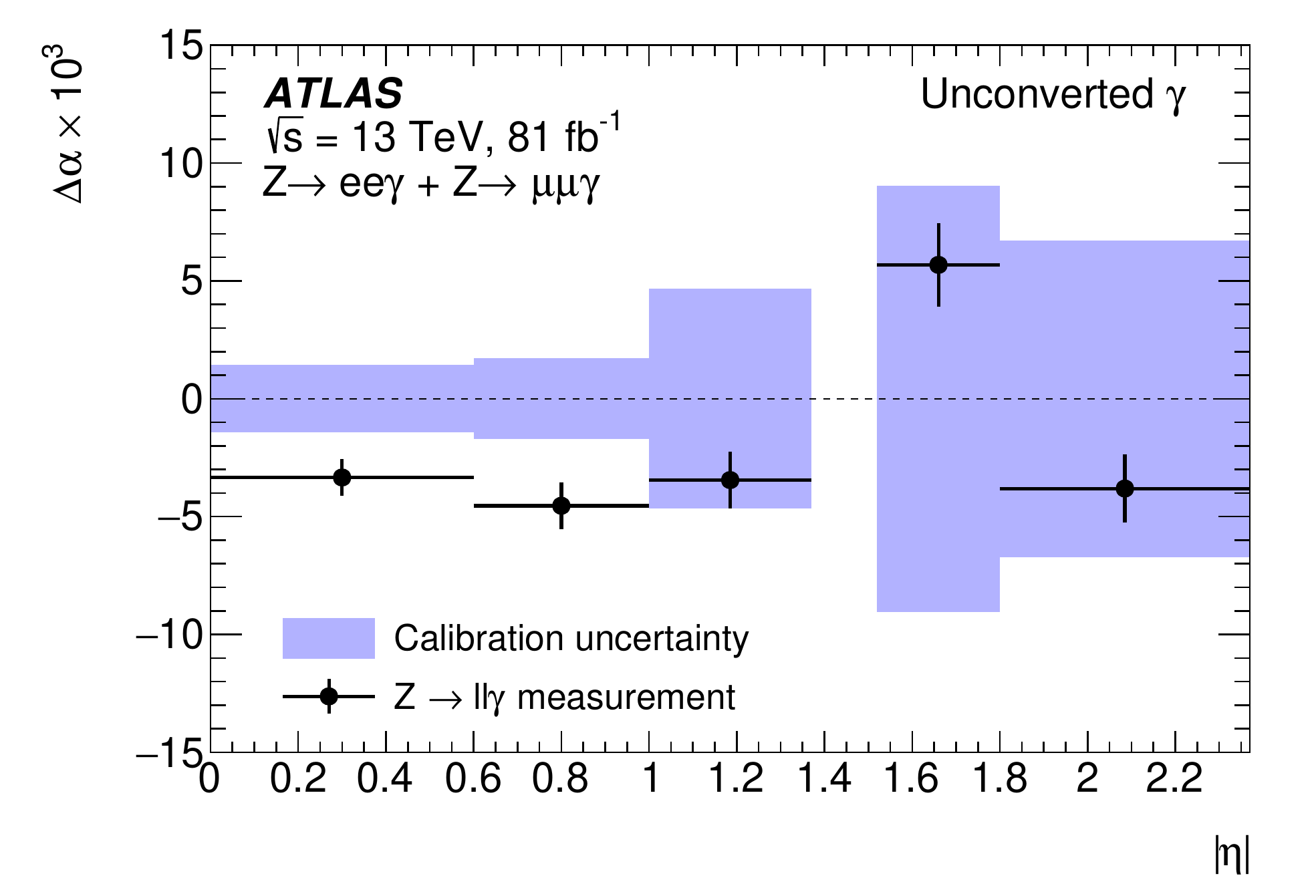}
\qquad
\includegraphics[width=0.49\textwidth]{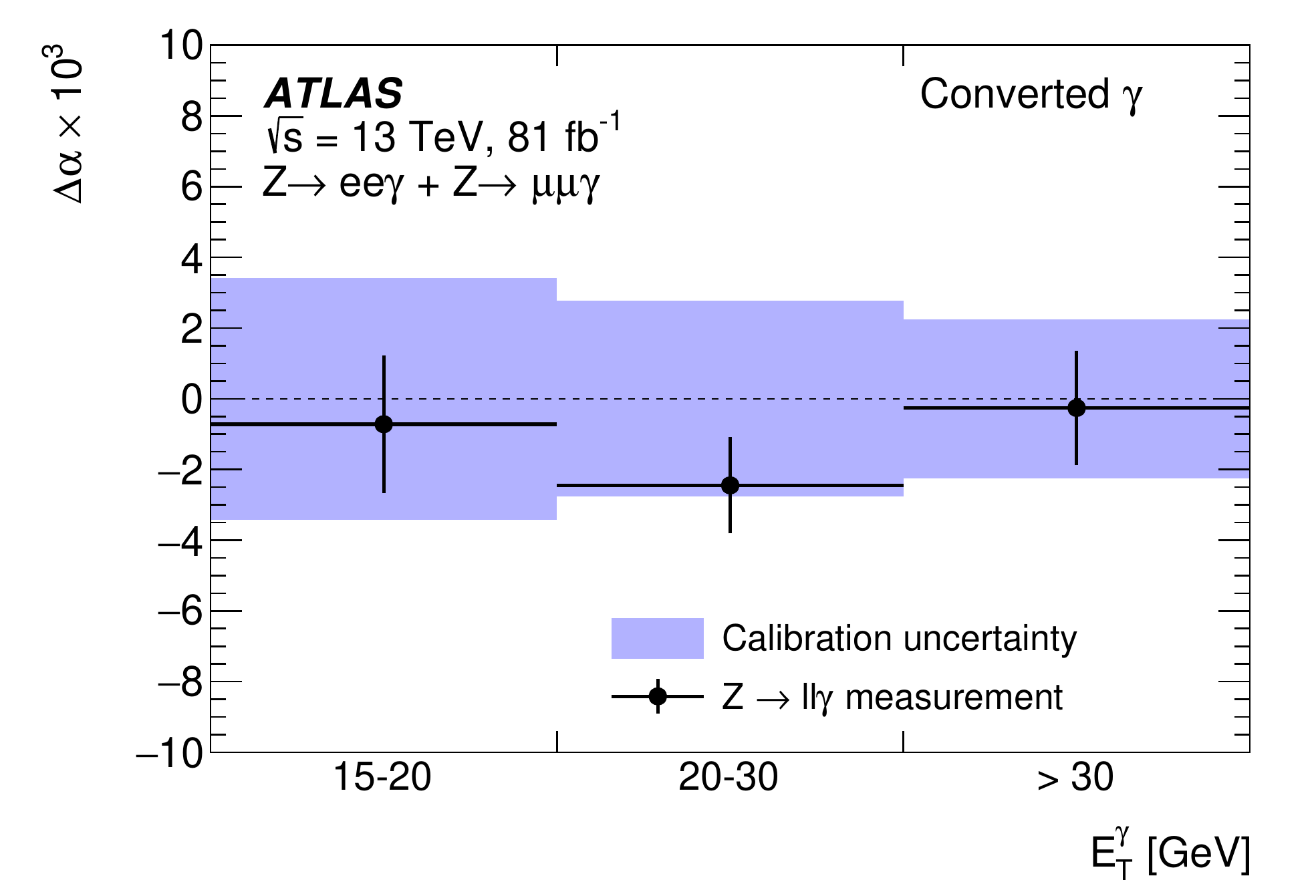}
\includegraphics[width=0.49\textwidth]{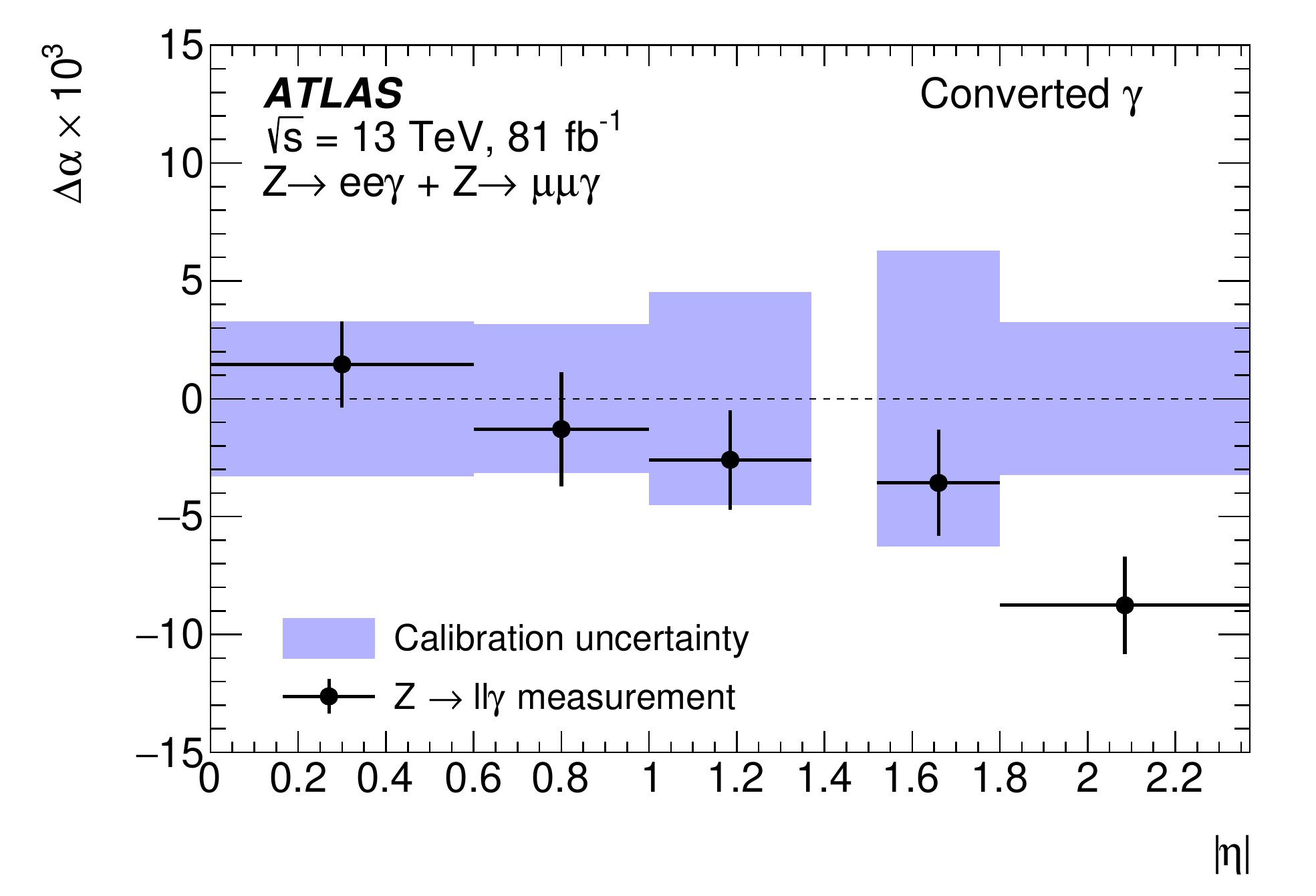}
 
\caption{Residual photon energy scale factors, $\Delta\alpha$, for unconverted (top) and converted (bottom) photons as a function of the photon transverse energy \et (left) and pseudorapidity $|\eta|$ (right), respectively. The points show the measurement with its total uncertainty and the band represents the full energy calibration uncertainty for photons from $Z \rightarrow \ell\ell\gamma$ decays.}
\label{fig:alphaZlly}
\end{figure}
 
\subsection{Energy scale and resolution corrections in low-pile-up data}
Special data with low pile-up were collected in 2017
at 13 \tev, as described in Section~\ref{sec:samples}. Energy scale factors are derived for this sample using the baseline method, described in Section~\ref{sec:calibScale}. The measurement is done in
24 $\eta$ regions given the small size of the sample.
 
An alternative approach, used for validation, consists of measuring the energy scale factors using high-pile-up data and extrapolating the results to the low-pile-up conditions. Two main effects are considered in the
extrapolation, namely the explicit dependence of the energy corrections on \muhat, and differences between the clustering thresholds used for the two samples; other effects are sub-leading and are treated as systematic uncertainties.
 
To evaluate the first effect, the high-pile-up energy scale corrections are measured in five intervals of \muhat\ in the range $20<\muhat<60$, in each of the 24 $\eta$ regions considered for
the low-pile-up sample. The results are parameterized using a linear function, which is extrapolated to $\muhat=2$. Over this range, the energy correction is found to vary by about 0.01\% in the
barrel, and by about 0.1\% in the endcap. The statistical uncertainty in the extrapolation is about 0.05\% in each $\eta$ region. The procedure is illustrated in Figure~\ref{fig:lowmuextrap}, for representative $\eta$ regions in the barrel and in the endcap.
 
\begin{figure}[tbp]
\centering
\includegraphics[width=0.49\textwidth]{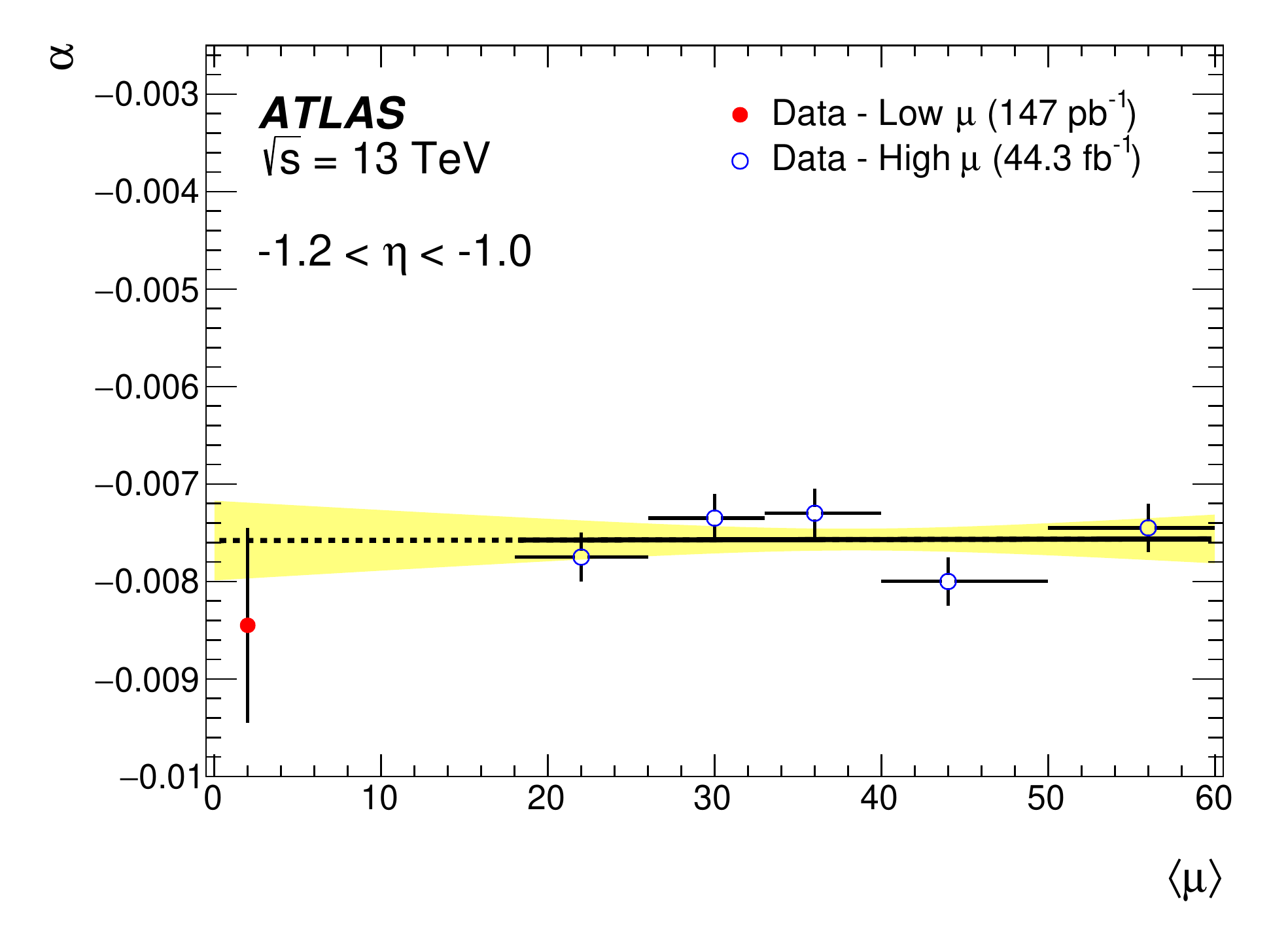}
\includegraphics[width=0.49\textwidth]{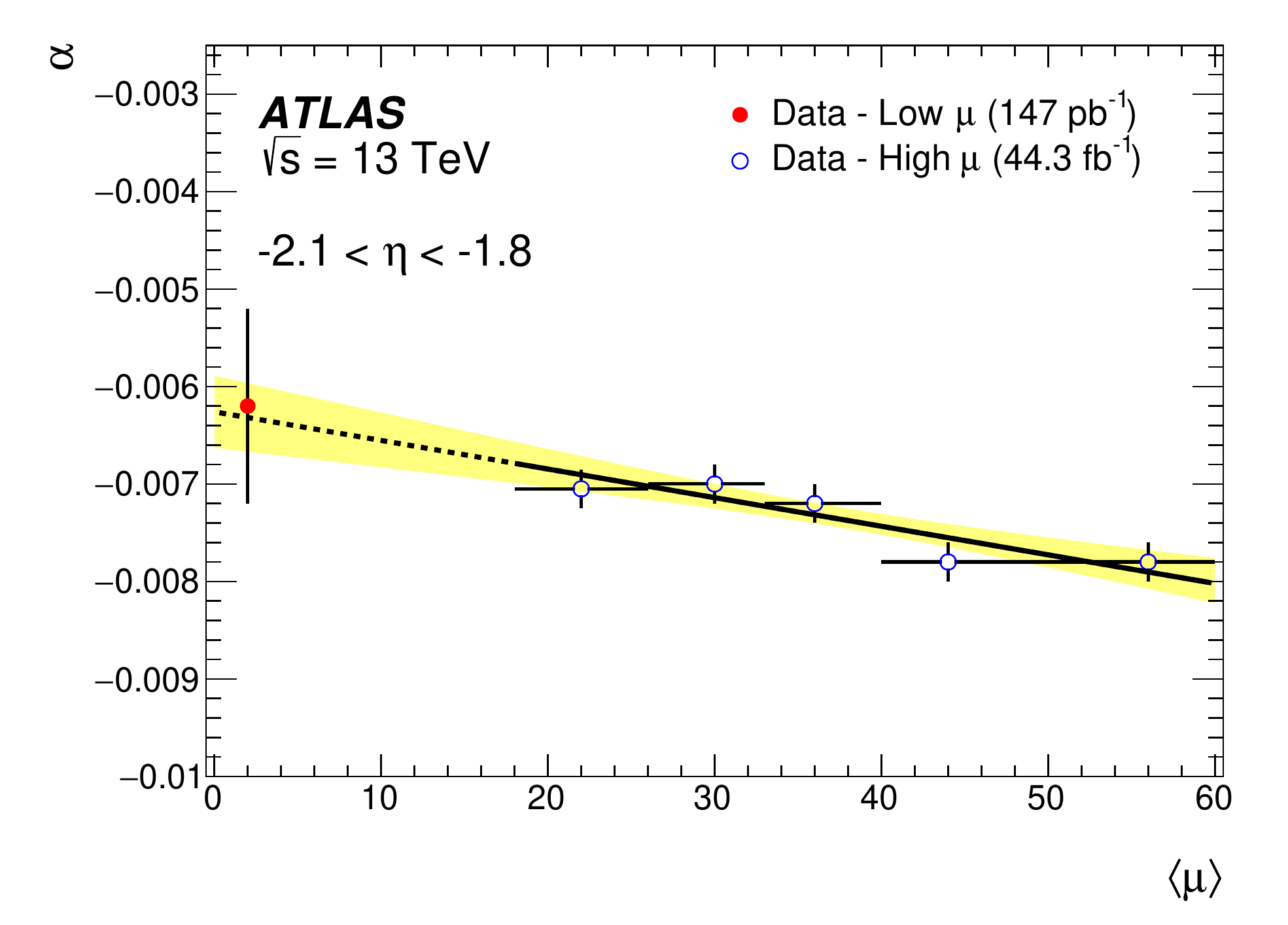}
\caption{Examples of the energy scale extrapolation from high pile-up to low pile-up in the barrel (left) and endcap (right). The blue points show the energy scale factors $\alpha$ for the high-pile-up dataset as a function of \muhat, the black lines show the extrapolation to $\muhat \sim 2$ using a linear function and five intervals of \muhat, the band represents the uncertainty in the extrapolation. The extrapolation results are compared with the energy scale factors extracted from the low-pile-up dataset, represented by the red point.\label{fig:lowmuextrap}}
\end{figure}
 
Secondly, as described in Section~\ref{sec:reconstruction}, the low-pile-up data were reconstructed with topo-cluster noise thresholds corresponding to $\mu=0$, while the standard runs used thresholds
corresponding to $\mu=40$. This results in an increased cluster size and enhanced energy response for the low-pile-up samples. The difference between the enhancements in data and simulation is measured using $Z$-boson decays, and a correction applied. The correction amounts to about $2\times 10^{-3}$ in the barrel and $4\times 10^{-3}$ in the endcap, with a typical uncertainty of $3\times 10^{-4}$.
 
Figure~\ref{fig:alpha-highmu-lowmu-thresh-corr} shows the comparison between the energy scale factors derived from low-pile-up data and extrapolated from high-pile-up data after correcting for the noise threshold effect. The observed difference is of the order of $0.1\%$ in the barrel region and increases to $0.5\%$ in the endcap region.
 
Different systematic uncertainties were considered for the extrapolation approach. In addition to the systematic uncertainties in high-pile-up data discussed in Section~\ref{sec:calibSys}, systematic
uncertainties related to the functional form chosen for the extrapolation or the number of $\mu$ intervals considered were evaluated and are of the order of a few $10^{-4}$. The changes of the LAr temperature, in the absence of collisions, between the low-pile-up and high-pile-up data-taking periods, was found to induce a variation of the energy scale by 0.006\%.
A systematic uncertainty in the energy scale is also added for the non-linear variation of the LAr temperature with $\mu$ and amounts to a few times $10^{-4}$ in the barrel
and $10^{-3}$ in the endcap. The total uncertainty in the extrapolated energy scale factors is about
0.05\% in the barrel, and on average 0.15\% in the endcap, as shown in Figure~\ref{fig:alpha-lowmu-uncer}.
 
\begin{figure}[htbp]
\begin{center}
\hspace{-0.3cm}\subfloat[]{\includegraphics[width=0.48\textwidth]{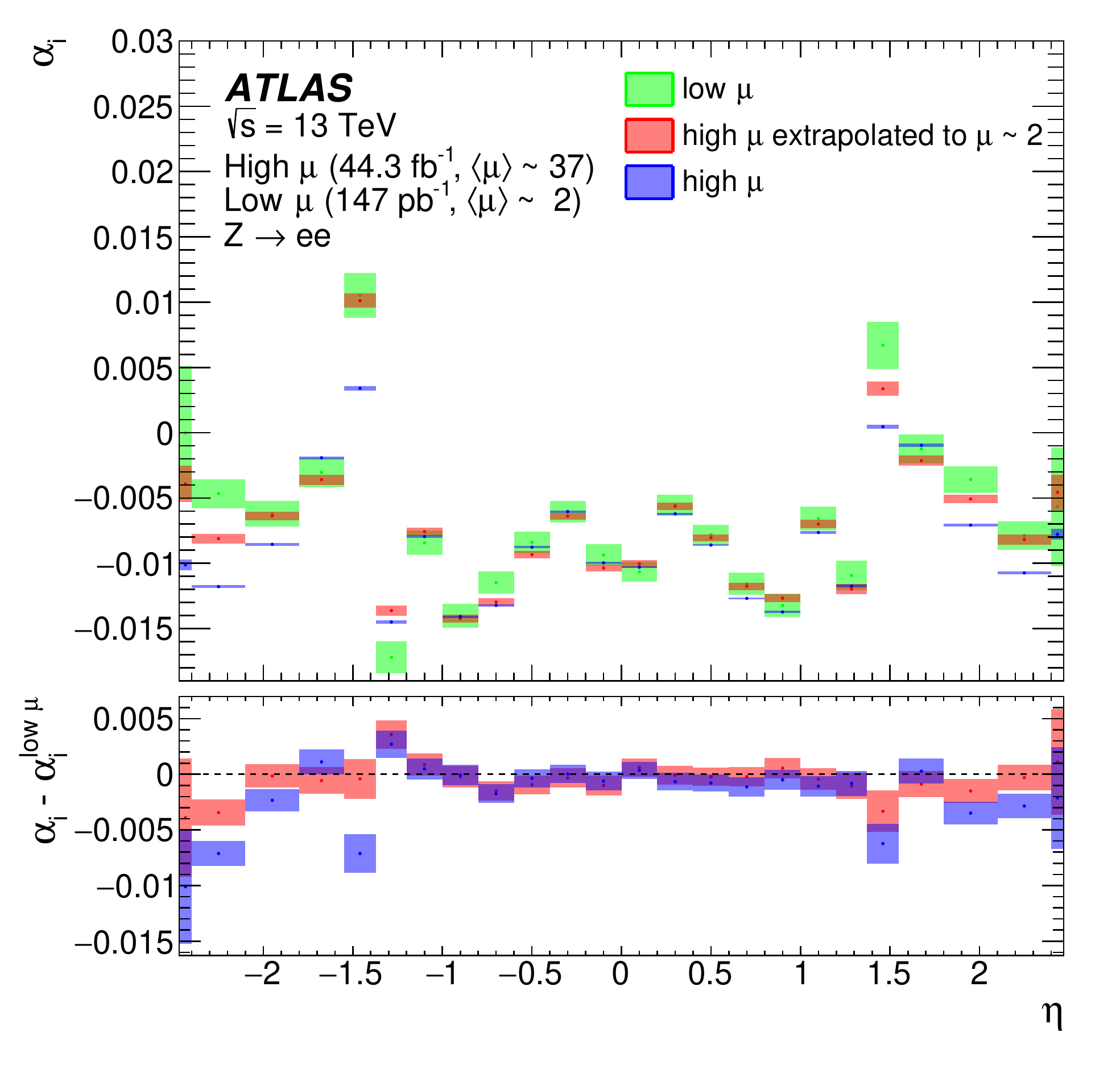}\label{fig:alpha-highmu-lowmu-thresh-corr}}
\hspace{-0.36cm}\subfloat[]{\raisebox{0.5cm}{\includegraphics[width=0.5\textwidth]{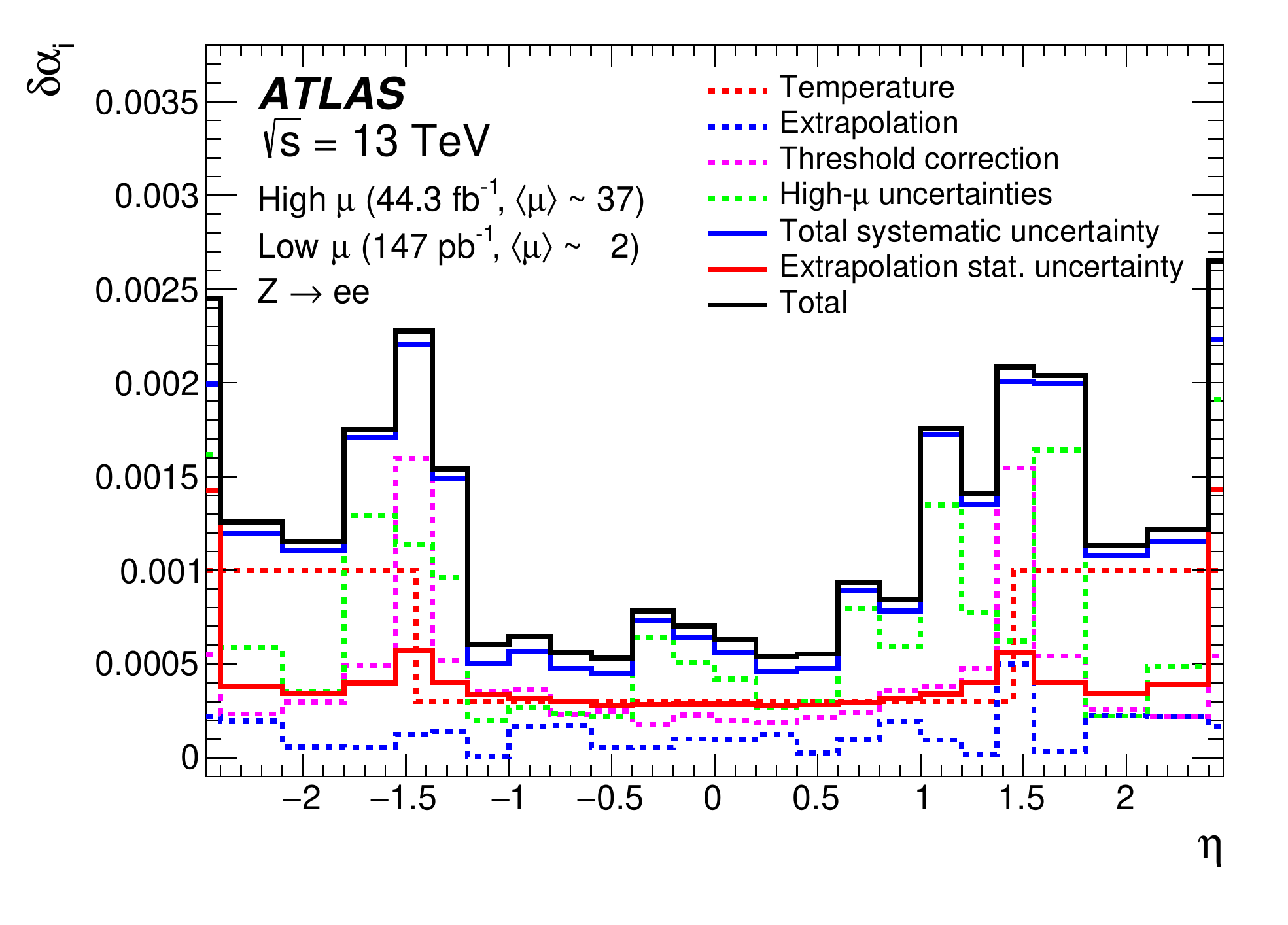}\label{fig:alpha-lowmu-uncer}}}
\caption{(a) Energy scale corrections derived from \Zee\ candidate events as a function of $\eta$ for the low-pile-up data, high-pile-up data and the extrapolated high-pile-up data after correction for the topo-cluster noise threshold difference. The shaded areas correspond to the statistical uncertainties. The bottom panel shows the differences between the energy scale corrections measured in the 2017 high-$\mu$ dataset without any correction or extrapolated to $\mu = 2$ and the measurements using 2017 low-$\mu$ data only. (b) Uncertainties in the energy scale corrections as a function of $\eta$ for the low-pile-up data.}
 
\end{center}
\end{figure}
 
 
\section{Electron identification}
\label{sec:eID}
 
Further quality criteria, called `identification selections' below, are used to improve the purity of selected electron and photon objects.
The identification of prompt electrons relies on a likelihood discriminant constructed from quantities measured in the inner detector, the calorimeter and the combined inner detector and calorimeter. A detailed description is given in Ref.~\cite{PERF-2017-01}. Recent changes implemented as a result of the migration to the supercluster reconstruction algorithm and adjustments made in parallel are discussed in the following. The identification criteria apply to all reconstructed electron candidates (see \Sect{\ref{sec:reconstruction}}).
 
\subsection{Variables in the electron identification}
 
The quantities used in the electron identification are chosen according to their ability to discriminate prompt isolated electrons from energy deposits from hadronic jets, from converted photons and from genuine electrons produced in the decays of heavy-flavour hadrons.
The variables can be grouped into properties of the primary electron track, the lateral and longitudinal development of the electromagnetic shower in the EM calorimeter, and the spatial compatibility of the primary electron track with the reconstructed cluster. They are described in \Tab{\ref{tab:IDcuts}} and summarized here.
 
The primary electron track is required to fulfil a set of quality requirements, namely hits in the two inner tracking layers closest to the beam line, as well as a number of hits in the silicon-strip detectors. The transverse impact parameter of the track and its significance are used to construct the likelihood discriminant. Furthermore, \deltapoverp\ and particle identification in the TRT are used.
 
The lateral development of the electromagnetic shower is characterized with variables calculated separately in the first and second layer of the electromagnetic calorimeter.
To reject clusters from multiple incident particles, \wtot\ is used (see Table~\ref{tab:IDcuts}). 
The lateral shower development is measured with $\Rphi$ and $\Reta$.
All lateral shower shape variables are calculated by summing energy deposits in calorimeter cells relative to the cluster's most energetic cell, and no significant difference between fixed-size EM clusters and superclusters is expected in these variables, as shown in \Fig{\ref{fig:elID:releasecomparison:rphi}} for $\Rphi$.
 
For the longitudinal shower shape variables, the numbers of cells contributing to the energy measurement in each layer are chosen dynamically in the supercluster approach, compared with fixed numbers of cells in fixed-size clusters.
The supercluster approach inherently suppresses noise in the calorimeter cells, resulting in lower values and narrower distributions. The electron identification uses $\fI$ and $\fIII$ (see Table~\ref{tab:IDcuts}).
The distribution of $\fIII$ is compared for fixed-size clusters and superclusters in \Fig{\ref{fig:elID:releasecomparison:f3}}. The significant differences between data and simulation are caused by a known mismodelling of calorimeter shower shapes in the \textsc{Geant4} detector simulation. These are accounted for in the optimisation of the electron identification (see \Sect{\ref{eID:efficiency}}) and corrected with data-to-simulation efficiency ratios in analyses.
Further discrimination against hadronic showers is achieved with $\Rhad$. 
 
The reconstructed track and the EM cluster are matched using  $\deltaeta$ and $\deltaphires$.
 
\begin{figure}[t]
\centering
\subfloat[]{
\includegraphics[width=.49\textwidth]{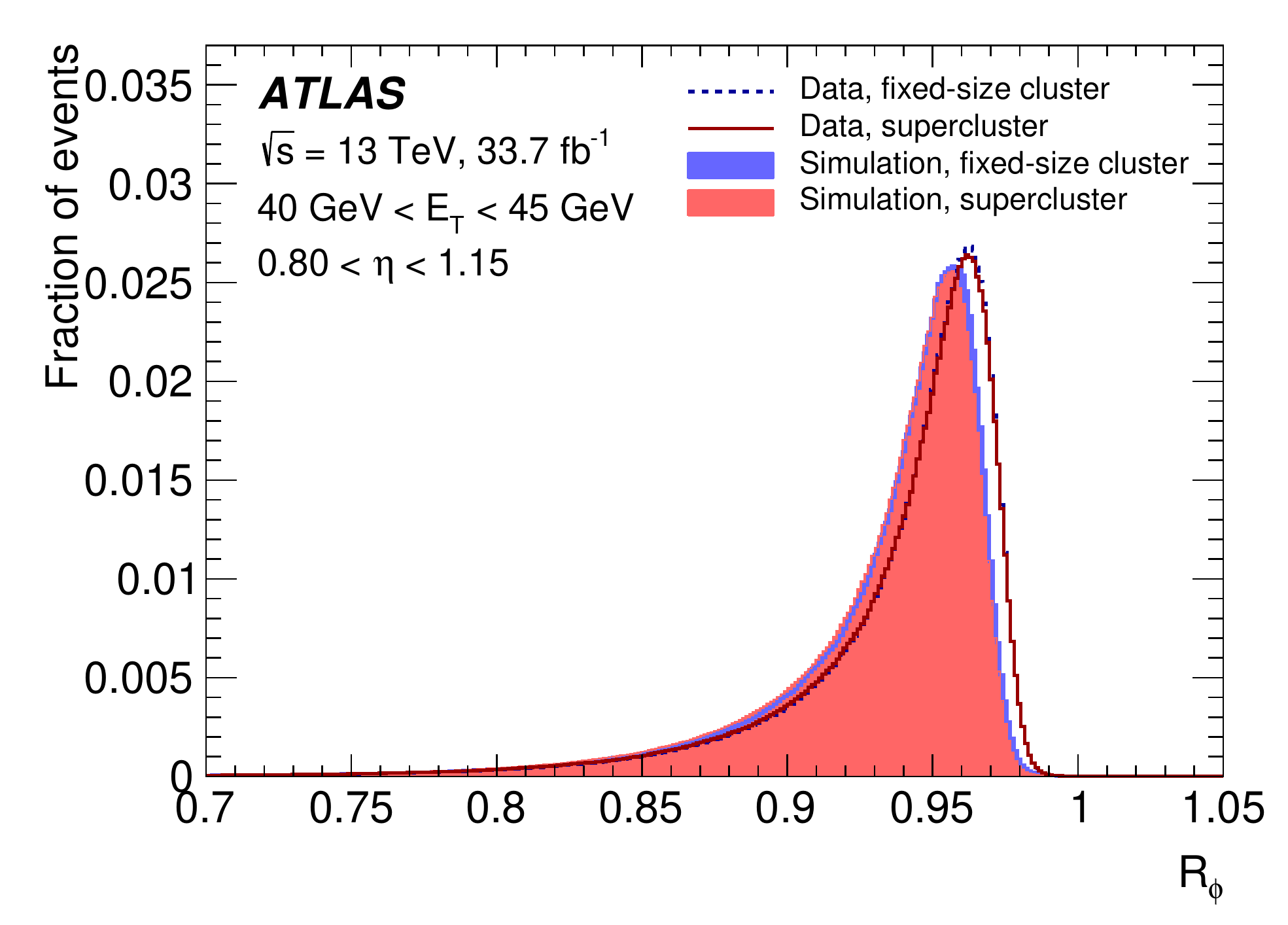}
\label{fig:elID:releasecomparison:rphi}
}
\subfloat[]{
\includegraphics[width=.49\textwidth]{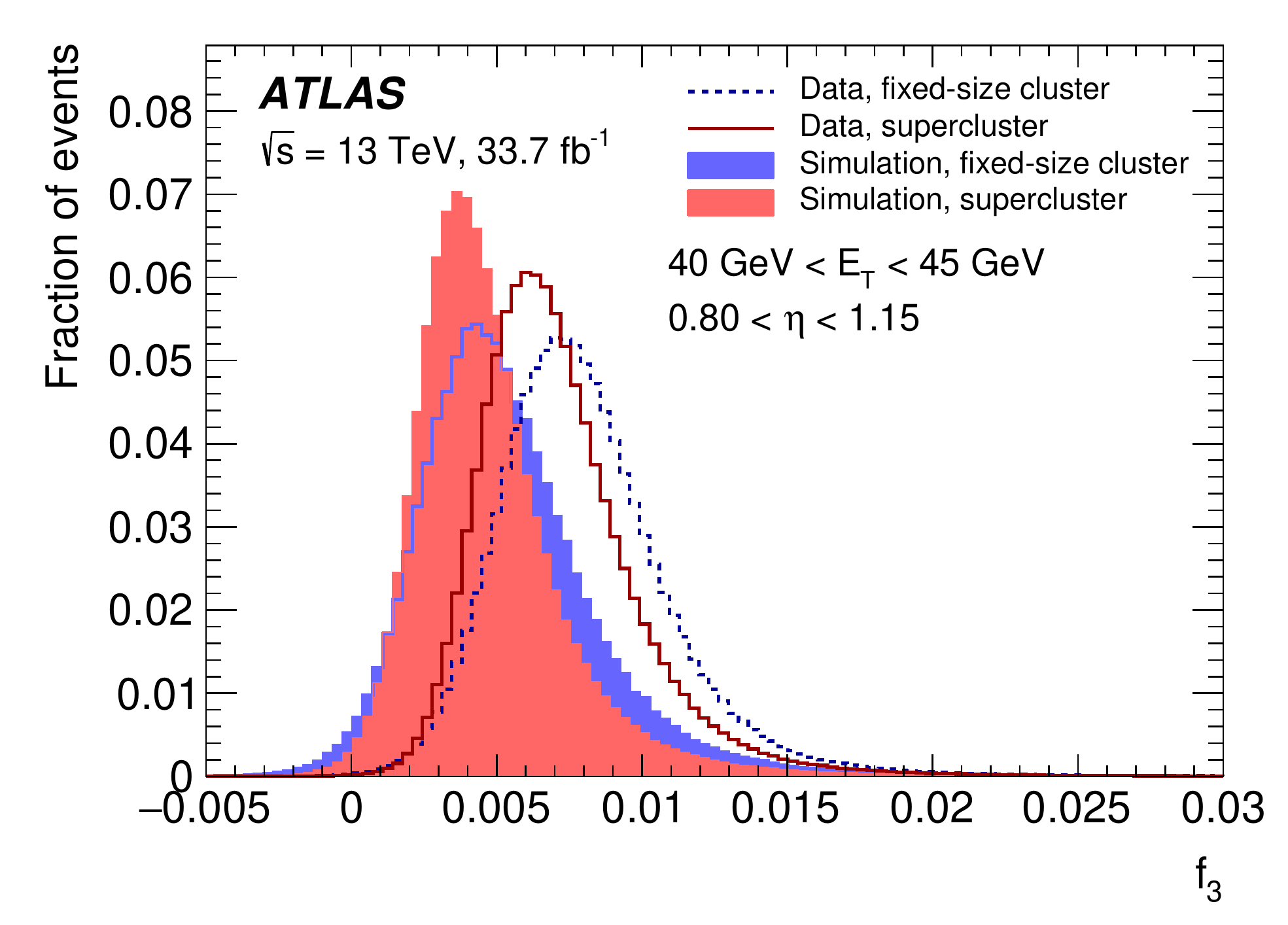}
\label{fig:elID:releasecomparison:f3}
}
\caption{
The distributions of (a) $\Rphi$ and (b) $\fIII$ obtained from $33.7$~\ifb\ of data recorded in 2016 at $\sqrt{s}=13$~\TeV\ and simulation for prompt electrons that satisfy $40<\et<45$~\GeV\ and
$0.80 < \left|\eta\right| < 1.15$. The variables are shown for fixed-size EM clusters and superclusters. The detector simulation of the corresponding distributions is performed with the \textsc{Geant4} versions 4.9.6 and 4.10, respectively.  The distributions for both the simulation and the data are obtained using the \Zee\ tag-and-probe method and KDE smoothing has been applied.
}
\end{figure}

\subsection{Likelihood discriminant}
A discriminant is formed from the likelihoods for a reconstructed electron to originate from signal, $L_{S}$, or background, $L_{B}$. They are calculated from  probability density functions (pdfs), $P$, which are created by smoothing histograms of the $n$ (typically 13) discriminating variables with an adaptive kernel density estimator (KDE~\cite{KDE}) as implemented in TMVA~\cite{TMVA}, separately for signal and background and in 9 bins in $\left|\eta\right|$ and 7 bins of \et:
\begin{equation}
L_{S(B)}(\mathbf{x}) = {\displaystyle \prod_{i=1}^{n}P_{S(B),i}(x_{i})}.
\nonumber
\end{equation}
For signal and background the pdfs take the values $P_{S,i}(x_{i})$ and $P_{B,i}(x_{i})$, respectively, for the quantity $i$ at value $x_i$.
The likelihood discriminant $d_{L}$ is defined as the natural logarithm of the ratio of $L_{S}$ and $L_{B}$.
 
The pdfs for signal were derived from \Zee\ (for $\et>15$~\GeV) and $J/\psi\rightarrow ee$ events (for $\et<15$~\GeV) prior to the 2017 data-taking period in $36.9$~\ifb\ of data recorded in the years 2015 and 2016. A reconstructed electron is selected in these events using a tag-and-probe method~\cite{PERF-2016-01}.
One of the electrons must satisfy a strict requirement on the likelihood discriminant of the previous electron identification~\cite{PERF-2017-01} and the other electron serves as a probe.
To reduce the background contamination in the selected data, probe electrons are required to satisfy a very loose requirement on the likelihood discriminant. This requirement rejects approximately 95\% of the background with a signal efficiency of 97\%, causing only a mild distortion of the likelihood pdfs.
Events with at least one reconstructed electron are selected to derive the pdfs for background. This sample primarily contains dijet events; contributions from genuine electrons, mainly from $W\rightarrow e\nu$ and \Zee\ decays, are suppressed to a negligible level using dedicated selection criteria.
Deriving the likelihood pdfs in data is an improvement compared to the previous likelihood-based identification, which used simulation.
Compared to the mismodelling in simulation, the selection applied in data and differences in the run conditions between the years 2015, 2016, 2017 and 2018 cause only mild differences in the pdfs.
 
The electron likelihood identification imposes a selection on the likelihood discriminant and some additional requirements.
The variable \fIII\ exhibits a dependence on the electron \et\ and $\eta$ that cannot adequately be captured by the seven and nine bins, respectively, in which the pdfs are determined. It is therefore only used for electrons with $\left|\eta\right|<2.37$ and $\et<80$~\GeV.
Electrons are also rejected if a two-track silicon conversion vertex was reconstructed with a momentum closer to the cluster energy than that of the primary electron track. To pass the Tight operating point, electrons must moreover satisfy $E/p<10$ and their primary track must satisfy $\pt>2$~\GeV. These additional criteria aim to reject background from converted photons. For very high \et\ the energy dependence of the shower shape variables can cause a degradation of efficiencies for very strict requirements on the likelihood discriminant. To avoid efficiency losses in the Tight identification,
the cuts on $d_L$ are chosen to be identical to the Medium identification for
$\et>150$~\GeV, and the operating points differ only in the additional requirements and an $\eta$-dependent requirement on the shower width in the first calorimeter layer, applied to Tight electrons.
 
\subsection{Efficiency of the electron identification}
\label{eID:efficiency}
\begin{figure}[t]
\centering
\includegraphics[width=.49\textwidth]{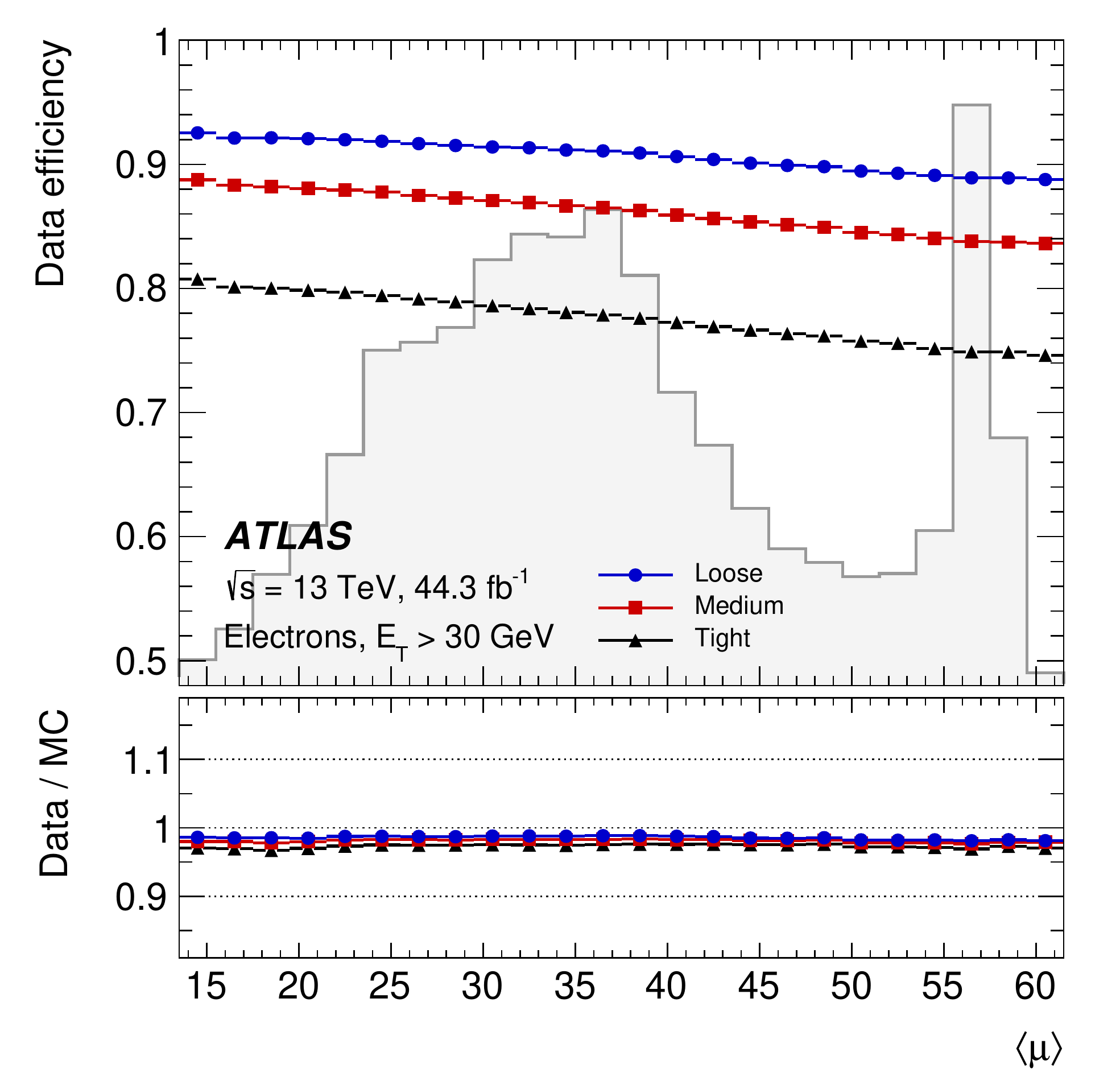}
\caption{The electron identification efficiency in data for electrons with $\et>30$~\GeV\ as a function of the average number of interactions per bunch crossing for the Loose, Medium and Tight operating points. The efficiencies are measured in \Zee\ events in data recorded in the year 2017. The shape of the \muhat\ distribution is shown as a shaded histogram. The bottom panel shows the data-to-simulation ratios. The total uncertainties are shown.
}
\label{fig:elID:pileup}
\end{figure}

The operating points Loose, Medium and Tight are each optimized in 9 bins in $\left|\eta\right|$ and 12 bins in $\et$ such that reconstructed electrons meet the requirements on the likelihood discriminant with some predefined efficiency. The values of these requirements are determined in simulated events.
For that purpose, the electromagnetic shower quantities and the combined track--cluster variables are shifted and adjusted in width such that the resulting distribution
of the likelihood discriminant of the simulated electrons closely matches that in data.
The discriminant threshold is adjusted linearly as a function of pile-up level to yield a stable rejection of background electrons.
The number of reconstructed vertices $n_{\mathrm{vtx}}$ serves as a measure for pile-up. Due to the deterioration of the discriminating power with pile-up, the approximately constant background rejection is accompanied by a reduction of signal efficiency as a function of the average number of interactions per bunch crossing, as shown in \Fig{\ref{fig:elID:pileup}} for a pure sample of electrons from $Z$-boson decays.
 
The target efficiencies are the same as in the previous identification~\cite{PERF-2017-01}, as these have proven to suit a wide range of analyses and topologies. For typical electroweak processes they are, on average, 93\%, 88\% and 80\% for the Loose, Medium, and Tight operating points and gradually increase from low to high \et. The reduced efficiency of the Medium and Tight operating points is
accompanied by an improved rejection of background processes by factors of approximately 2.0 and 3.5, respectively, in the range $20~\GeV<\et<50$~\GeV. The background efficiency was
evaluated in QCD two-to-two processes simulated as described in \Sect{\ref{sec:data_set}}.
\Fig{\ref{fig:elID:Eteta}} shows the resulting efficiencies in data.
With increasing \et, the identification efficiency varies from 58\% at $\et = 4.5$~\GeV\ to 88\% at $\et = 100$~\GeV\  for the Tight operating point, and from 86\% at $\et = 20$~\GeV\ to 95\% at $\et = 100$~\GeV\ for the Loose operating point.
In 2015, a different gas mixture was used in the TRT causing higher efficiencies. Similar efficiencies are obtained for the data recorded in the years 2016 and 2017 and residual differences are caused by their dependence on pileup.
The discontinuity in the efficiency curve at $\et=15$~\GeV\ is caused by a known mismodelling of the variables used in the likelihood discriminant at low \ET:
performing the optimization of the discriminant cuts using simulated events leads to a higher efficiency in data in this region, resulting in the rise at low \ET\ observed in
the lower panels of \Fig{\ref{fig:elID:Eteta}}.
 
The uncertainties in the efficiency are $\pm 7\%$ at $\et = 4.5~\GeV$ and decrease with transverse energy, reaching better than $\pm 1\%$ for $30~\GeV < \et < 250~\GeV$. The systematic uncertainties in
the measurements are dominated by background subtraction uncertainties at low \ET, and are derived as decribed in Ref.~\cite{PERF-2017-01}. For larger values of \et, additional systematic uncertainties of $\pm0.5\%$, $\pm1.0\%$, $\pm1.5\%$ assigned due to variations in the electron efficiency with \et\ for Loose, Medium and Tight identification, respectively,  limit the precision.
 
\begin{figure}[t]
\centering
\includegraphics[width=.49\textwidth]{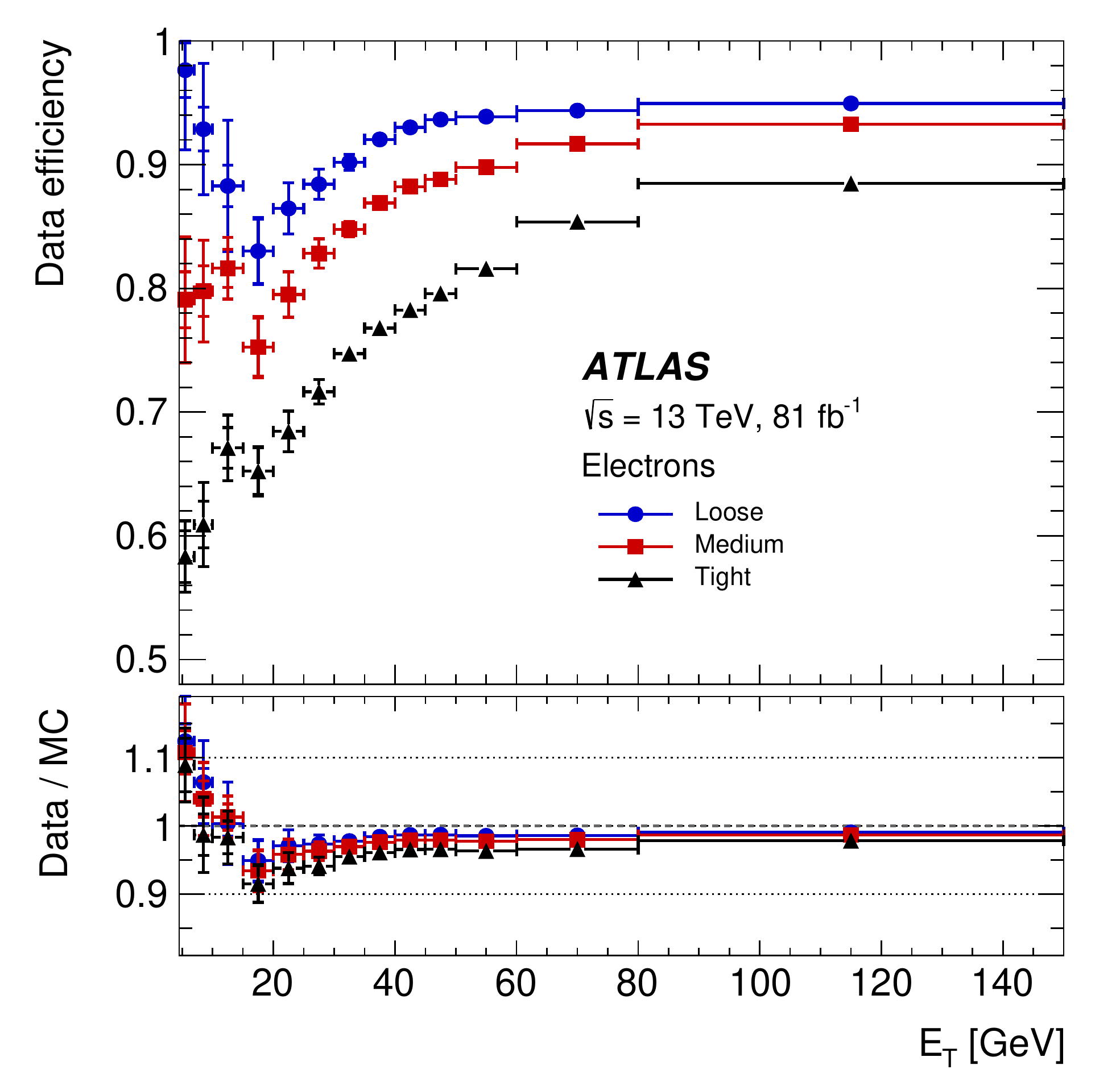}
\includegraphics[width=.49\textwidth]{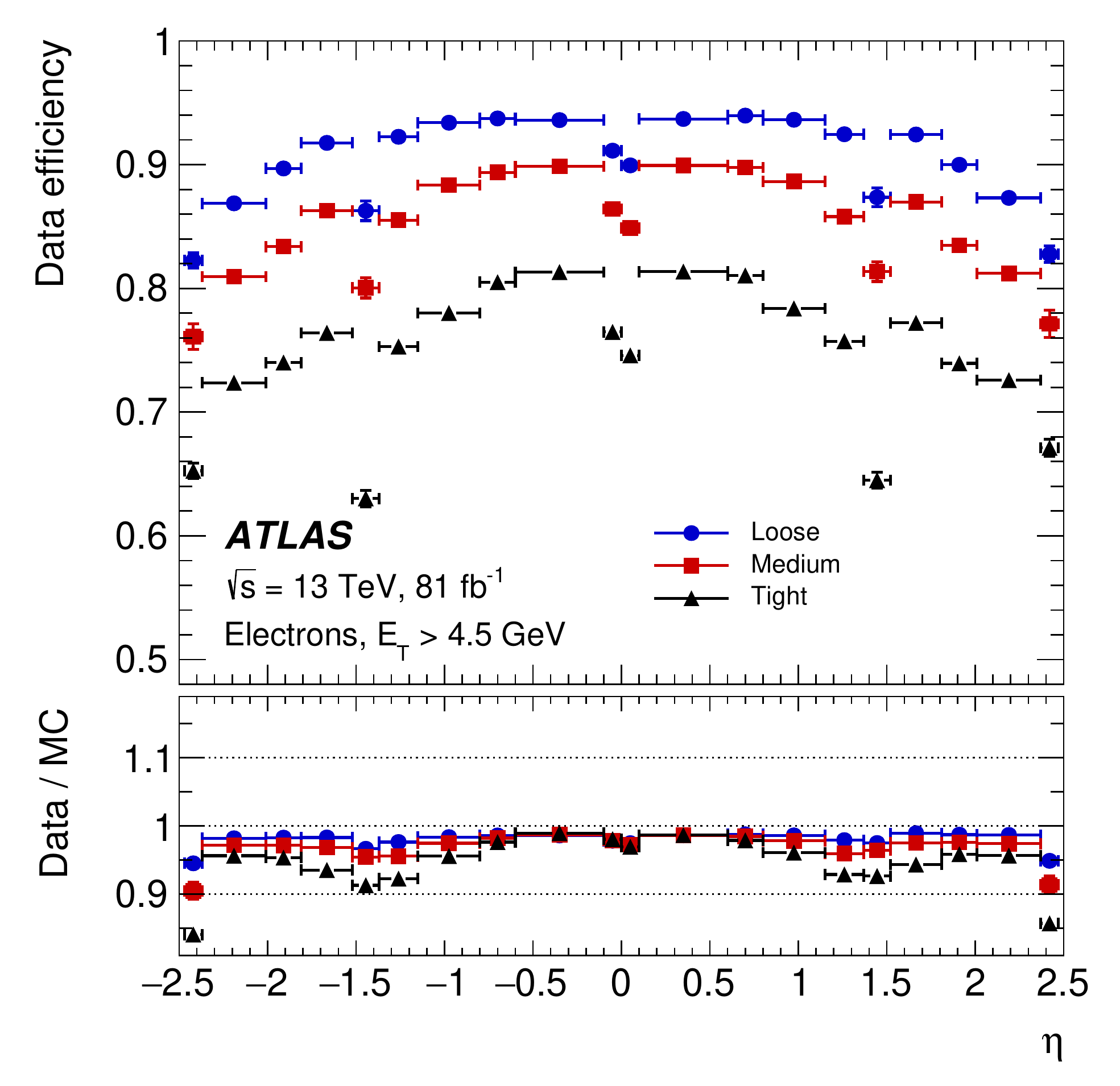}
\caption{The electron identification efficiency in \Zee\  events in data as a function of \et\ (left) and as a function of $\eta$ (right) for the Loose, Medium and Tight operating points. The efficiencies are obtained by applying data-to-simulation efficiency ratios measured in $J/\psi\rightarrow ee$ and \Zee\ events to \Zee\ simulation. The inner uncertainties are statistical and the total uncertainties are the statistical and systematic uncertainties in the data-to-simulation efficiency ratio added in quadrature. For both plots, the bottom panel shows the data-to-simulation ratios.}
\label{fig:elID:Eteta}
\end{figure}
 
 
\section{Photon identification}
\label{sec:pID}
 
\subsection{Optimization of the photon identification}
 
The photon identification criteria are designed to efficiently select prompt, isolated photons and
reject backgrounds from hadronic jets.
The photon identification is constructed from one-dimensional selection criteria, or a {\itshape cut-based
selection}, using the shower shape variables described in Table~\ref{tab:IDcuts}.
The variables using the EM first layer play a particularly important role in rejecting $\pi^0$ decays
into two highly collimated photons.
 
The primary identification selection is labelled as Tight, with less restrictive selections
called Medium and Loose, which are used for trigger algorithms. The Loose
identification criteria have remained unchanged since the beginning of Run~2, and Loose was the main
selection used in the triggering of photon and diphoton events in 2015 and 2016. It uses the $\Rhad$, $\Rhadone$,
$\Reta$, and $\wetatwo$ shower shape variables.
The Medium selection, which adds a loose cut on $\Eratio$,
became the main trigger selection in the beginning of 2017, in order to maintain an acceptable trigger rate.
Because the reconstruction of photons in the ATLAS trigger system does not differentiate between
converted and unconverted photons, the Loose and Medium identification criteria are the
same for converted and unconverted photons.
The Tight identification criteria described in this paper are designed to select a subset of the photon candidates passing the
Medium criteria.
Because the shower shapes vary due to the geometry of the calorimeter, the cut-based selection of
Loose, Medium and Tight are optimized separately in bins of $|\eta|$.
The Tight identification presented here is also optimized in separate bins of \et, and compared
with an earlier version of the Tight identification that makes an \et-independent selection.
 
The Tight identification is optimized using TMVA, and performed
separately for converted and unconverted photons.
The shower shapes of converted photons differ from unconverted photons due to the opening angle of
the $e^+e^-$ conversion pair, which is amplified by the magnetic field, and from the additional
interaction of the conversion pair with the material upstream of the calorimeters.
 
The Tight identification is optimized using a series of MC samples
that provide prompt photons and representative backgrounds at different transverse momenta. For photons with
$10<\et<25$~\gev, the $Z\to\ell\ell\gamma$ MC sample with the selection described in Section~\ref{sec:data_set} is used as a signal.
The corresponding background sample is obtained from data consisting of $Z$+jets events collected using a similar event selection,
but with relaxed requirements on the dilepton and dilepton+photon invariant masses $m_{\ell\ell}$ and $m_{\ell\ell\gamma}$. Above $\et=25$~\gev,
the inclusive-photon production
MC sample described in Section~\ref{sec:monte_carlo} is compared with a dijet background MC sample that is
enriched in high-\et\ energy deposits using a generator-level filter. No isolation selection is applied to
the training samples, and the shower shape variables are corrected to match the shower shapes observed
in data using the correction procedure described in Ref.~\cite{PERF-2017-02}.
 
Figures~\ref{fig:photonID:optimization} and ~\ref{fig:photonID:bkgs} show the result of the Tight identification optimization
in terms of the efficiencies as a function of \et for the signal and background MC training samples. The optimized selection,
labelled \et{\itshape -dependent}, is compared with a reference selection that uses criteria that do not change with
\et\ (\et{\itshape -independent}). The new, \et-dependent Tight identification allows the efficiencies of
low- and high-\et\ photon regions to be tuned separately. The Tight identification is tuned
to give a $\sim$20\% higher efficiency at low \et, and an improved background rejection at high \et.
The \muhat\ dependence of the photon identification is depicted in Figure~\ref{fig:photonID:muDep}
for photons from $Z\to\ell\ell\gamma$ decays.
 
\begin{figure}[t]
\centering
\includegraphics[width=.49\textwidth]{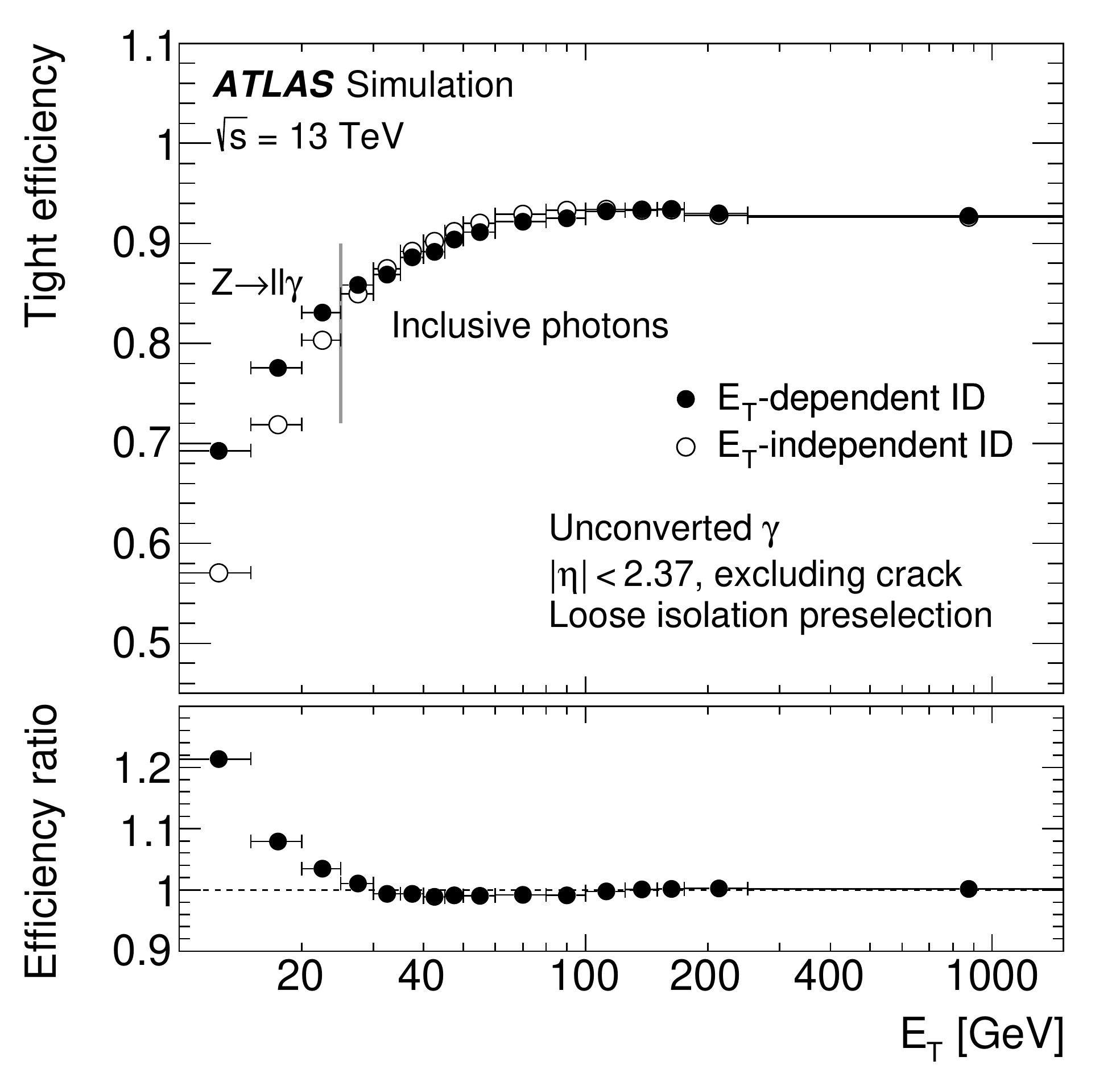}
\includegraphics[width=.49\textwidth]{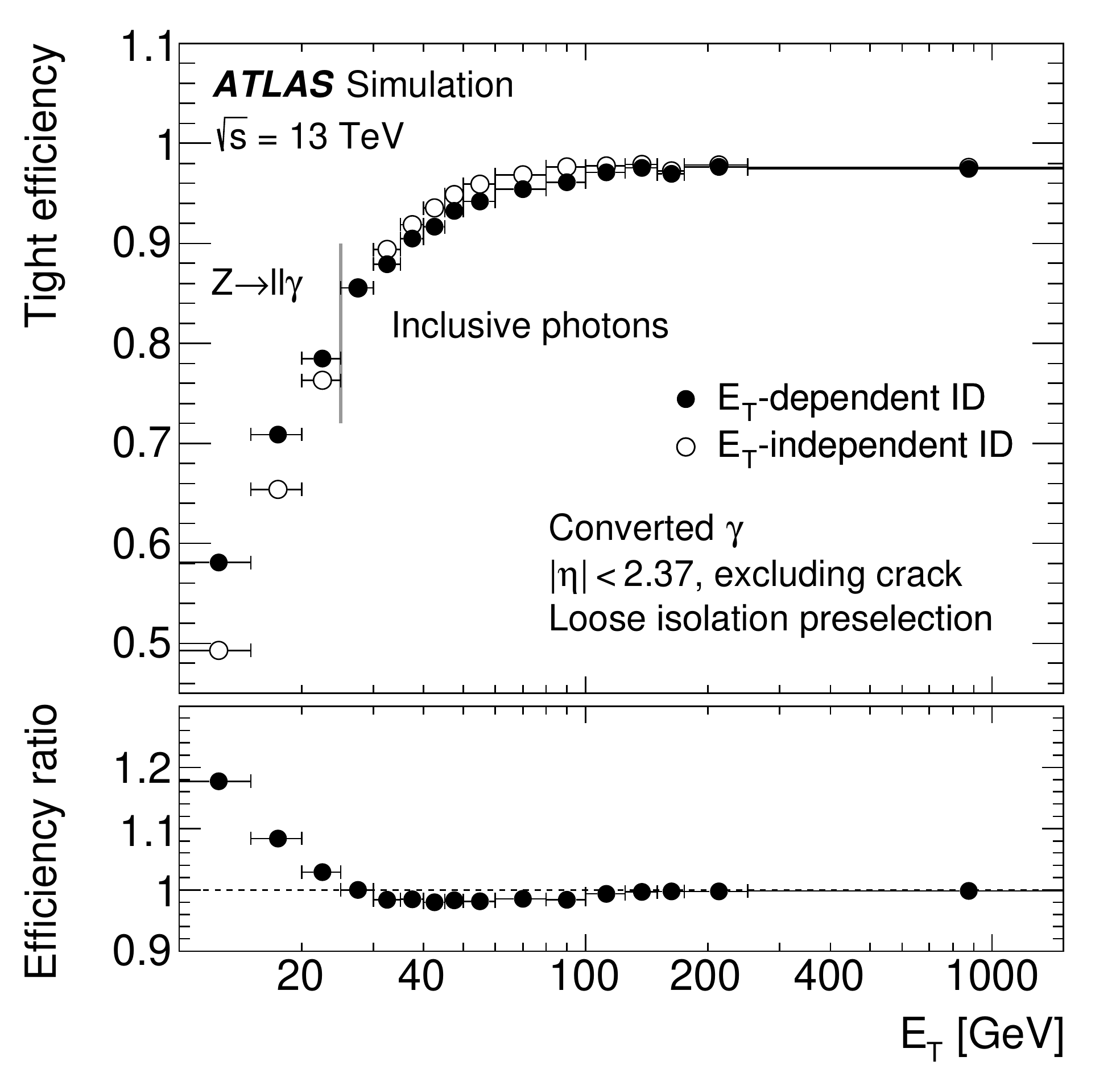}
\caption{
Efficiencies of the Tight photon identification for unconverted (left) and converted (right) signal photons,
plotted as a function of photon \et.
The signal events are taken from the sample of $Z{\rightarrow}\ell\ell\gamma$ photons with $\et<25$~\gev,
and from inclusive-photon production above 25~\gev.
In each case, the \et-independent and \et-dependent selections are compared.
The Loose isolation (see Section~\ref{sec:iso_ph_WPs_Eff}) is applied as a preselection.
For both plots, the bottom panel shows the ratios between the \et-dependent and the \et-independent identification efficiencies.
}
\label{fig:photonID:optimization}
\end{figure}
 
\begin{figure}[t]
\centering
\includegraphics[width=.49\textwidth]{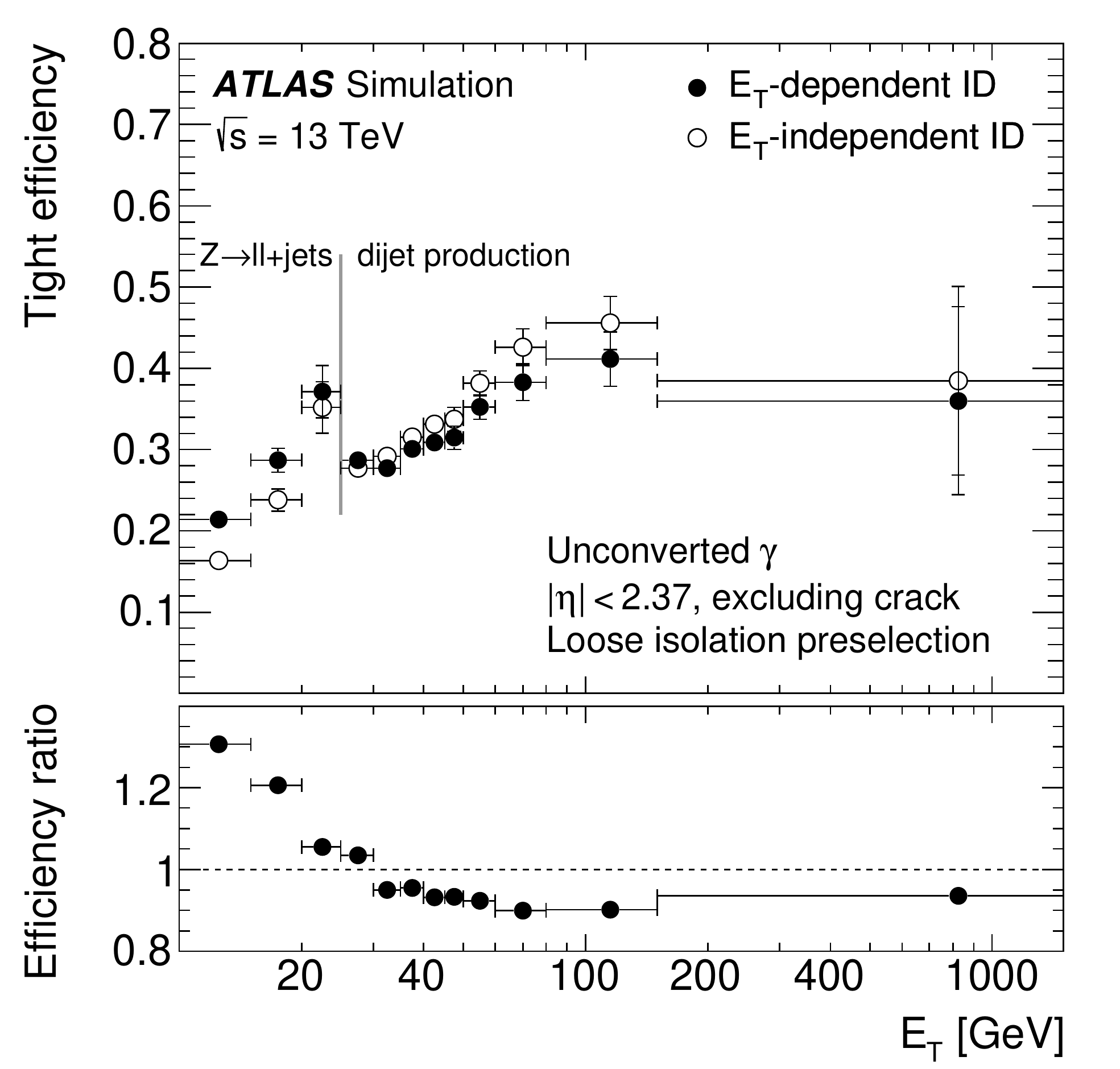}
\includegraphics[width=.49\textwidth]{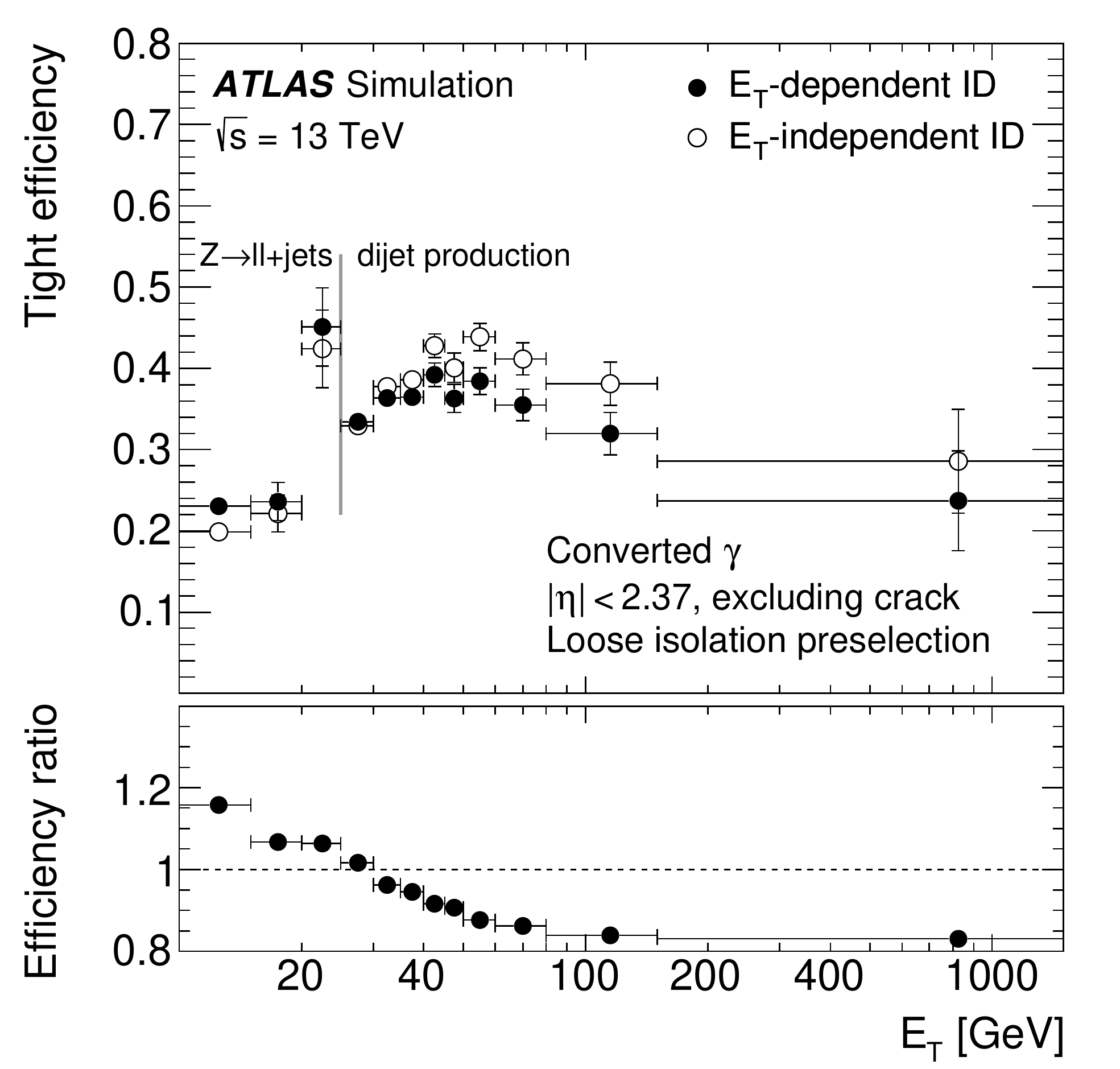}
\caption{
Efficiencies of the Tight photon identification for unconverted (left) and converted (right) background photons from jets,
plotted as a function of photon \et.
The background is taken from $Z{\rightarrow}\ell\ell+$jets
production below 25~\gev, and filtered dijet production above 25~\gev.
In each case, the \et-independent and \et-dependent selections are compared.
The Loose isolation (see Section~\ref{sec:iso_ph_WPs_Eff}) is applied as a preselection.
For both plots, the bottom panel shows the ratios between the \et-dependent and the \et-independent identification efficiencies.
}
\label{fig:photonID:bkgs}
\end{figure}

\begin{figure}[t]
\centering
\includegraphics[width=.49\textwidth]{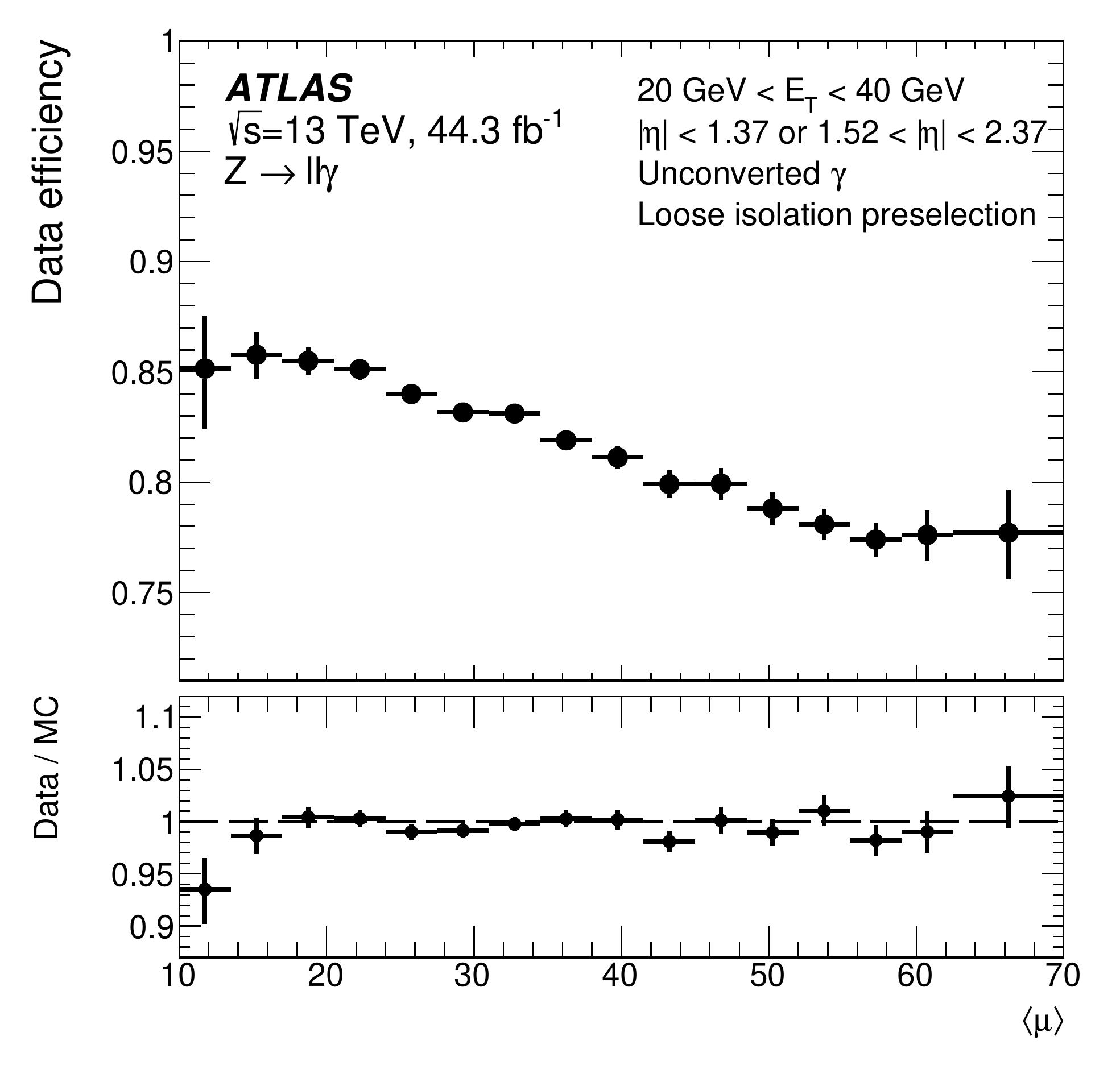}
\includegraphics[width=.49\textwidth]{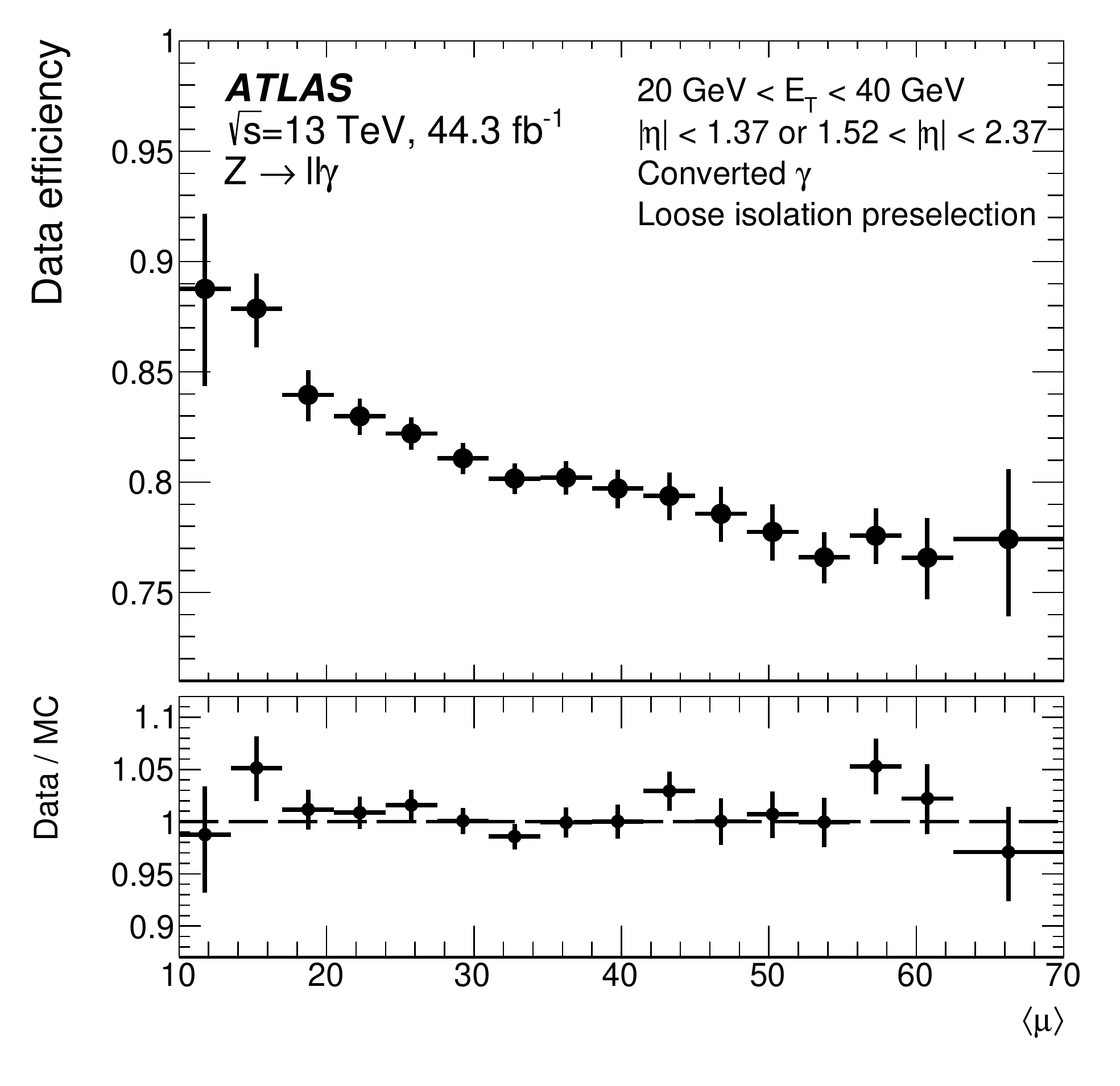}
\caption{Photon identification efficiency as a function of \muhat\ for unconverted (left) and converted photons (right), as
measured by the radiative $Z$ method, for photons with $20<\et<40$~\gev. Backgrounds, which are not
subtracted in this plot, are estimated to be below 1\%. The error bars show the statistical uncertainties.
For both plots, the bottom panel shows the data-to-simulation ratios.}
\label{fig:photonID:muDep}
\end{figure}

\subsection{Efficiency of the photon identification}
 
To assess the performance of the (\et-dependent) Tight photon identification on data,
three photon efficiency measurements are performed using distinct data samples. The first uses
an inclusive-photon production data selection, the second uses photons radiated from leptons in
$Z \to \ell\ell\gamma$ decays, and the third uses electrons from \Zee\ decays, with
a method that transforms the electron shower shapes to resemble the photon shower shapes.
These efficiency measurements are described in detail in Ref.~\cite{PERF-2017-02}, and summarized below.
All three procedures measure photons that are isolated, using the Loose working-point
definition (see Section~\ref{sec:iso_ph_WPs_Eff}).
 
The three measurements use a common method to characterize the imperfect
modelling of shower shapes in simulated samples, in order to estimate its impact on the efficiency
measurement in data. Nominally, the MC shower shapes are compared with data in control regions enriched
in real photons and corrected by applying a simple shift to the distributions, whose magnitude is
determined by a $\chi^2$ minimization procedure. However, some data--MC differences cannot be corrected
by this procedure, such as the widths of the distributions. In order to estimate any residual data--MC
differences, the $\chi^2$ minimisation is repeated considering only the tail of the distribution, defined
as the region containing 30\% of the distribution on the side closer to the identification cut value.
The shift value obtained when comparing the data and simulation tails is used to define a systematic uncertainty in the modelling of the shower shapes, and is derived for all variables for which a
mismodelling is observed. Four variations are defined using sets of correlated variables;
the variables within each set are shifted together:
\{$\Rhad$\}, \{$\Rphi$\}, \{$\Reta$,$\wetatwo$\}, and \{$\wthree$,$\Fside$,$\wtot$\}.
The result is equivalent to four sets of MC simulated samples, which can be used to assign systematic
uncertainties for mismodelling effects that impact the data measurement, and which are considered to
be uncorrelated variations.
 
The method using $Z\to\ell\ell\gamma$ decays selects data as
described in Section~\ref{sec:data_set}. Additional requirements on the invariant mass of the
three-body system, $80<m_{\ell\ell\gamma}<100$~\gev, and on the lepton-pair invariant mass, $40<m_{\ell\ell}<83$~\gev, select radiative $Z$-boson decays while rejecting backgrounds from $Z+\gamma$ and
$Z+$jets production. The efficiency and purity of the samples with and without the Tight identification
requirement are determined from fits of signal and background templates, extracted from simulated
$Z\to\ell\ell\gamma$ and $Z+$jets events, to the observed three-body invariant-mass distribution.
 
The systematic uncertainties in the photon efficiency measurement using $Z\to\ell\ell\gamma$ decays
include a closure test using simulated signal and background samples to assess the validity of the
measurement. To assess the impact of simulation mismodelling, the measurement is repeated comparing the
\textsc{Powheg-Pythia8} and \textsc{Sherpa} $Z \to \ell\ell$ samples and the difference is taken as a systematic
uncertainty. The shower shape correction uncertainties are considered by repeating the
measurement with each of the four sets of modified simulation samples, and the observed differences are
added in quadrature. Finally, as a test of the background description, the fit range of the $m_{\ell\ell\gamma}$
distribution is varied from its nominal value of [65,105]~\gev\ using two variations, [45,95]~\gev\ and [80,120]~\gev, and the
efficiency differences are assigned as a systematic uncertainty.
 
The method to extract the photon efficiency using inclusive-photon production relies on data
collected with prescaled photon triggers that feature a Loose identification requirement,
as described in Section~\ref{sec:data_set}. This
data sample contains a mixture of real photons and backgrounds from jet production, and a matrix method is
used to extract the photon efficiency. The matrix method constructs four regions by categorizing Loose photon candidates according to whether they pass or fail the Tight identification, and
whether they pass or fail track-based isolation cuts. The four regions contain eight unknowns (i.e. the numbers of
signal and background events in each region); if the isolation efficiencies for signal and background from each
region are known, the efficiency for Loose photons to pass the Tight identification can be extracted. The
isolation efficiencies for loosely and tightly identified signal photons are determined from the Monte Carlo samples, and the isolation efficiencies for backgrounds are obtained in a jet-enriched
control region constructed by inverting identification criteria. Finally, the efficiency for reconstructed photon candidates to pass the Loose identification is determined from simulation, as this
contribution is not measured in data by this method. The magnitude of the correction is typically less than 5\%, and smaller at high \et.
 
Systematic uncertainties assigned to the matrix method include a closure uncertainty
that quantifies the agreement between
the background isolation efficiencies derived in the data control region and in the
regions to which they are applied. This effect is estimated using simulation, and is
the largest source of uncertainty in the measurement. The robustness of the method is tested by
varying the track-based isolation requirement, and assigning any difference in measured efficiency
as a systematic uncertainty. The impact of uncertainties in the shower shape corrections is estimated using simulation; the effects of the four shower shape variations described above are added in
quadrature. Finally, an uncertainty is assigned for a potential mismodelling in the MC-based correction to extrapolate from Loose
to reconstructed photons. This uncertainty is based on the Loose identification efficiency measured with
radiative photons in $Z\rightarrow\ell\ell\gamma$ events.
 
Photon efficiencies can be estimated in a data sample of electrons from \Zee\ decays whose
shower shape variables have been modified to resemble photon shower shapes, a technique referred to as
the electron extrapolation method. This efficiency measurement, described in
Ref.~\cite{PERF-2017-02}, uses the \Zee\ sample defined in Section~\ref{sec:data_set}, with the photon
Loose isolation requirement applied to the electron candidates. Electron shower shape
variables are modified using a Smirnov transform~\cite{Smirnov} derived from simulated \Zee\ and
inclusive-photon production samples. The candidate electrons in data contain a small background from
$W+$jets and multijet production; this background is subtracted
by fitting simulated signal samples and background templates derived from data
control regions to the $m_{ee}$ data distributions. The electron candidates are counted for events in the range $70<m_{ee}<110$~\gev, and the efficiencies are measured using the tag-and-probe method
as described in Section~\ref{sec:eID}.
 
The systematic uncertainties in the electron extrapolation method are as follows. First, a closure test
is performed to determine whether the transformed electrons can reproduce the expected photon
efficiency, using the simulation and in the absence of background. The difference in relative efficiency,
which can be as high as 3\%, is applied as a correction to the measured data efficiency, and the magnitude
of the correction is assigned as the systematic uncertainty.
Systematic effects that affect the Smirnov transformations include the fraction of fragmentation
photons in the simulated inclusive-photon sample, which is varied by $\pm50$\%, and the predicted
fraction of true converted photons, which is varied by $\pm10$\%, to assess the impact of the imperfect
simulation on the efficiency measurement.
The uncertainty in the modelling of identification variables in simulation is assessed by
defining Smirnov transformations for each of the four sets of variations of the shower shape modelling,
recalculating the efficiency for each case; the total modelling uncertainty is taken as the
sum in quadrature of the individual variations.
The uncertainty due to the limited size of the MC samples used to derive the Smirnov transformations
is assessed using the bootstrap method.
Finally, the uncertainty associated with the subtraction of the $W+$jets and multijet backgrounds in the signal region is tested
by reducing the level of background through a restriction of the selected invariant-mass range to $80<m_{\ell\ell}<100$~\gev, and repeating the measurement procedure. The resulting difference in the measured efficiency is taken as the systematic uncertainty.

The three efficiency measurements are compared with MC simulation in order to obtain scale factors, in bins
of \et\ and $|\eta|$,
that are
used to correct the MC simulations so that the simulations
closely resemble data.
Before determining these scale factors, the shower shapes in these MC simulations were
corrected to match data using the procedure described in Ref.~\cite{PERF-2017-02}.

Figures~\ref{fig:photonEffs:unconverted} and \ref{fig:photonEffs:converted} depict the Tight identification efficiencies for unconverted and
converted photons as measured with the three efficiency methods.
The data/MC scale factors are also shown for each measurement separately. The three efficiency measurements
are performed using different processes, with different event topologies that may impact the photon efficiency. Despite this fact,
the efficiency measurements are compatible within their statistical and systematic uncertainties.
 
\begin{figure}[t]
\centering
\includegraphics[width=.49\textwidth]{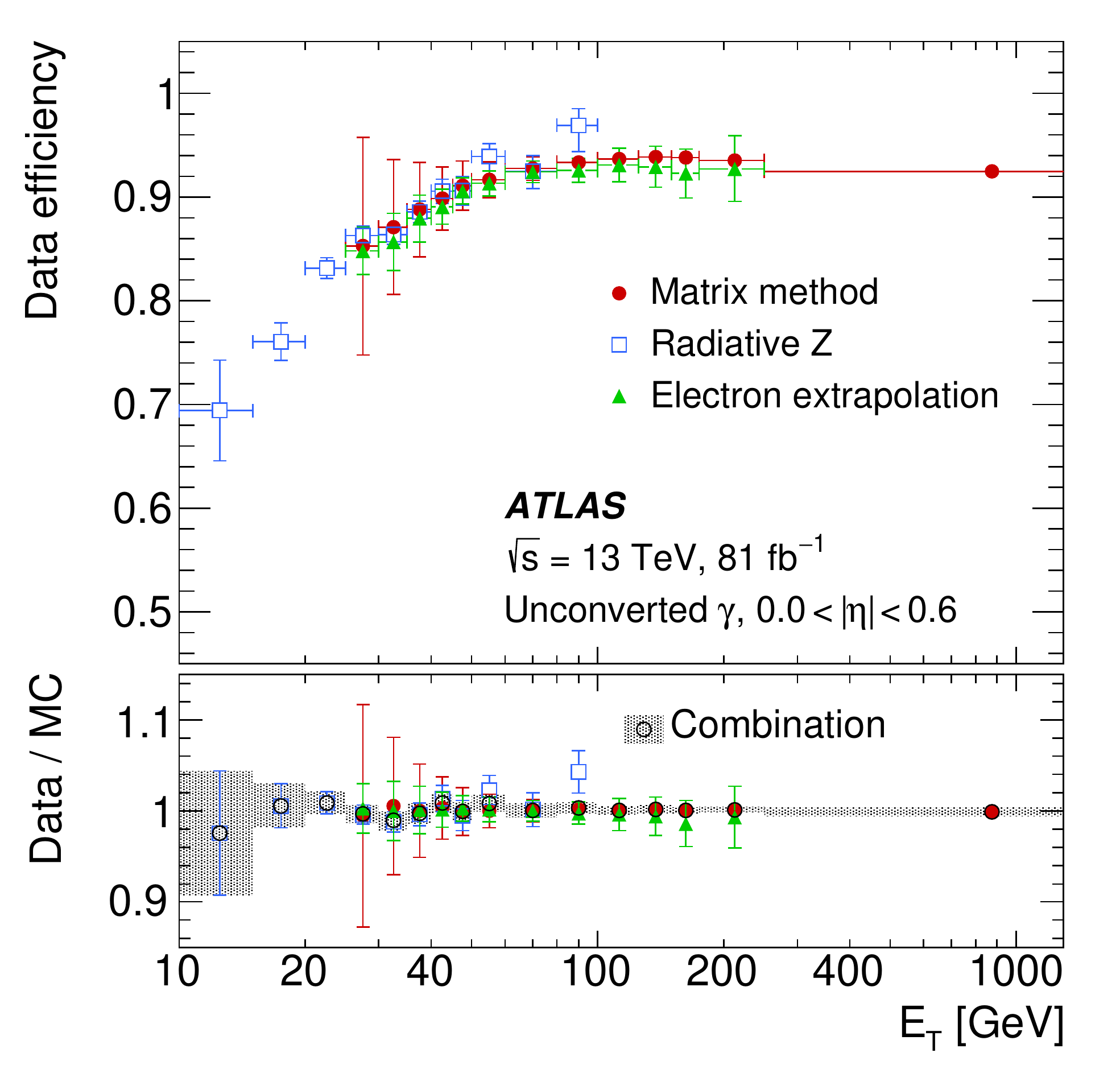}
\includegraphics[width=.49\textwidth]{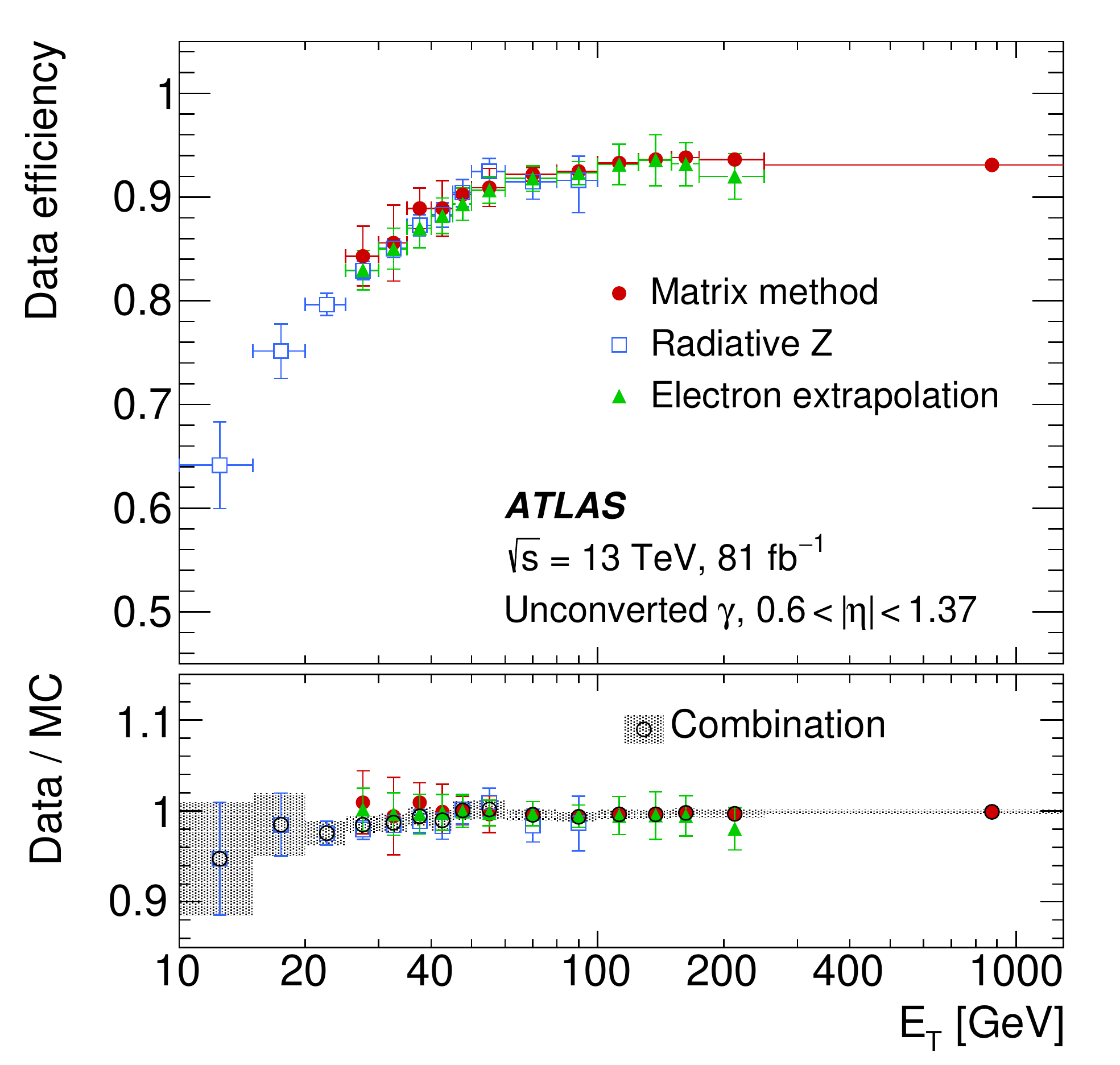}\\
\includegraphics[width=.49\textwidth]{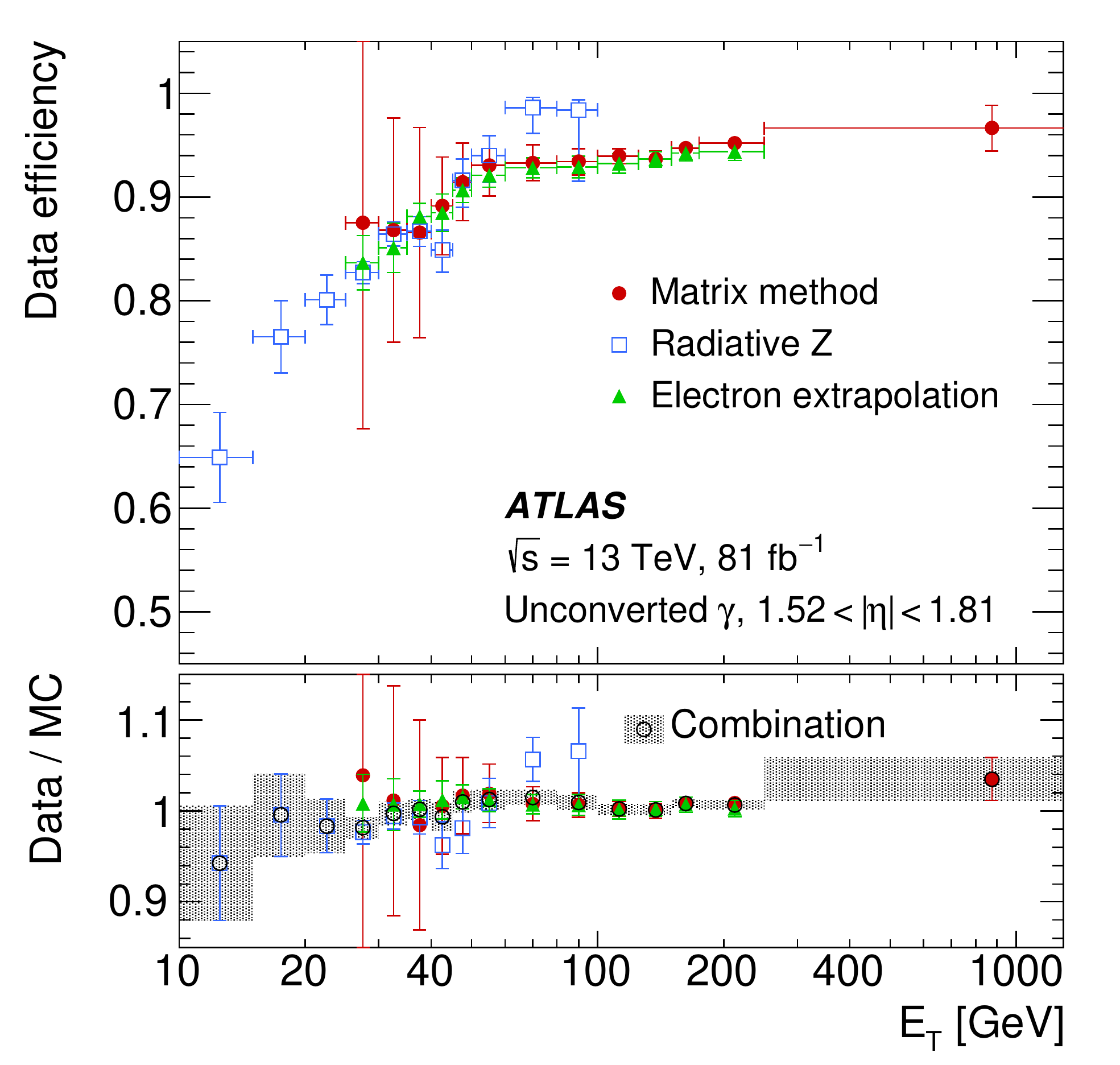}
\includegraphics[width=.49\textwidth]{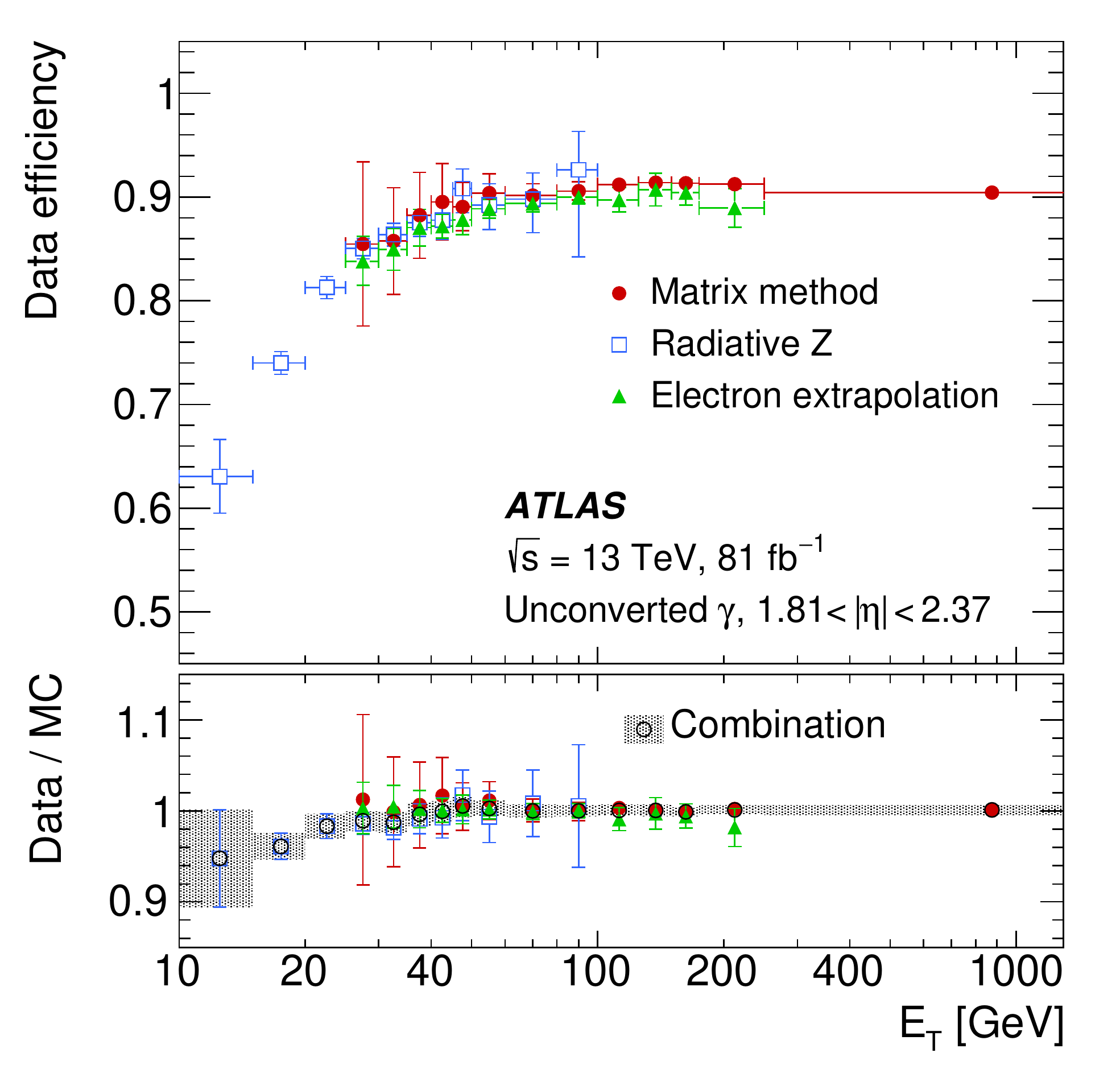}
\caption{The photon identification efficiency, and the ratio of data to MC efficiencies, for unconverted photons
with a Loose isolation requirement applied as preselection, as a function of \et in four different $|\eta|$ regions.
The combined scale factor, obtained using a weighted average of scale factors from the individual measurements, is also presented; the band represents the total uncertainty.
}
\label{fig:photonEffs:unconverted}
\end{figure}
 
\begin{figure}[t]
\centering
\includegraphics[width=.49\textwidth]{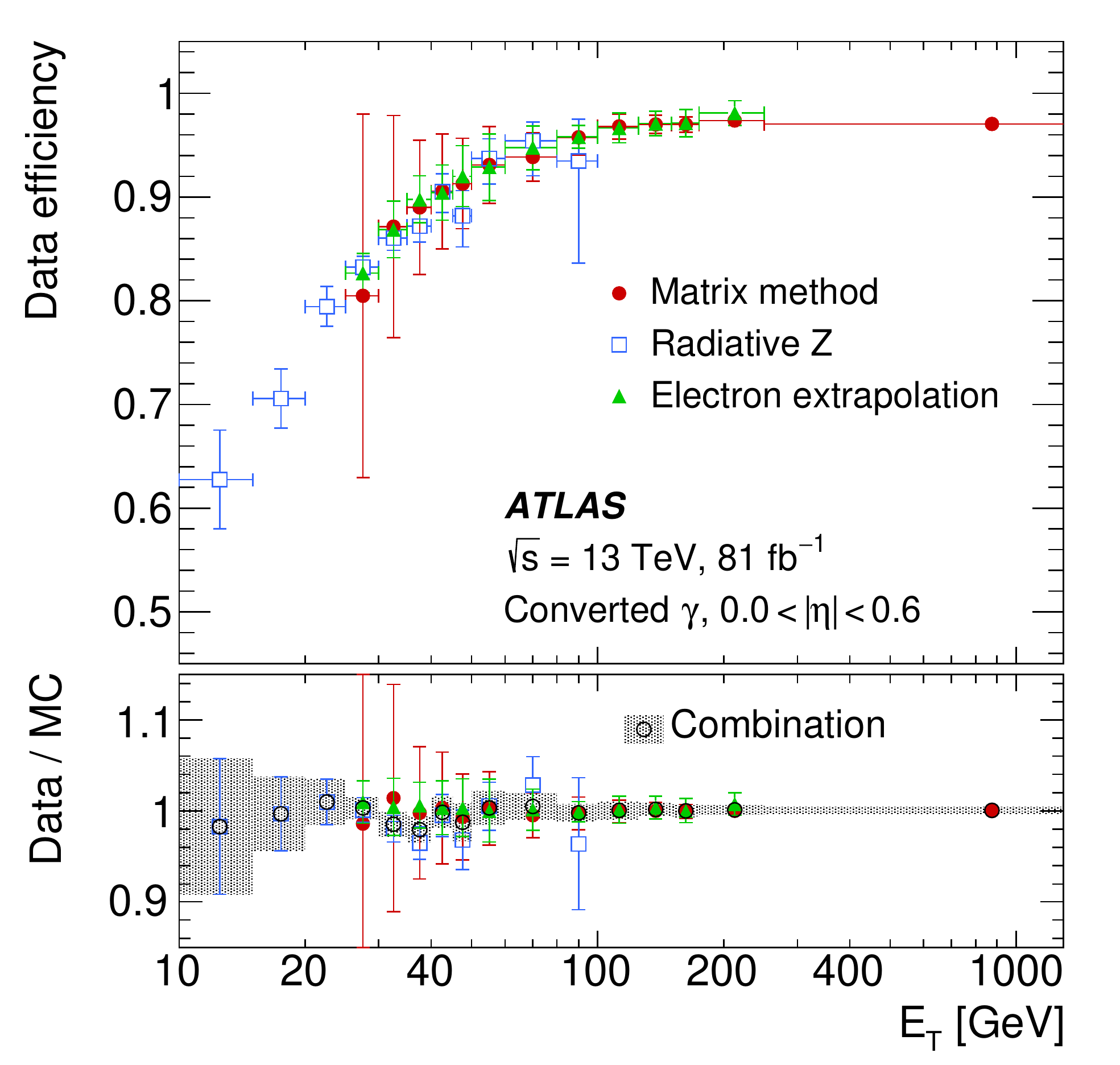}
\includegraphics[width=.49\textwidth]{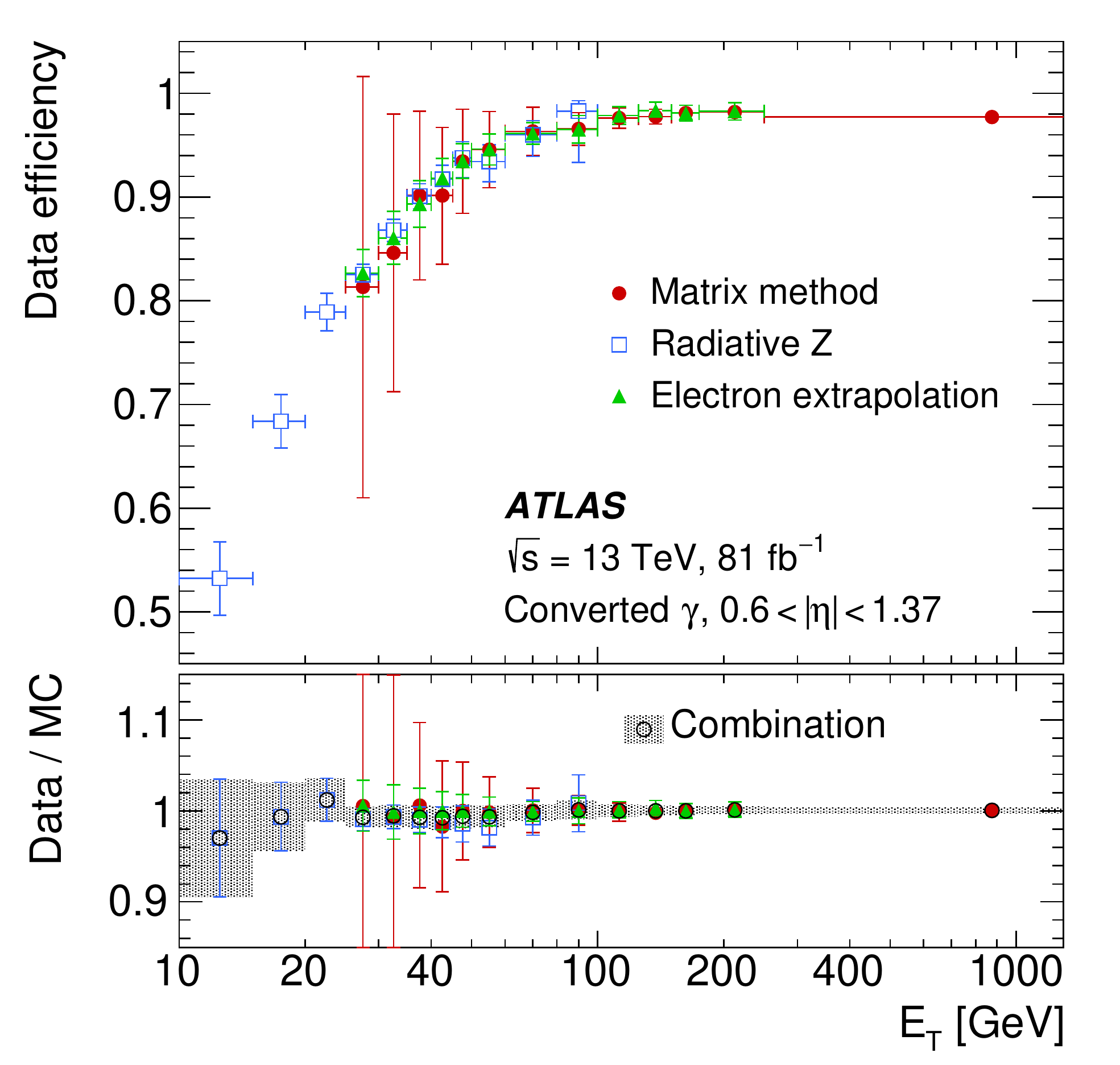}\\
\includegraphics[width=.49\textwidth]{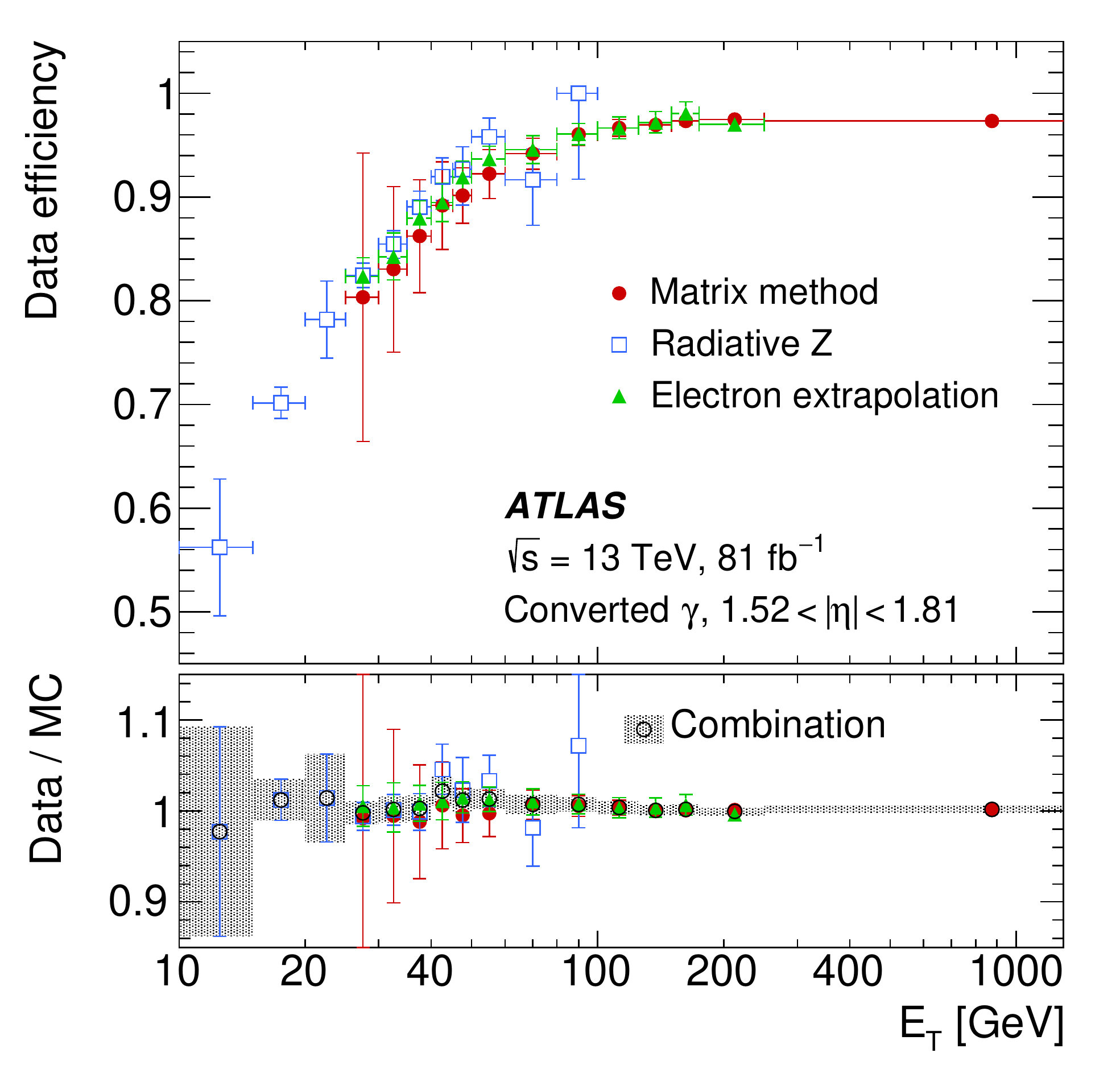}
\includegraphics[width=.49\textwidth]{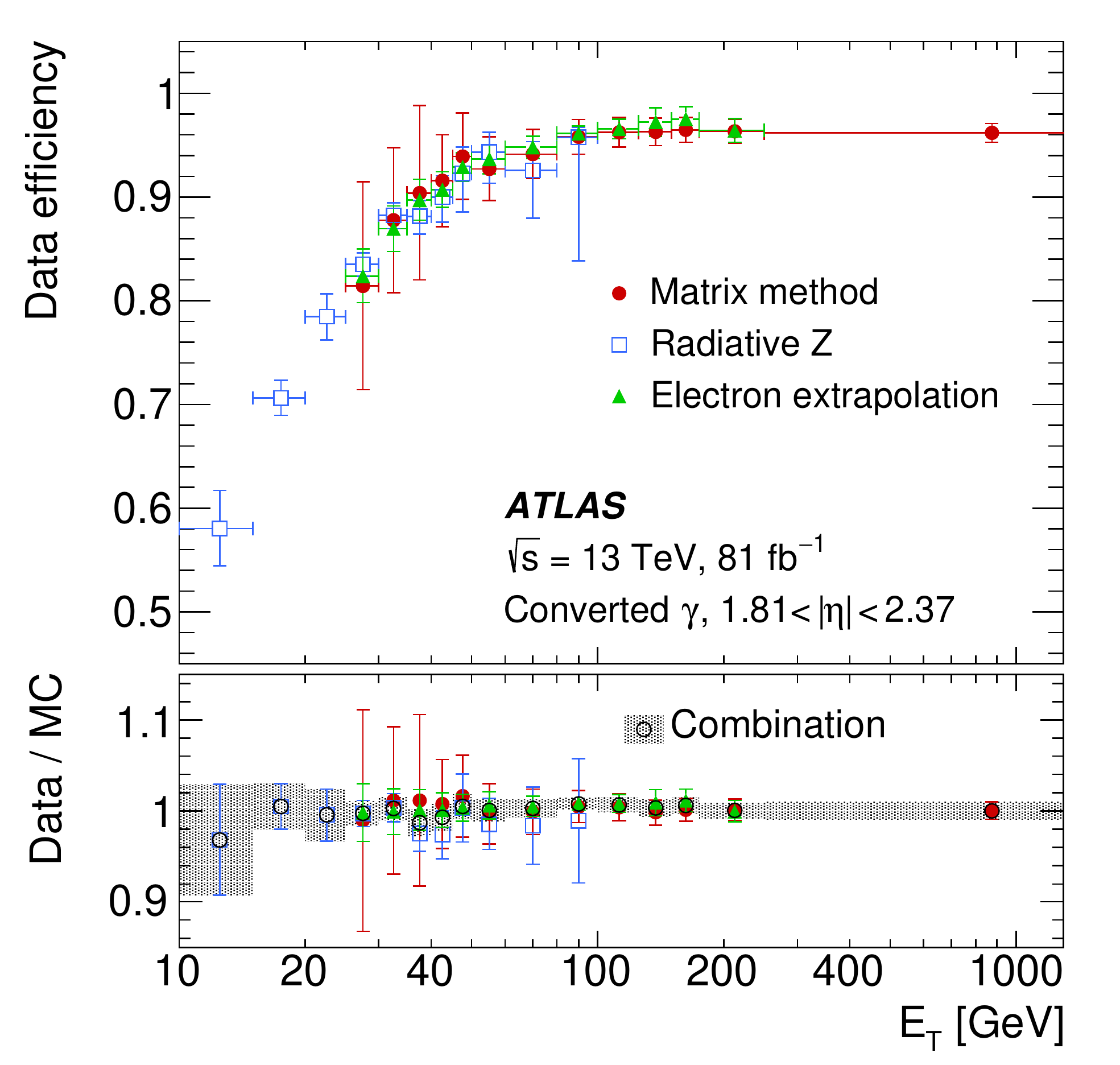}
\caption{The photon identification efficiency, and the ratio of data to MC efficiencies, for converted photons
with a Loose isolation requirement applied as preselection, as a function of \et in four different $|\eta|$ regions.
The combined scale factor, obtained using a weighted average of scale factors from the individual measurements, is also presented; the band represents the total uncertainty.
}
\label{fig:photonEffs:converted}
\end{figure}
 
The scale factors from each of the three efficiency measurements are combined using a weighted average.
The statistical and systematic
uncertainties are assumed to be uncorrelated between the methods. The total uncertainty of the combined scale factors
ranges between 7\% at low \et\ and 0.5\% at high \et\ for unconverted photons, and between
12\% (low \et) and less than 1\% (high \et) for converted photons. For \et > 1.5 \tev, where no measurement
is performed, the scale factor measured in the \et bin [0.25,1.5] \tev\ is used, with the same uncertainty.
 
\section{Electron and photon isolation}
\label{sec:iso}
The activity near leptons and photons can be quantified from the tracks of nearby charged particles, or from energy deposits in the calorimeters, leading to two classes of isolation variables.
 
The raw calorimeter isolation~\cite{PERF-2017-01} ($E_{\mathrm{T, raw}}^{\mathrm{isol}}$) is built by summing the transverse energy of positive-energy topological clusters whose barycentre falls within a cone centred around the electron or photon cluster barycentre.
The topological cluster energy scale is the EM scale. The raw calorimeter isolation includes the EM particle energy ($E_{\mathrm{T,core}}$), which is subtracted by removing the energy of the EM calorimeter cells contained in a $\Delta\eta \times \Delta\phi = 5\times7$ (in EM-middle-layer units) rectangular cluster around the barycentre of the EM particle cluster. The advantage of this simple method is a stable subtraction for real or fake/non-prompt objects for any transverse momentum and pile-up. The disadvantage is that it does not subtract all the EM particle energy and an additional leakage correction is needed. This leakage is parameterized as a function of \et\ and $|\eta|$ using MC samples of single electrons or photons without pile-up. Additionally, a correction for the pile-up and underlying-event contribution to the isolation cone is also estimated~\cite{pusub}.
 
Finally, the fully corrected calorimeter isolation variable is computed as:
\begin{equation}
\etcx = E_{\mathrm{T,raw}}^{\mathrm{isol\mathrm{XX}}} - E_{\mathrm{T,core}} - E_{\mathrm{T,leakage}}(\et,\eta,\Delta R) - E_{\mathrm{T,\textrm{pile-up}}}(\eta,\Delta R),
\nonumber
\end{equation}
where $\mathrm{XX}$ refers to the size of the employed cone, $\Delta R = \mathrm{XX}/100$. A cone size $\Delta R=0.2$ is used for the electron working points whereas cone sizes $\Delta R=0.2$ and 0.4 are used for photon working points.
 
The track isolation variable (\ptcx) is computed by summing the transverse momentum of selected tracks within a cone centred around the electron track or the photon cluster direction.
Tracks matched to the electron or converted photon are excluded.
Since for electrons produced in the decay of high-momentum heavy particles, other decay products can be very close to the electron direction, the track isolation for electrons is defined with a variable cone size (\ptvcx) -- the cone size shrinks for larger transverse momentum of the electron:
\begin{equation}
\Delta R = \min \left( \frac{10}{\pt[\gev]}, \Delta R_{\mathrm{max}}\right),
\nonumber
\end{equation}
where $\Delta R_{\mathrm{max}}$ is the maximum cone size (typically 0.2).
 
The tracks considered are required to have $\pt > 1\ \gev$ and $|\eta| < 2.5$, at least seven silicon (Pixel + SCT) hits, at most one shared hit
(defined as $n^{\text{sh}}_{\text{Pixel}}+n^{\text{sh}}_{\text{SCT}}/2$, where $n^{\text{sh}}_{\text{Pixel}}$ and $n^{\text{sh}}_{\text{SCT}}$ are the numbers of hits assigned to several tracks in the Pixel and SCT detectors), at most two silicon holes (i.e. missing hits in the pixel and SCT detectors) and at most one pixel hole.
In addition, for electron isolation, the tracks are required to have a loose vertex association, i.e. the track was used in the primary vertex fit, or it was not used in any vertex fit but satisfies $|\Delta z_0|\sin\theta < 3$~mm, where $|\Delta z_0|$ is
the longitudinal impact parameter relative to the chosen primary vertex; for photon isolation, all selected tracks satisfying
$|\Delta z_0|\sin\theta < 3$ mm are used.
 
In this section, the isolation efficiency measurements are illustrated with the data recorded in 2017; nevertheless, the measurements are performed for the full high-$\mu$ dataset described in Section~\ref{sec:data_set}.
 
\subsection{Electron isolation criteria and efficiency measurements}
 
The implementation of isolation criteria is specific to the physics analysis needs, as it results from a compromise between a highly-efficient identification of prompt electrons, isolated or produced in a busy environment, and a good rejection of electrons from heavy-flavour decays or light hadrons misidentified as electrons.
The different electron-isolation working points used in ATLAS are presented in Table~\ref{tab:ele_wps}.
 
The working points can be defined in two different ways, targeting a fixed
value of efficiency or with fixed cuts on the isolation variables. The Gradient
working point is designed to give an efficiency of 90\% at $\pt = 25\ \gev$ and
$99\%$ at $\pt=60\ \gev$, uniform in $\eta$. The requirements on \etcs\ and \ptvcs\ (cut maps) for this working point
are derived from $J/\psi \to ee$ ($\et < 15\ \gev$) and \Zee\ ($\et > 15\ \gev$) MC simulations and Tight identification requirements. The three other working points, HighPtCaloOnly, Loose and Tight, have a fixed requirement on the calorimeter and/or the track isolation variables.
 
\begin{table}[!ht]
\centering
\caption{Definition of the electron isolation working points and isolation efficiency $\epsilon$.
In the Gradient working point definition, the unit of \pt is \gev. All working points use a cone size
of $\Delta R = 0.2$ for calorimeter isolation and $\Delta R_{\mathrm{max}} = 0.2$ for track isolation.}
\resizebox{\textwidth}{!}{
\begin{tabular}{lcc}
\hline \hline
Working point & Calorimeter isolation  & Track isolation    \\ \hline
 
Gradient         & $\epsilon=0.1143\times \pt+92.14\%$ (with \etcs)       & $\epsilon=0.1143\times \pt+92.14\%$ (with \ptvcs) \\ \hline
HighPtCaloOnly & $\etcs < \text{max}(0.015\times \pt, 3.5~ \text\gev)$ & -                                                            \\
Loose          & $\etcs/\pt < 0.20$                                     & $\ptvcs /\pt < 0.15$                \\
Tight          & $\etcs/\pt < 0.06$                                     & $\ptvcs /\pt < 0.06$                \\
\hline
\hline
\end{tabular}
}
\label{tab:ele_wps}
\end{table}

\begin{figure}[th!]
\centering
\includegraphics[width=0.49\textwidth]{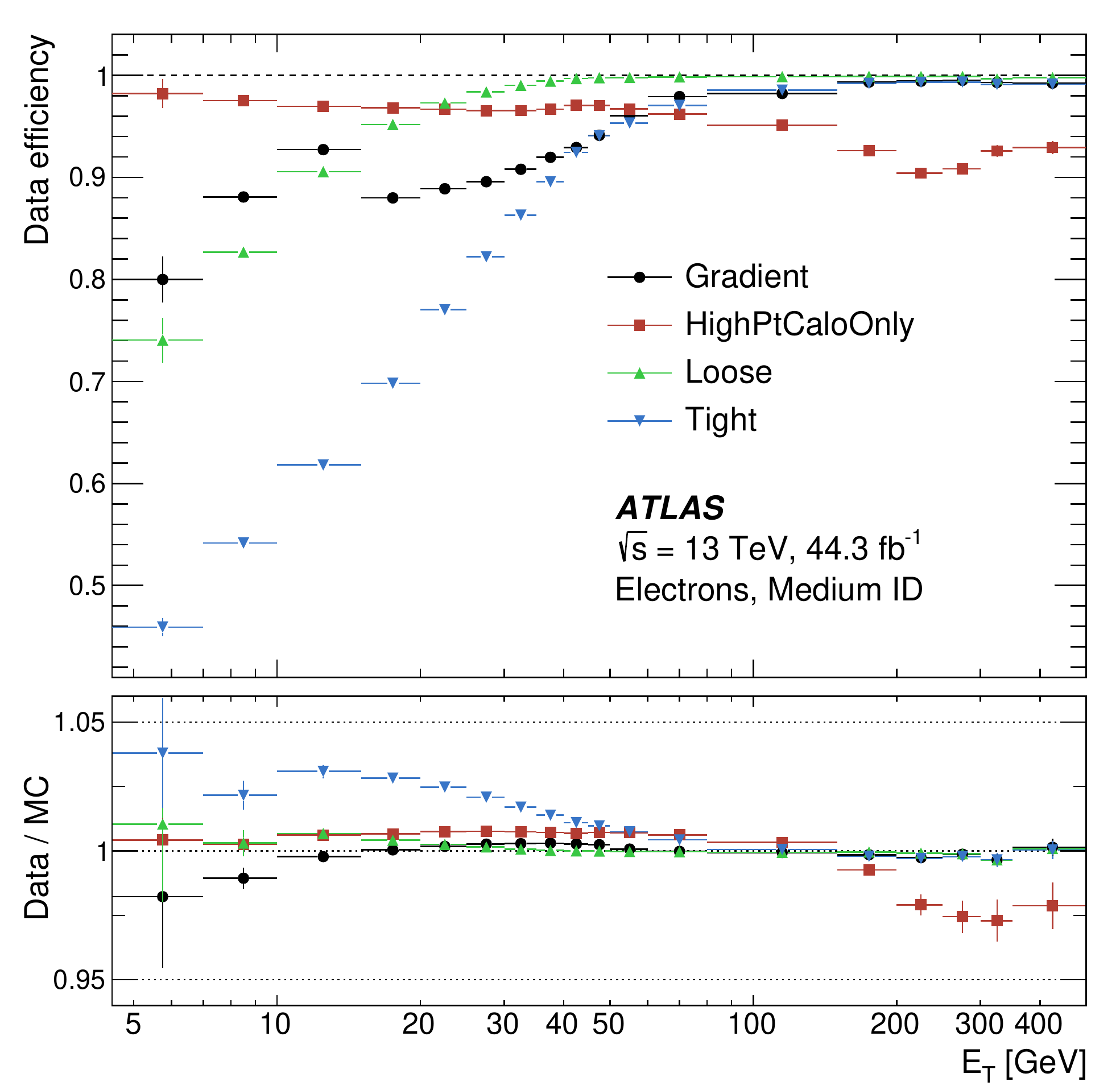}
\includegraphics[width=0.49\textwidth]{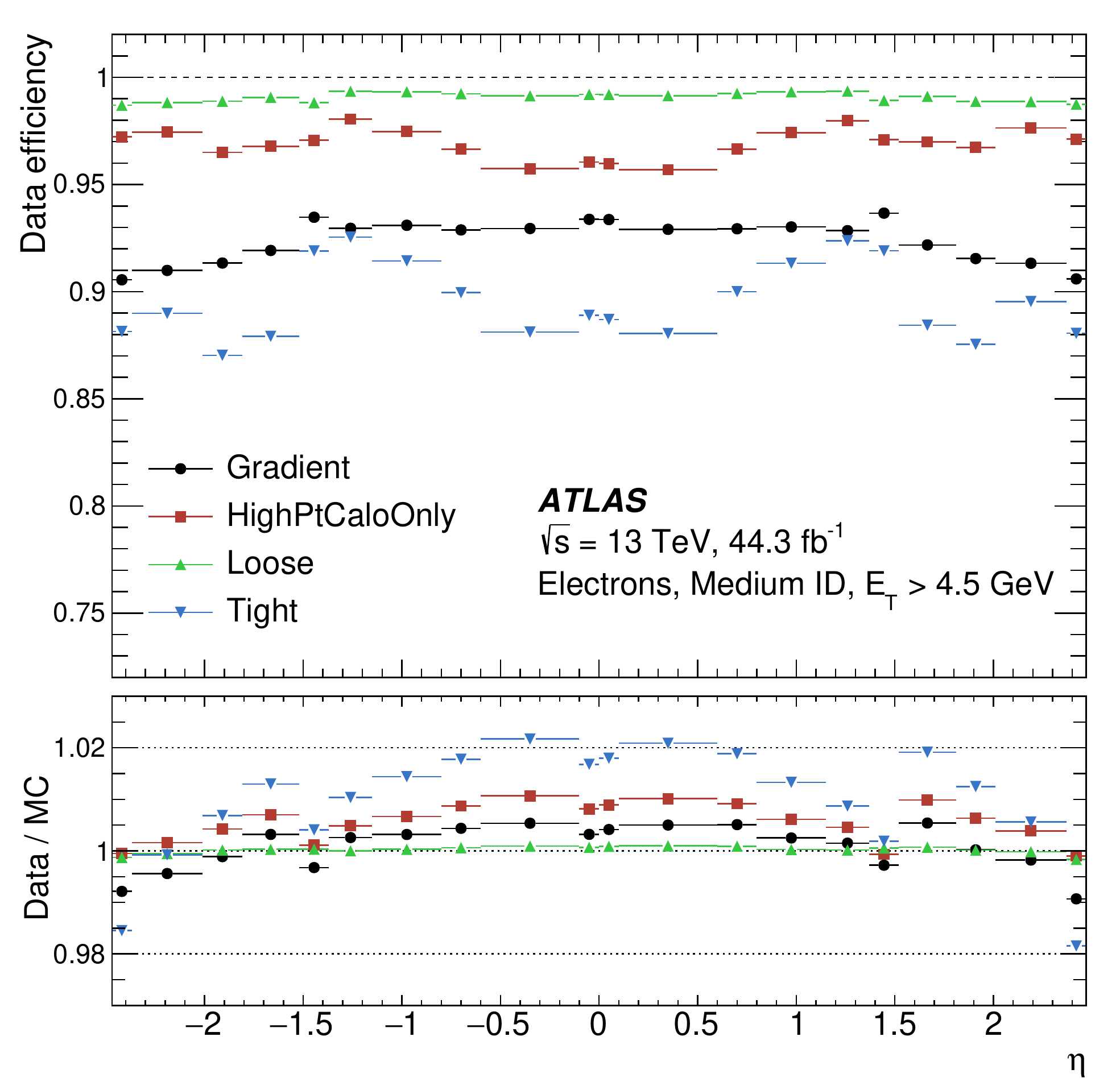}
\includegraphics[width=0.49\textwidth]{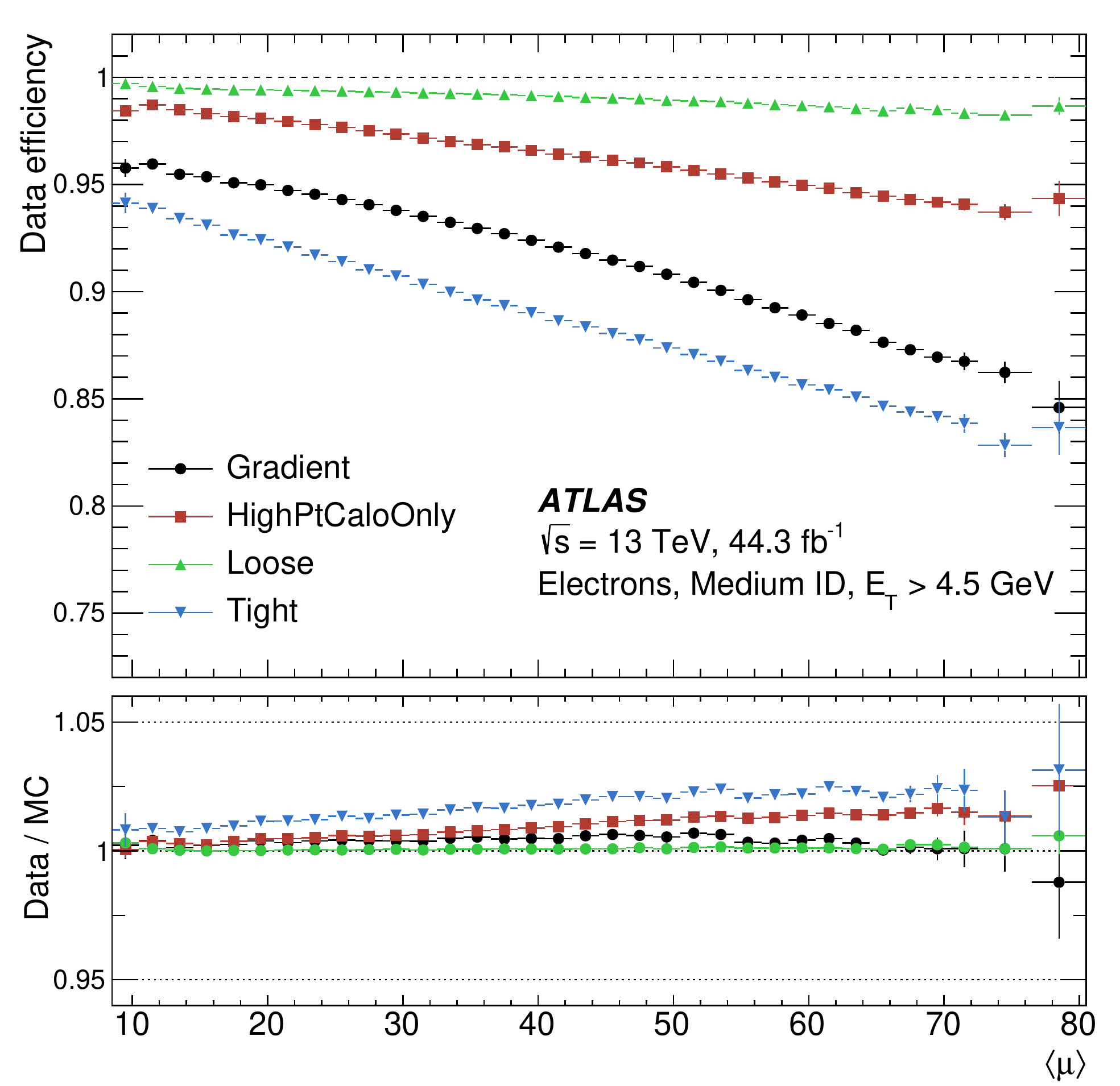}
\caption{Efficiency of the different isolation working points for electrons
from inclusive \Zee\ events as a function of the electron $\ET$ (top left),
electron $\eta$ (top right) and the number of interactions per bunch crossing \muhat\ (bottom).
The electrons are required to fulfil the Medium selection from the likelihood-based electron
identification.
The lower panel shows the ratio of the efficiencies measured in data and in MC simulations. The total uncertainties are shown, including the statistical and systematic components.}
\label{fig:eleiso_inclusive}
\end{figure}

Figure \ref{fig:eleiso_inclusive} shows the electron isolation efficiency measured in data recorded in 2017 and the corresponding data-to-MC simulation ratios as a function of the electron \ET and $\eta$, and of the number of interactions per bunch crossing for the isolation working points summarized in Table~\ref{tab:ele_wps}. The pile-up correction to the calorimeter isolation is applied, and reduces the dependence of the isolation efficiency by about a factor of five.
These results are obtained using a sample enriched in \Zee\ events, where the electrons satisfy the Medium identification. The method used to compute the electron isolation efficiency and the associated uncertainties are described in Ref.~\cite{PERF-2017-01}.
For Gradient, a jump in the efficiency is observed at the transition point of 15~\gev\ because the value of the isolation efficiency is process dependent: the cut maps are optimized with $J/\psi \to ee$ events below 15~\gev, while the measurement is performed with \Zee\ events in the full range.
The Tight operating point gives the highest background rejection below 60~\gev\ and the most significant difference in shape in $\eta$.
As the name suggests, HighPtCaloOnly gives the highest rejection in the high-\ET region ($\et > 100~\gev$). The Gradient and Tight operating points give the highest pile-up
dependency, the isolation efficiency decreasing from $\sim$95\% at low \muhat\ to $\sim$85\% when \muhat\ is around 70--80.
 
The overall differences between data and MC simulation are less than approximately 1–5\% depending on the working point, with the largest difference observed for Tight isolation. For electrons
with \ET\ higher than $500~\gev$ no measurement can be performed because of the limited number of data events, and the results from the \et bin [300,500] \gev\ are used with an additional systematic
uncertainty varying betwen 0.1\% and 1.7\%, depending on the isolation working point. The overall scale factor uncertainties range from about 5\% for electrons with \ET below 7~GeV, to less than 0.5\% towards high \ET.

\subsection{Photon isolation criteria and efficiency measurements}
\label{sec:iso_ph_WPs_Eff}
 
\begin{table}[!ht]
\centering
\caption{Definition of the photon isolation working points.}
\begin{tabular}{lcc}
\hline \hline
Working point & Calorimeter isolation & Track isolation \\ \hline
 
Loose & $\etcs < 0.065 \times \et$ & $\ptcs/\et < 0.05$ \\
Tight & $\etcb < 0.022 \times \et + 2.45$ \gev & $\ptcs /\et < 0.05$ \\
TightCaloOnly & $\etcb < 0.022 \times \et + 2.45$ \gev & - \\
\hline
\hline
\end{tabular}
\label{tab:phWPs}
\end{table}
 
\begin{figure}[th!]
\centering
\includegraphics[width=0.49\textwidth,angle=0]{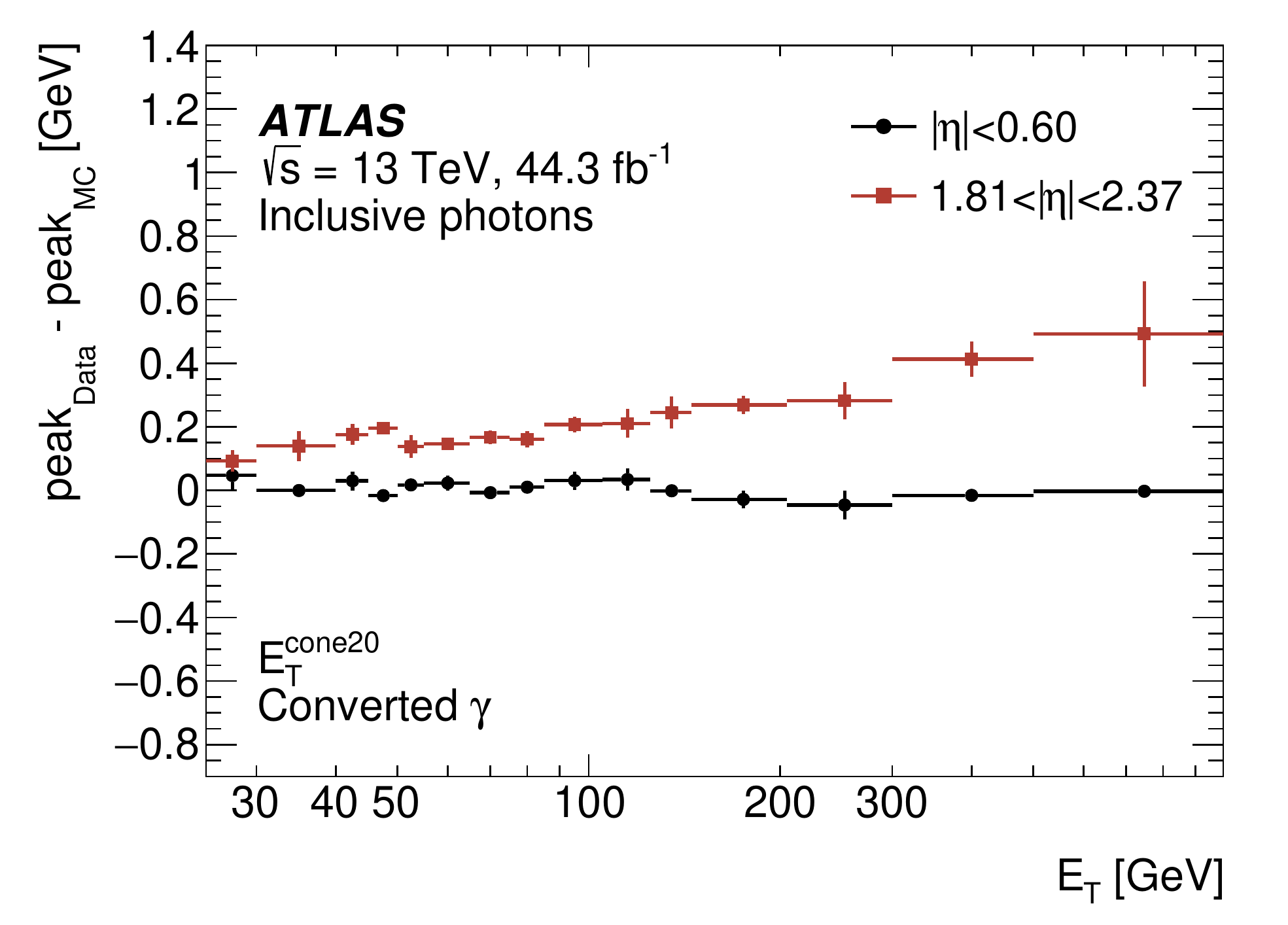}
\includegraphics[width=0.49\textwidth,angle=0]{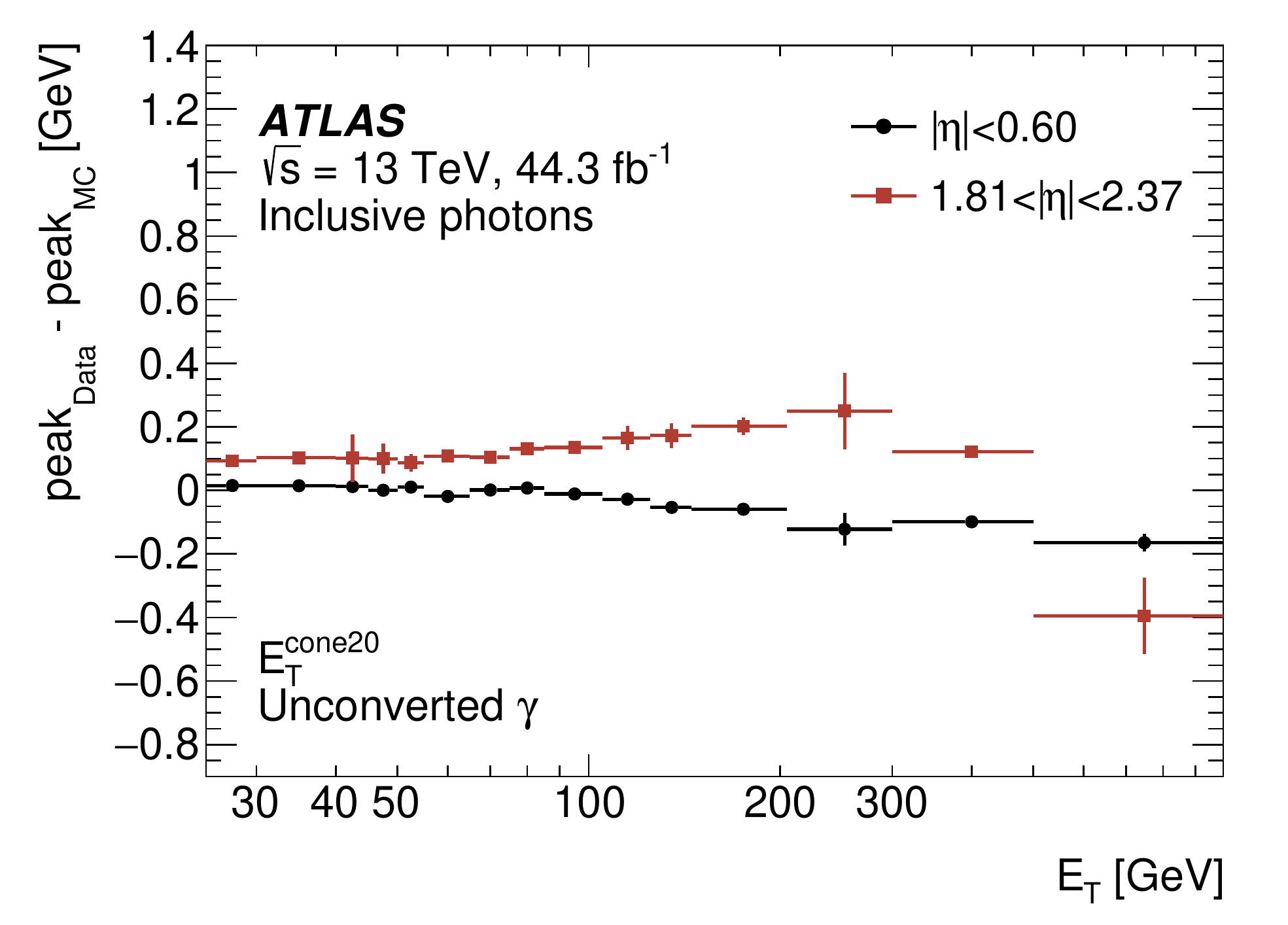}
\hspace{0mm}
\includegraphics[width=0.49\textwidth,angle=0]{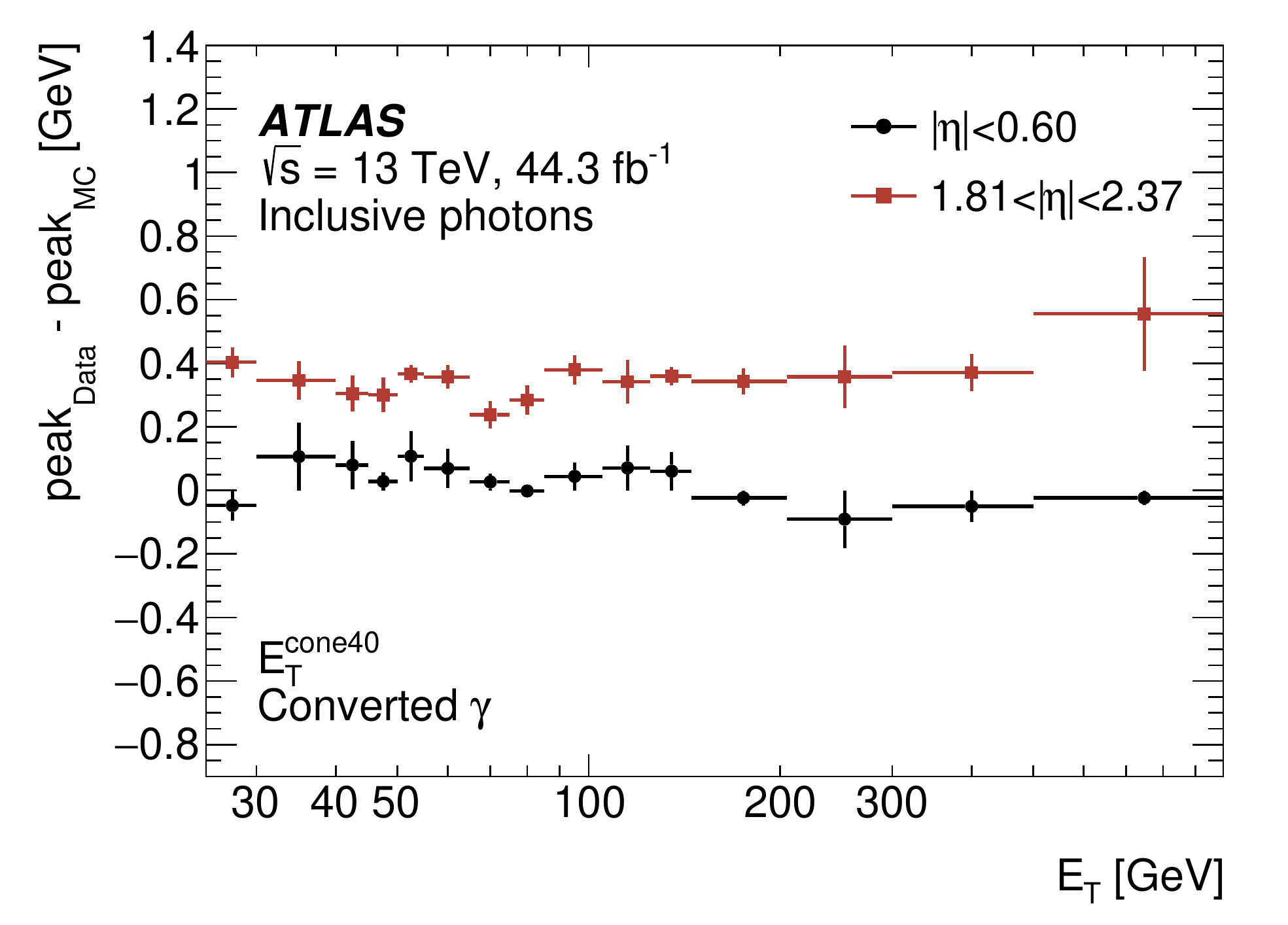}
\includegraphics[width=0.49\textwidth,angle=0]{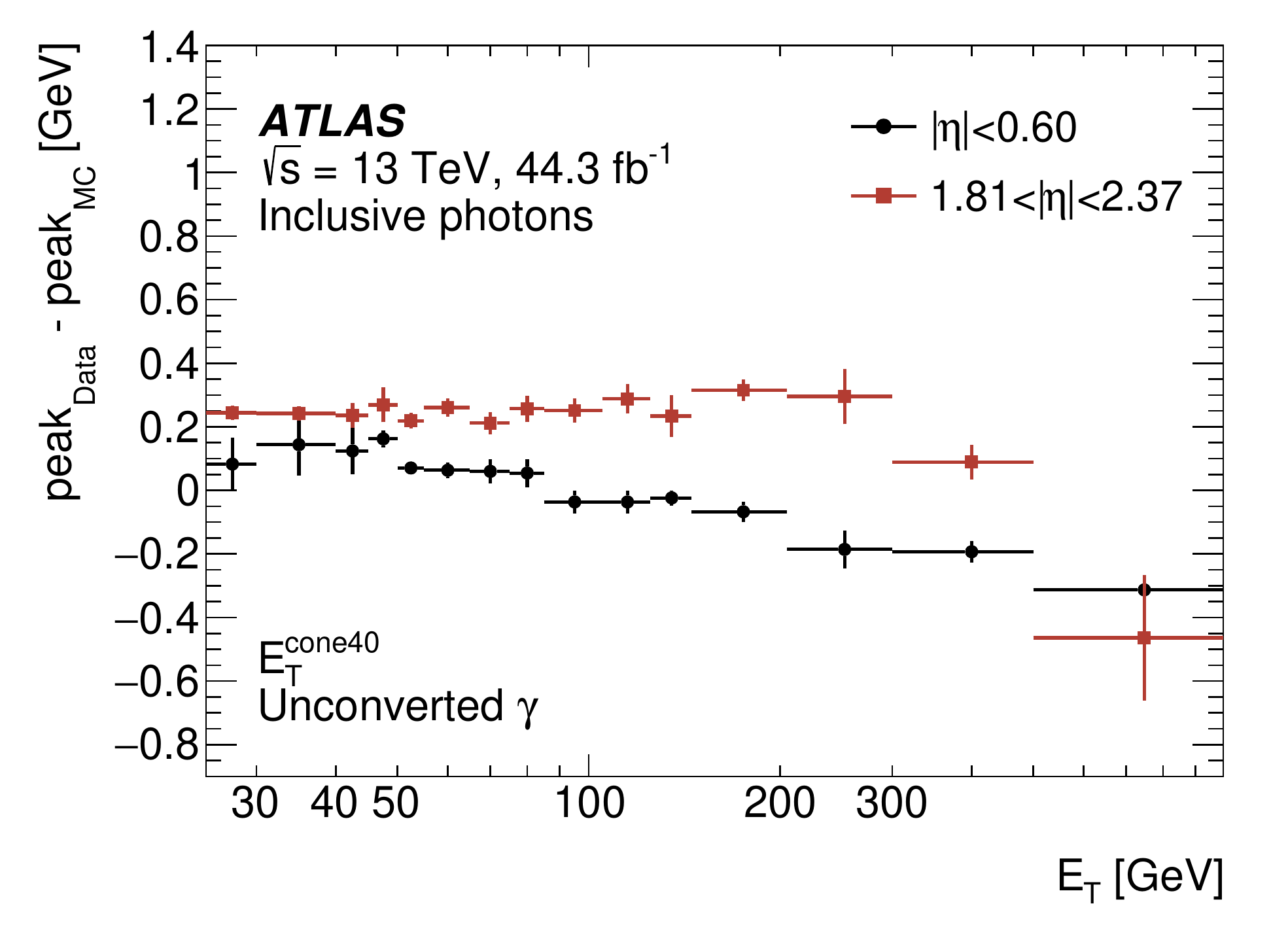}
\caption{The data-driven shifts for \etcs (top) and \etcb (bottom) obtained with 2017 data and \textsc{Pythia8} MC simulations; the Tight isolation working point is applied as preselection to decrease the level of background. The results are shown as a function of photon \ET, in two $\eta$ regions of the detector ($|\eta|<0.6$ and $1.81<|\eta|<2.37$), separately for converted (left) and unconverted (right) photons. Only the uncertainties associated with the fit parameters are shown.}
\label{fig:photon_SgPh_DDshifts}
\end{figure}
 
\begin{figure}[th!]
\centering
\includegraphics[width=0.49\textwidth,angle=0]{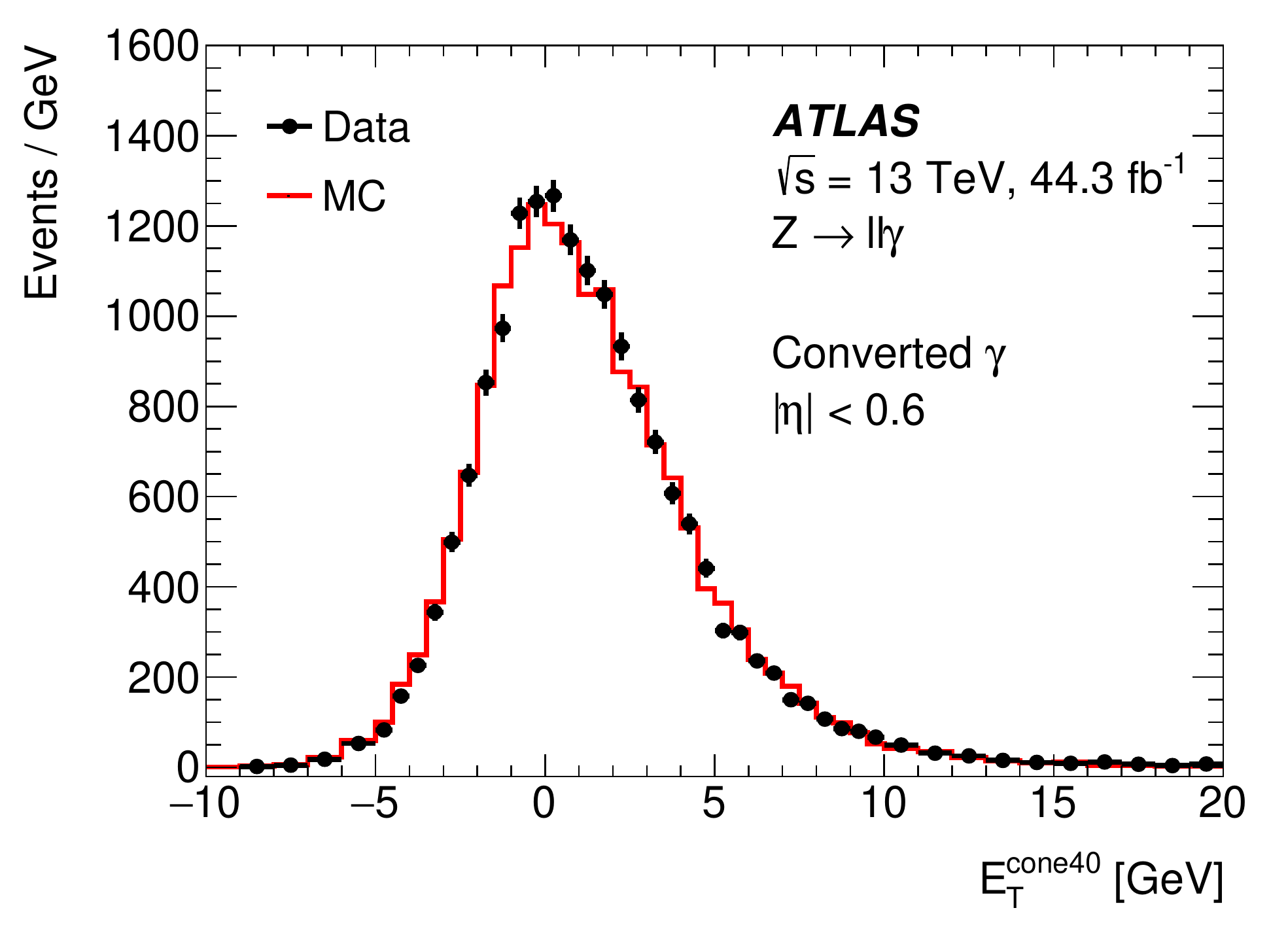}
\includegraphics[width=0.49\textwidth,angle=0]{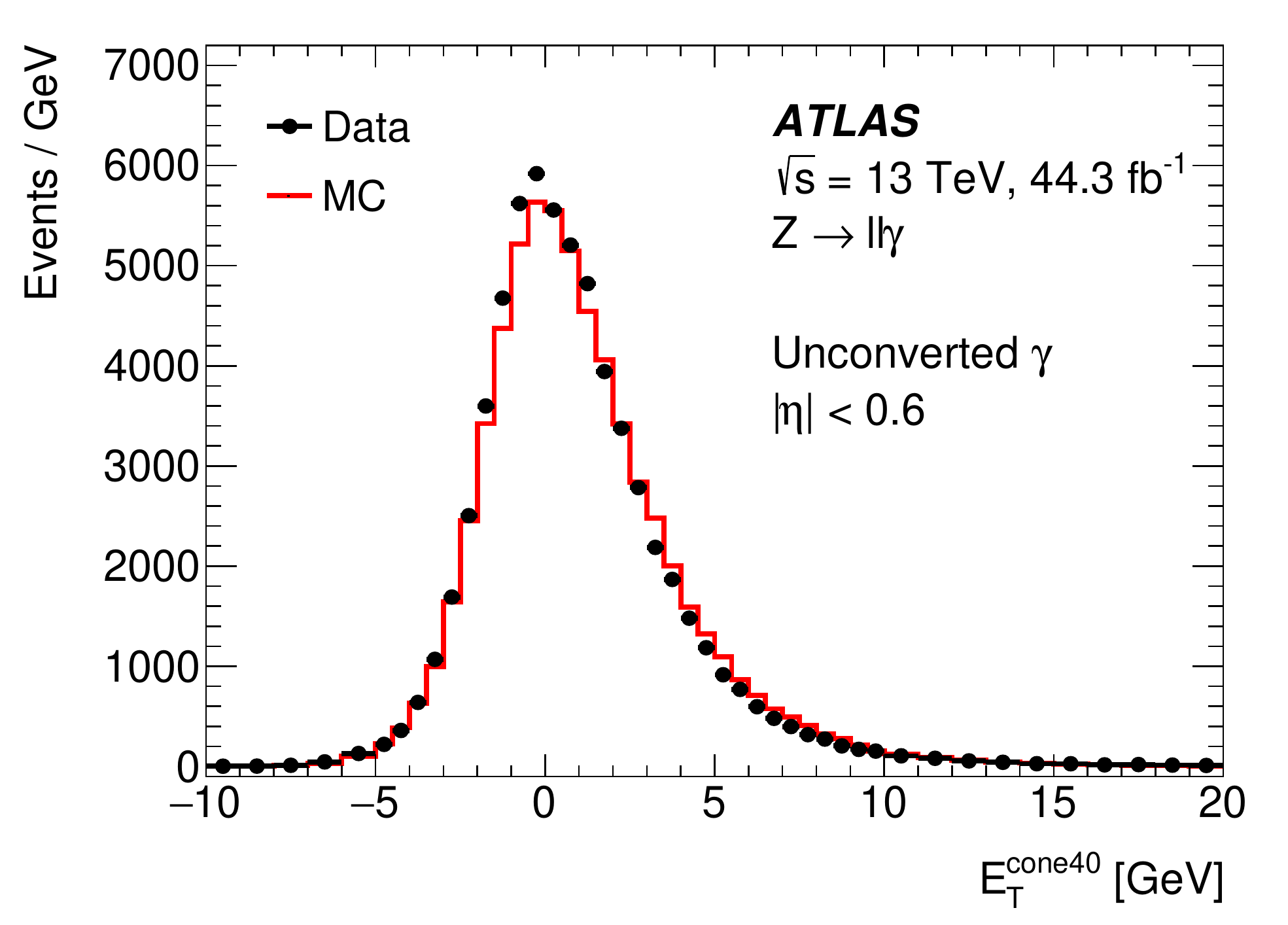}
\caption{Distribution of \etcb in data and simulation using $Z\to\ell\ell\gamma$ events, in the central region of the detector ($|\eta|<0.6$), separately for converted (left) and unconverted (right) photons after the data-driven shifts are applied. Only the statistical uncertainties are shown.}
\label{fig:photon_RadZ_cone40}
\end{figure}
 
Three photon isolation operating points are defined using requirements on the calorimeter and track isolation variables, as summarized in Table~\ref{tab:phWPs}. For the calorimeter-based photon isolation variables a discrepancy between the peak positions of their distributions in data and simulation has been observed since Run~1~\cite{PERF-2010-04}, pointing to a mismodelling in simulation of the lateral profile development of the electromagnetic showers. As a result, the photon isolation efficiencies in data and simulations disagree, leading to scale factors significantly different from 1.
 
These discrepancies are mitigated by applying data-driven shifts to the calorimeter isolation variables for photons in simulation. The shifts are obtained by performing fits to the calorimeter isolation variable distribution, using Crystal Ball pdfs~\cite{CB}, in regions dominated by real photons, in data and simulation.
The fits are performed in bins of photon $\eta$, \ET and conversion status, separately for \etcs and \etcb isolation variables.
The difference in the fitted peak values between data and simulation defines the shift value, which is added to the photon calorimeter isolation values in simulation.
Figure~\ref{fig:photon_SgPh_DDshifts} illustrates the data-driven shifts obtained with 2017 data and the \textsc{Pythia8} simulation for the \etcs and \etcb isolation variables in two $\eta$ regions.
Figure~\ref{fig:photon_RadZ_cone40} shows the distribution of the \etcb isolation variable in 2017 data and simulation, using $Z\to\ell\ell\gamma$ events after the data-driven shifts are applied.
 
The photon isolation efficiency is studied in two main signatures: radiative $Z$ decays (valid for $10 < \et < 100\ \gev$) and inclusive photons (used in the $25\ \gev < \ET < \, \sim$1.5 \tev\ range).

\subsubsection{Measurement of photon isolation efficiency with radiative $Z$ decays}
 
As detailed in Section~\ref{sec:pID}, final-state radiation in $Z$-boson decays provides a clean environment to probe photons in the low-\et range. Using the same method as for the photon identification, photon isolation efficiencies are measured for the operating points presented in Table~\ref{tab:phWPs}. The evolution of the isolation efficiency measured in 2017 data as a function of $\eta$ and \et is illustrated in Figure~\ref{fig:photon_radZ_isoEffvsetaEt}, together with the data-to-simulation efficiency ratio.
The overall differences between data and simulation are less than approximately 5\%.
The decrease of efficiency with increasing pile-up activity is shown in Figure~\ref{fig:photon_radZ_isoEffvsmu}.
A loss of efficiency of $\sim$10\% is measured when increasing \muhat\ from 15 to 60.
This loss is well described by the simulation.
 
\begin{figure}[h!]
\centering
\includegraphics[width=0.49\textwidth,angle=0]{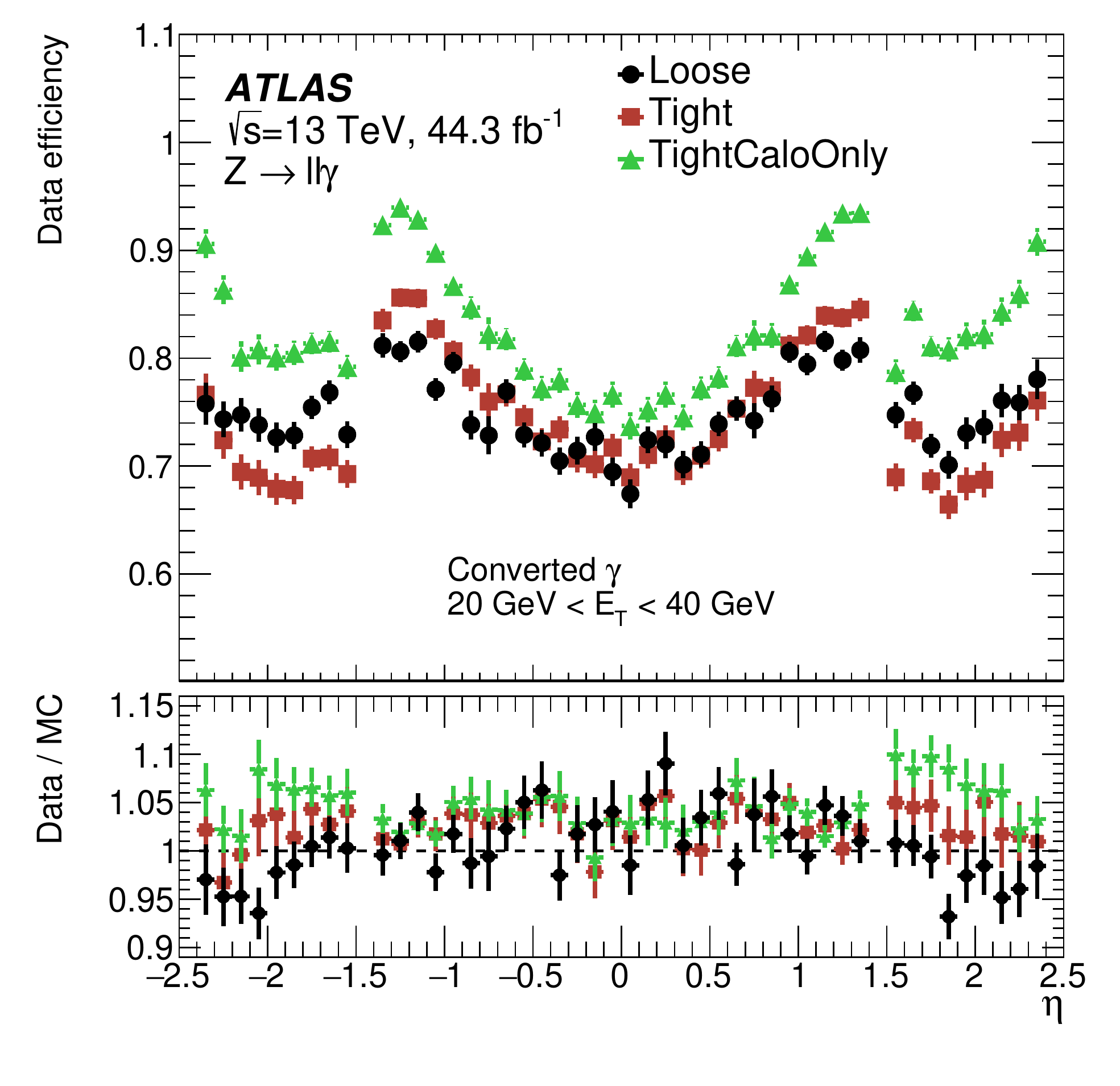}
\includegraphics[width=0.49\textwidth,angle=0]{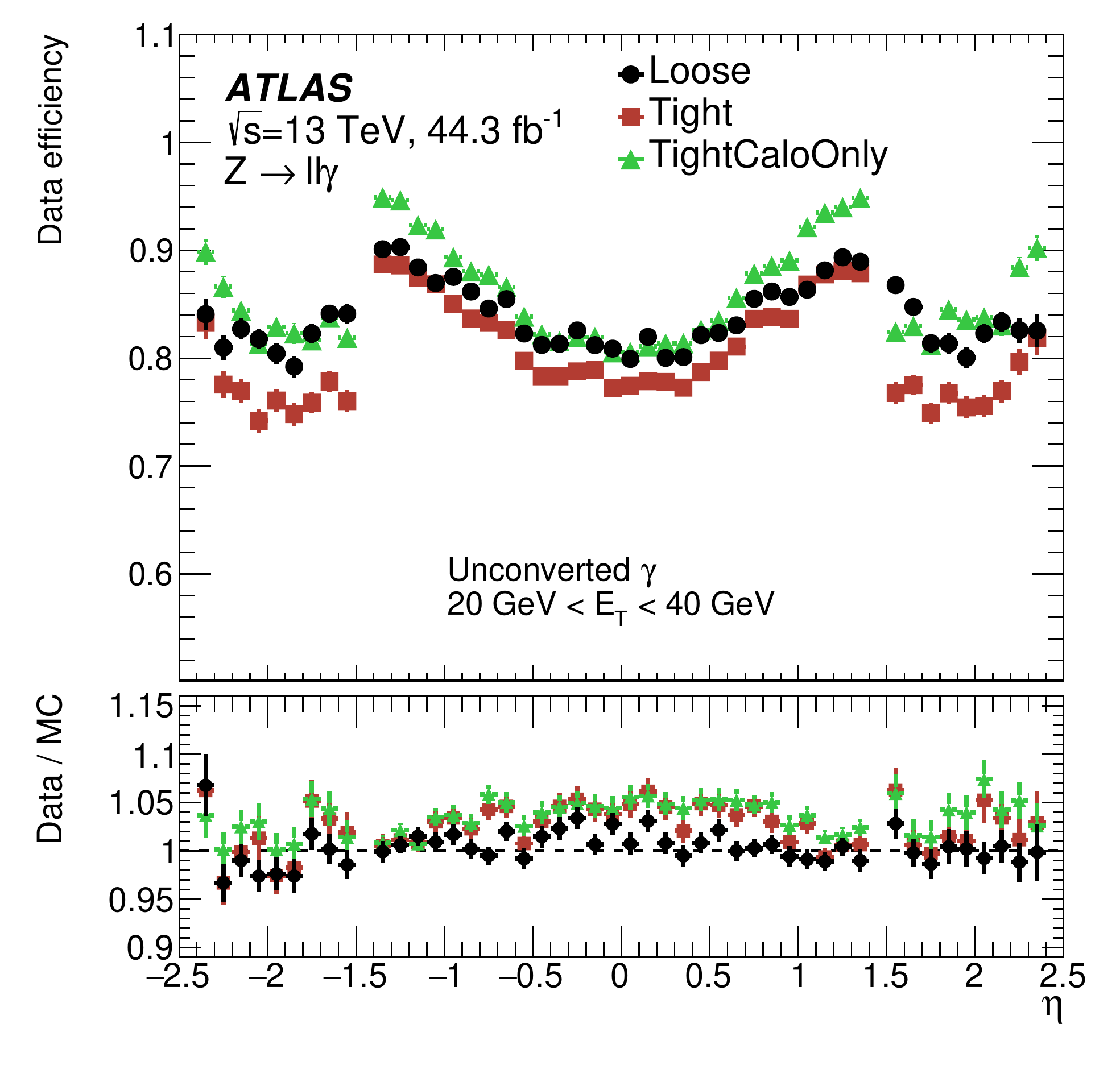}
\includegraphics[width=0.49\textwidth,angle=0]{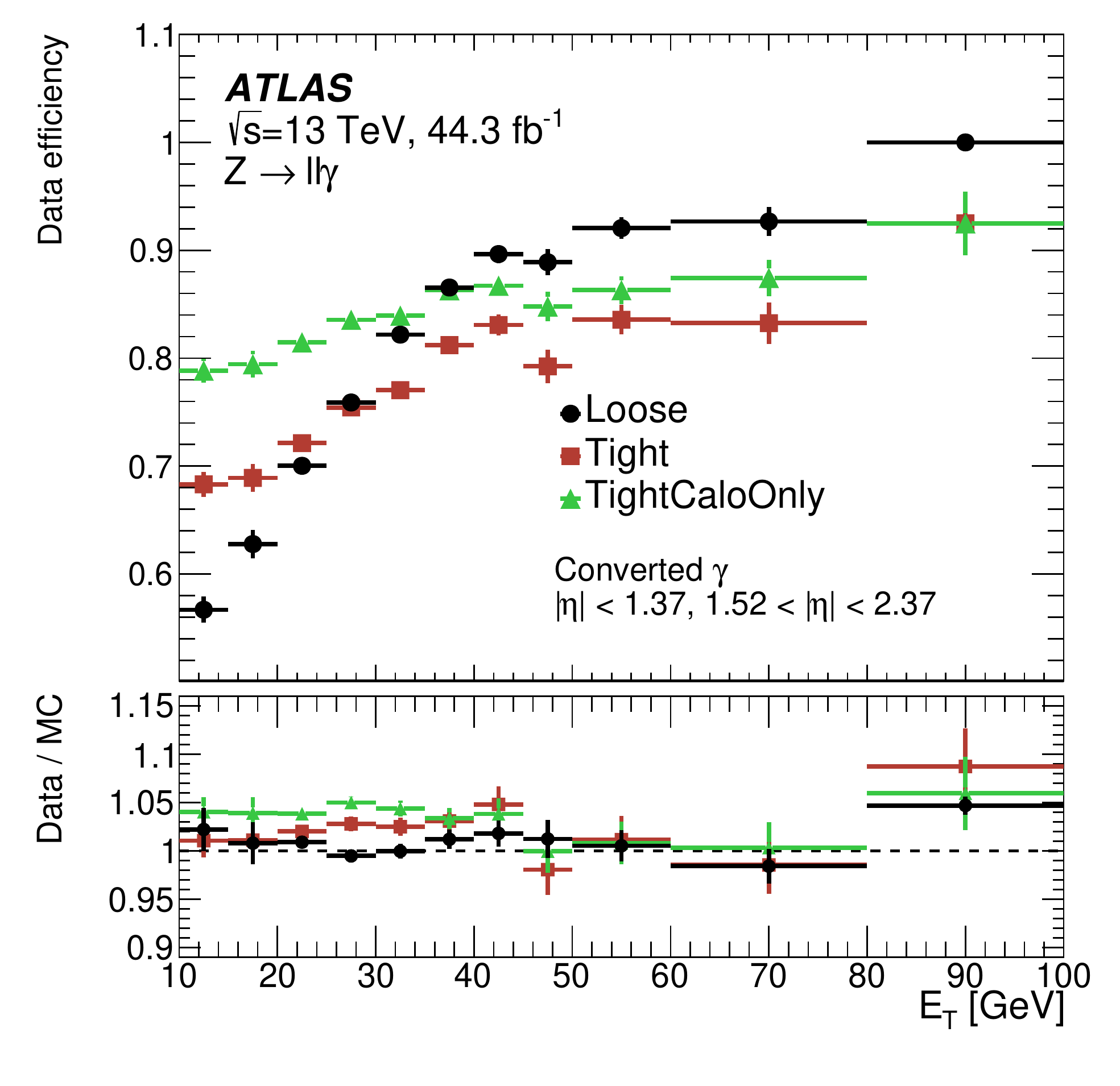}
\includegraphics[width=0.49\textwidth,angle=0]{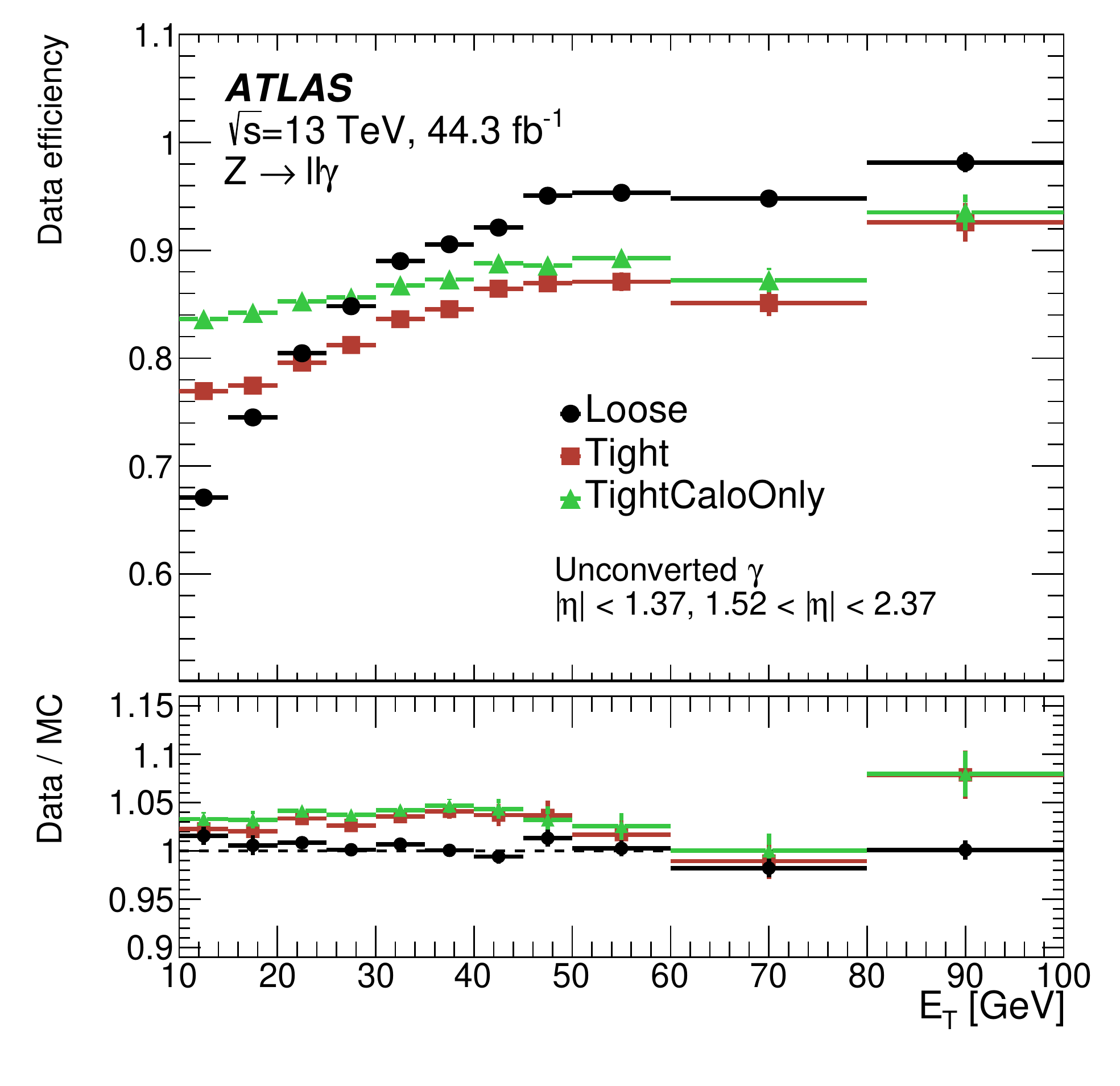}
\vspace{-0.3cm}
\caption{Efficiency of the isolation working points defined in Table~\ref{tab:phWPs}, using $Z\to\ell\ell\gamma$ events, for converted (left) and unconverted (right) photons as a function of photon $\eta$ (top) and \ET (bottom).
The lower panel shows the ratio of the efficiencies measured in data and in simulation. The total uncertainties are shown, including the statistical and systematic components.
}
\label{fig:photon_radZ_isoEffvsetaEt}
\end{figure}
 
\begin{figure}[h!]
\centering
\includegraphics[width=0.49\textwidth,angle=0]{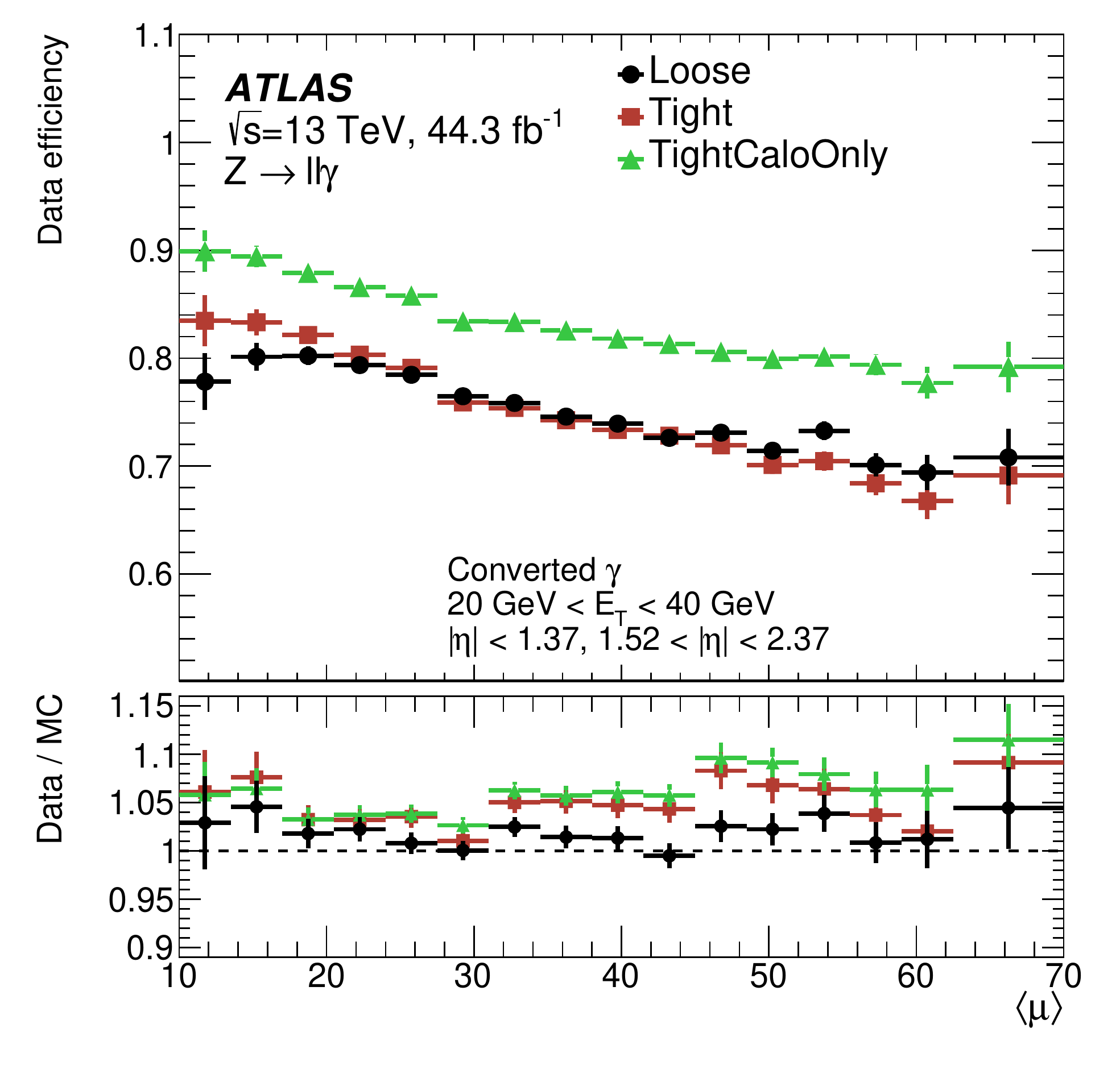}
\includegraphics[width=0.49\textwidth,angle=0]{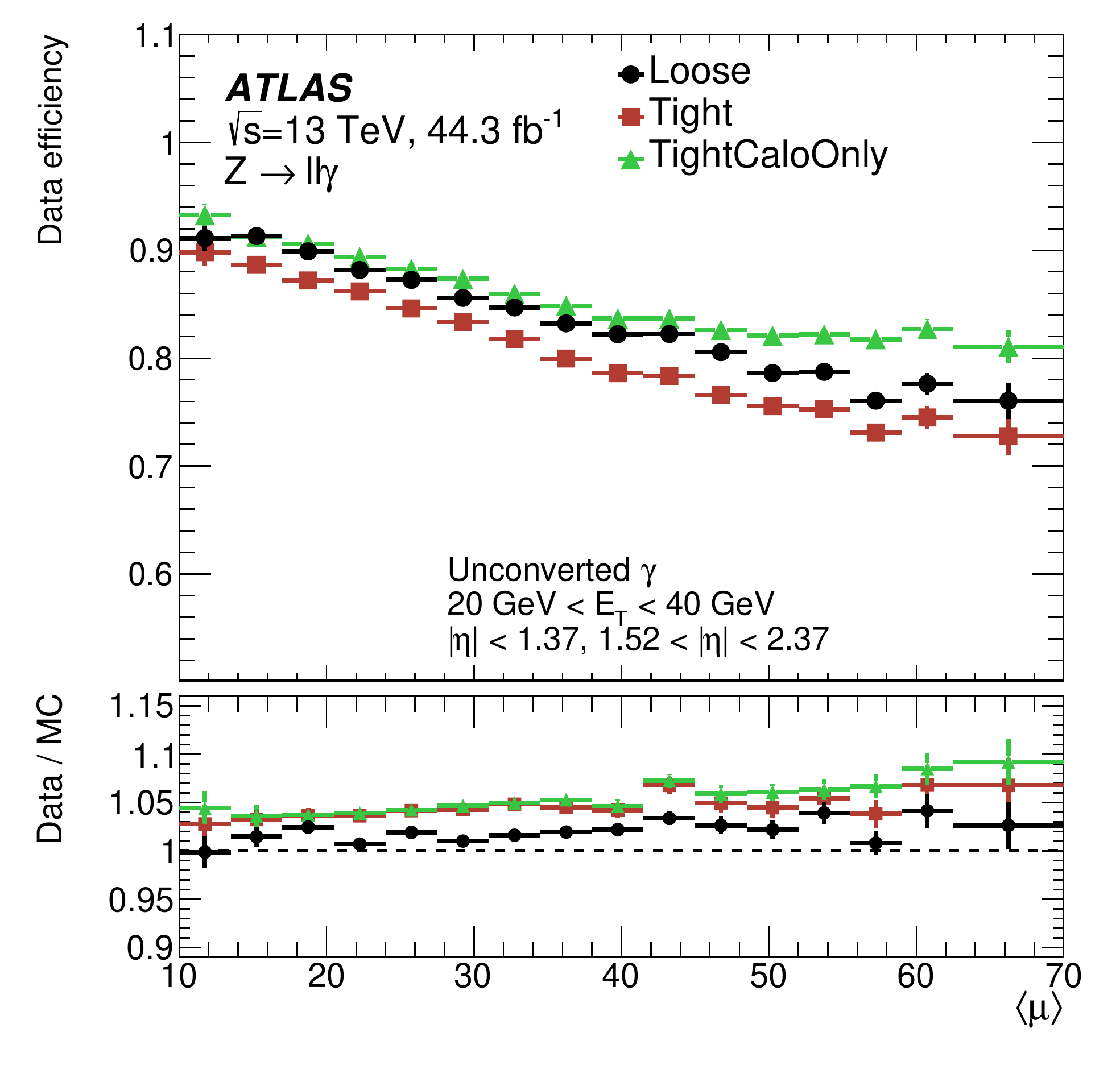}
\vspace{-0.3cm}
\caption{Efficiency of the isolation working points defined in Table~\ref{tab:phWPs}, using $Z\to\ell\ell\gamma$ events, for converted (left) and unconverted (right) photons as a function of \muhat.
The lower panel shows the ratio of the efficiencies measured in data and in simulation. The total uncertainties are shown, including the statistical and systematic components.
}
\label{fig:photon_radZ_isoEffvsmu}
\end{figure}

\subsubsection{Photon calorimeter isolation efficiency measurement with inclusive-photon events}
Photon isolation studies with inclusive-photon events are performed using two different methods for the calorimeter-based and track-based isolations. This is because the distribution of the track isolation variable shows a large peak at $\ptcs = 0$ followed by a 1~\gev\ gap, due to the selection of the tracks entering the $\ptcs$ computation, and by a small tail, and cannot be fitted with an analytic function. In consequence, the efficiency measurement is done separately for the track isolation and calorimeter isolation criteria applied to define the working points presented in Table~\ref{tab:phWPs}. When the measurement is performed for the track-based (calorimeter-based) isolation, the requirements on the calorimeter-based (track-based) isolation are applied at preselection level to reduce the background from jets.

The photon calorimeter isolation (\textit{calo-only}) efficiency with inclusive-photon events is obtained by fitting the distribution of the calorimeter isolation, \etcb or \etcs, minus the relevant \ET fraction ($0.022 \times \ET$ for Tight and $0.065 \times \et$ for Loose), hereafter simply called the isolation distribution.
The measurement is performed in bins of photon $\eta$, \ET, conversion status and data-taking period. The \textsc{Pythia8} inclusive-photon sample described in Section~\ref{sec:monte_carlo} is used for the true photon template.
 
A set of alternate selections is used to determine the isolation distributions for the backgound and their uncertainty. These criteria, denoted LoosePrime$N$, select photon candidates that pass the Loose identification but fail at least one out of $N$ shower shape cuts used in the Tight identification.~\footnote{The sets of criteria of which at least one is not satisfied are requirements on \{\wthree, $\Fside$\} for $N$=2; \{\wthree, \Fside, $\DeltaE$\} for $N$=3; \{\wthree, \Fside, \DeltaE, $\Eratio$\} for $N$=4 and \{\wthree, \Fside, \DeltaE, \Eratio, $\wtot$\} for $N$=5.} The nominal background template is obtained from photon candidates passing the LoosePrime4 identification.
As in the measurements of the data-driven shifts, the photon isolation efficiency is obtained by performing a set of fits in regions defined in simulation and data. Although background enriched,
the sample passing LoosePrime4 also contains true photons that fail the Tight identification requirement; these are defined as `leakage' photons and subtracted.
The sequence of fits proceeds as follows:
\begin{itemize}
\item[1.] A model for the isolation distribution for signal photons is defined from a fit, using a Crystal Ball function, to the isolation distribution obtained for tightly identified photons in simulation.
\item[2.] The corresponding model for leakage photons is defined from a fit to the isolation distribution obtained for LoosePrime4 photons in simulation.
\item[3.] The isolation distribution for background photons (i.e. the sum of fake and leakage photons) is parameterized using a two-component fit to the distribution observed for photons satisfying the
LoosePrime4 requirement in
data. An unconstrained Crystal Ball function is used to model the isolation distribution for fake photons, and the model for leakage photons is defined in point 2.
\item[4.] Finally, the number of signal photons is estimated from a two-component fit to the isolation distribution observed for tightly identified photons in data. The background component uses the model defined in point 3, and the signal photon component uses the model defined in point 1.
\end{itemize}
The fits described in points 3 and 4 above are performed twice. The first time, they are performed to estimate the number of leakage photons from the ratio of the number of photon candidates passing
the Tight and LoosePrime4 identification selection. When the fit in the LoosePrime4 sample is performed again (point 3), the number of leakage photons is constrained, allowing a better estimation of the fake photon isolation distribution. Finally the fit in the tightly identified sample is also redone, with the background component formed only by fake photons. Once the background component is subtracted, only real photons meeting the Tight identification criterion remain and are used for the isolation efficiency measurements.

Finally, the \textit{calo-only} isolation efficiency in data is obtained by integrating the background-subtracted isolation distribution for tightly identified photons in data, up to the working point cut-off of 0~\GeV\ (Loose) or 2.45~\GeV\ (Tight and TightCaloOnly).
Three sources of systematic uncertainty are considered: discrepancies between the fitted isolation distribution and that observed for photons in data; differences between results obtained using LoosePrime3 and LoosePrime5 instead of LoosePrime4 for the determination of the background templates; and uncertainties in the estimation of the number of leakage photons in the LoosePrime4 sample. A binomial statistical error in the scale factors is also calculated and added in quadrature to the systematic components.

\begin{figure}[ht!]
\centering
\includegraphics[width=0.49\textwidth,angle=0]{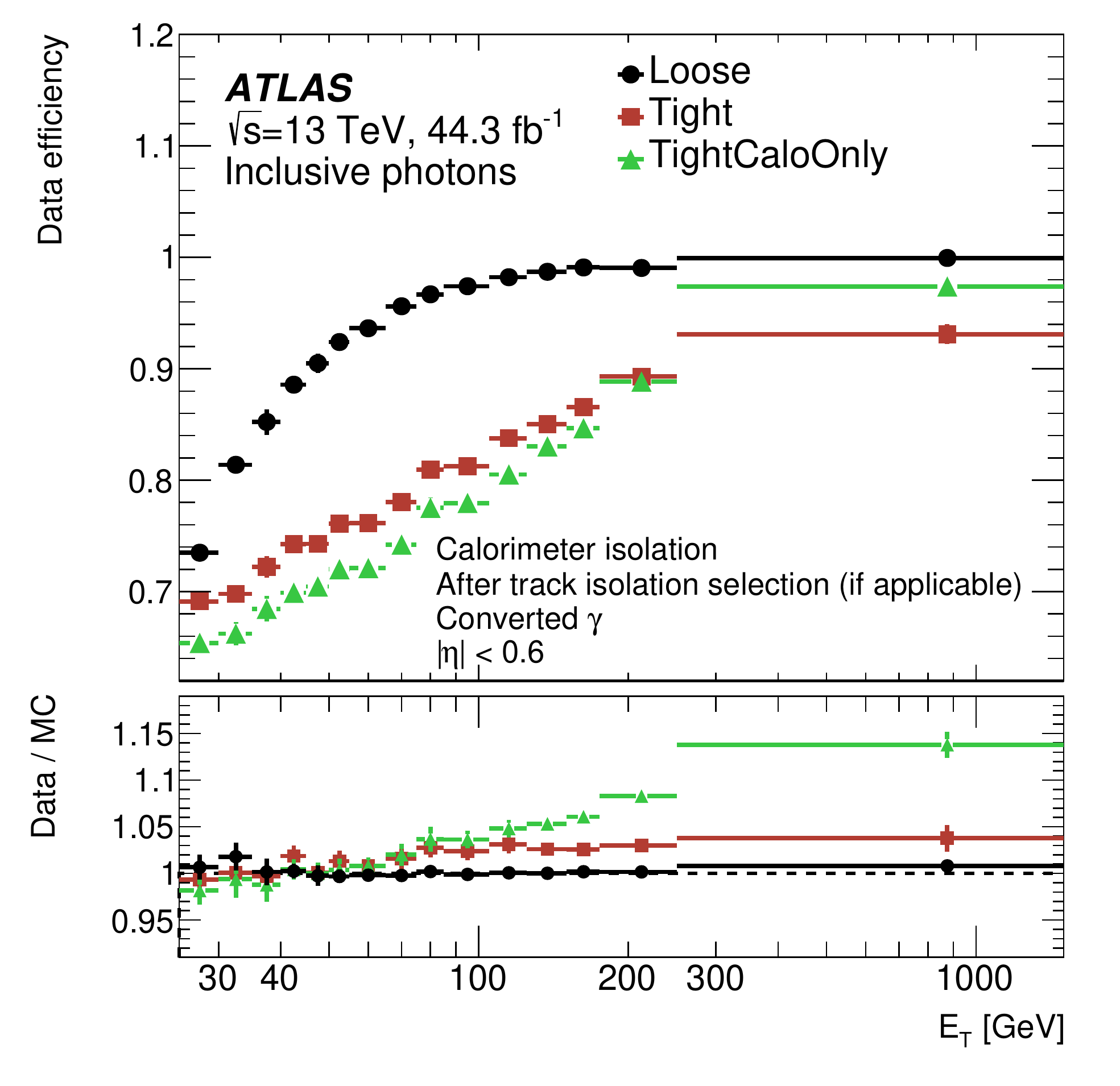}
\includegraphics[width=0.49\textwidth,angle=0]{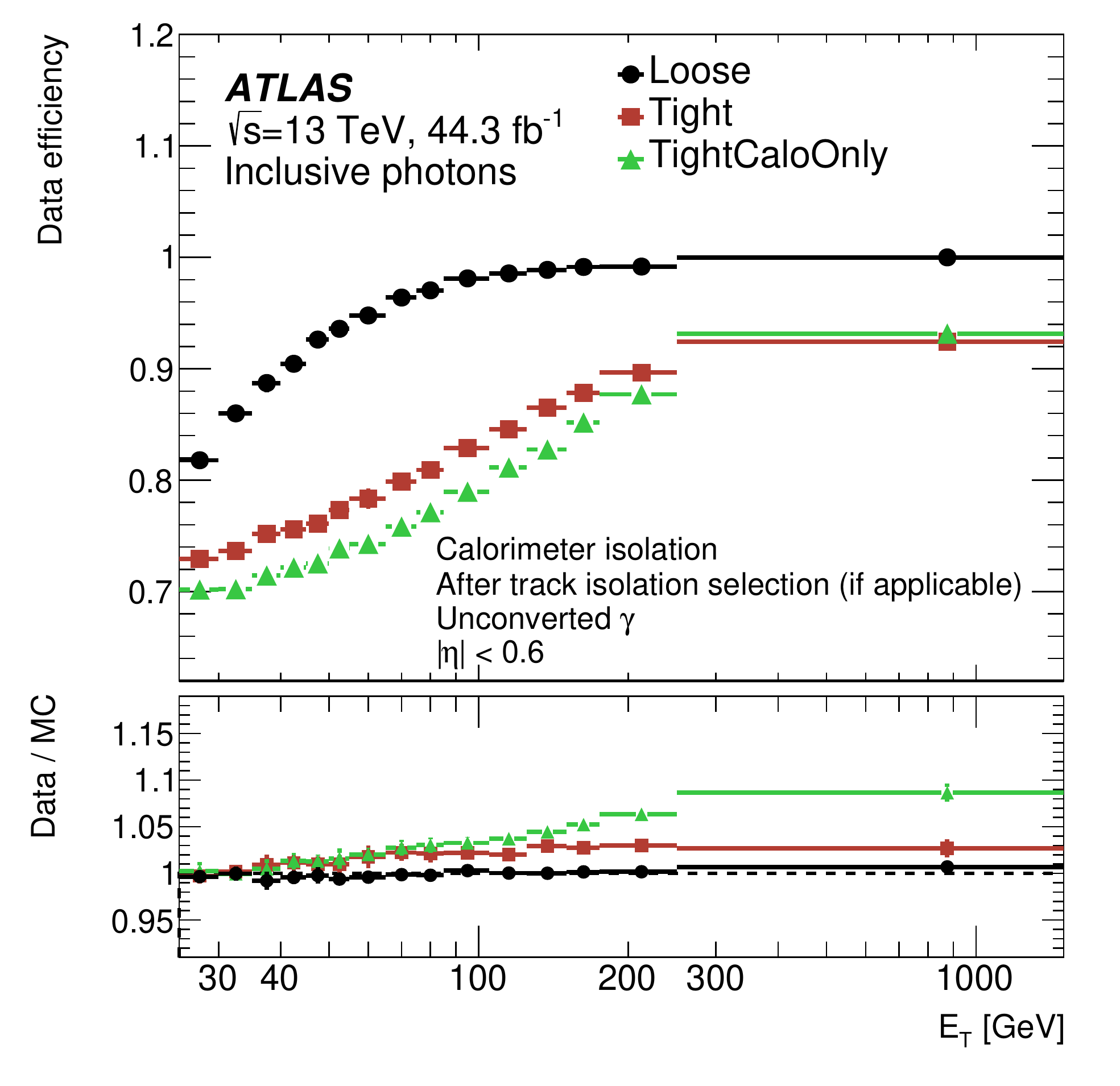}
\includegraphics[width=0.49\textwidth,angle=0]{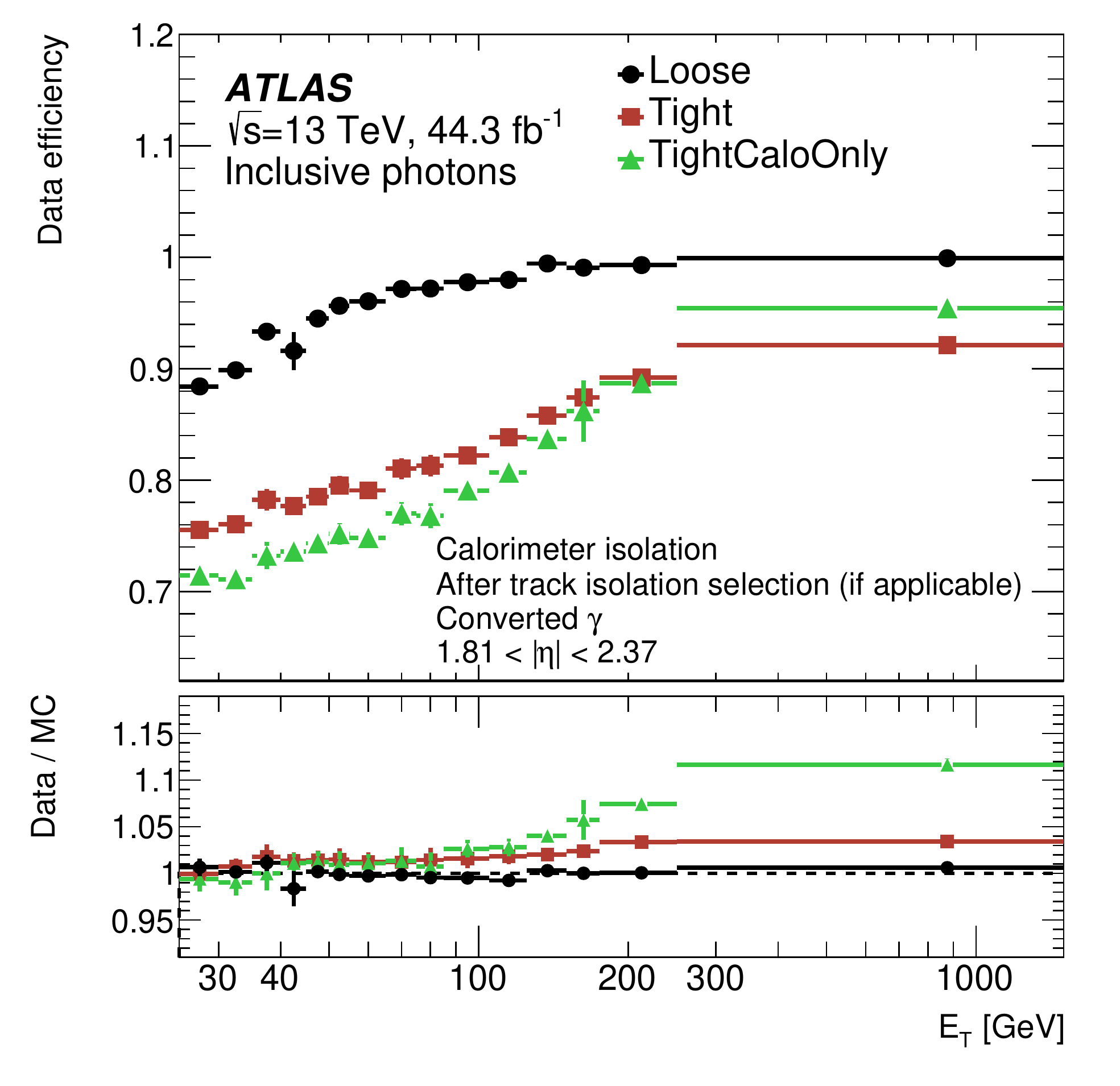}
\includegraphics[width=0.49\textwidth,angle=0]{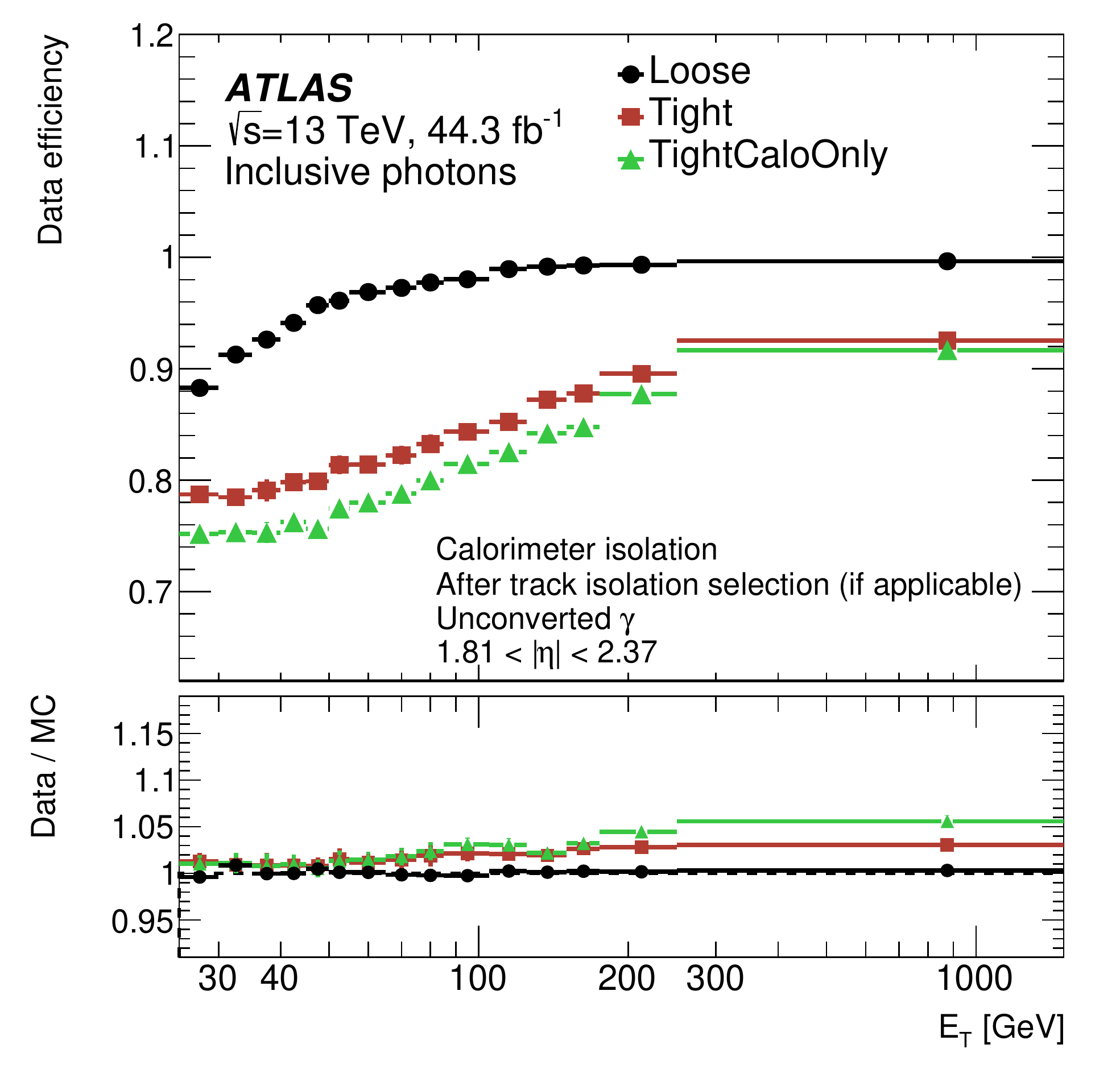}
\caption{Efficiency of the different \textit{calo-only} isolation working points for photons from inclusive-photon events, as a function of photon \ET in two $\eta$ bins ($|\eta|<0.6$ top, and $|\eta|>1.81$ bottom). The results are shown for converted (left) and unconverted (right) photons.
The lower panel shows the ratio of the efficiencies measured in data and in simulation. The total uncertainties are shown, including the statistical and systematic components.
}
\label{fig:photon_caloOnly_SgPh}
\end{figure}
 
The \textit{calo-only} isolation efficiencies measured with inclusive-photon events in 2017 data are shown in Figure~\ref{fig:photon_caloOnly_SgPh}.
The overall differences between data and simulation increase from a few percent in the low \ET region up to 15\% at high \ET ($> 200$~\GeV) for the TightCaloOnly working point, and only up to 5\% for Loose and Tight.

\subsubsection{Photon track-based isolation efficiency measurement with inclusive-photon events}
As in the measurement of the \textit{calo-only} photon isolation efficiency, the main source of background comes from jets misidentified as photons. This background is estimated with a template fit to the track isolation distribution, in a region enriched in background photons satisfying LoosePrime4 but failing the Tight identification criterion. The \textit{track-only} photon isolation efficiency is measured in a signal region enriched in tightly identified photons, after the background is subtracted. To assign the systematic uncertainties, the fit range is varied as well as the definition of the background template, where the photons are required to pass LoosePrime2, LoosePrime3 or LoosePrime5 instead of the LoosePrime4 criterion.
Efficiencies for each configuration are computed, and with them the corresponding scale factors.
Once the different scale factors are calculated, a bin-by-bin scan is performed, keeping the largest deviation from the nominal value among the considered variations.
The total uncertainty is obtained by adding the systematic and statistical components in quadrature.
 
\begin{figure}[ht!]
\centering
\includegraphics[width=0.49\textwidth,angle=0]{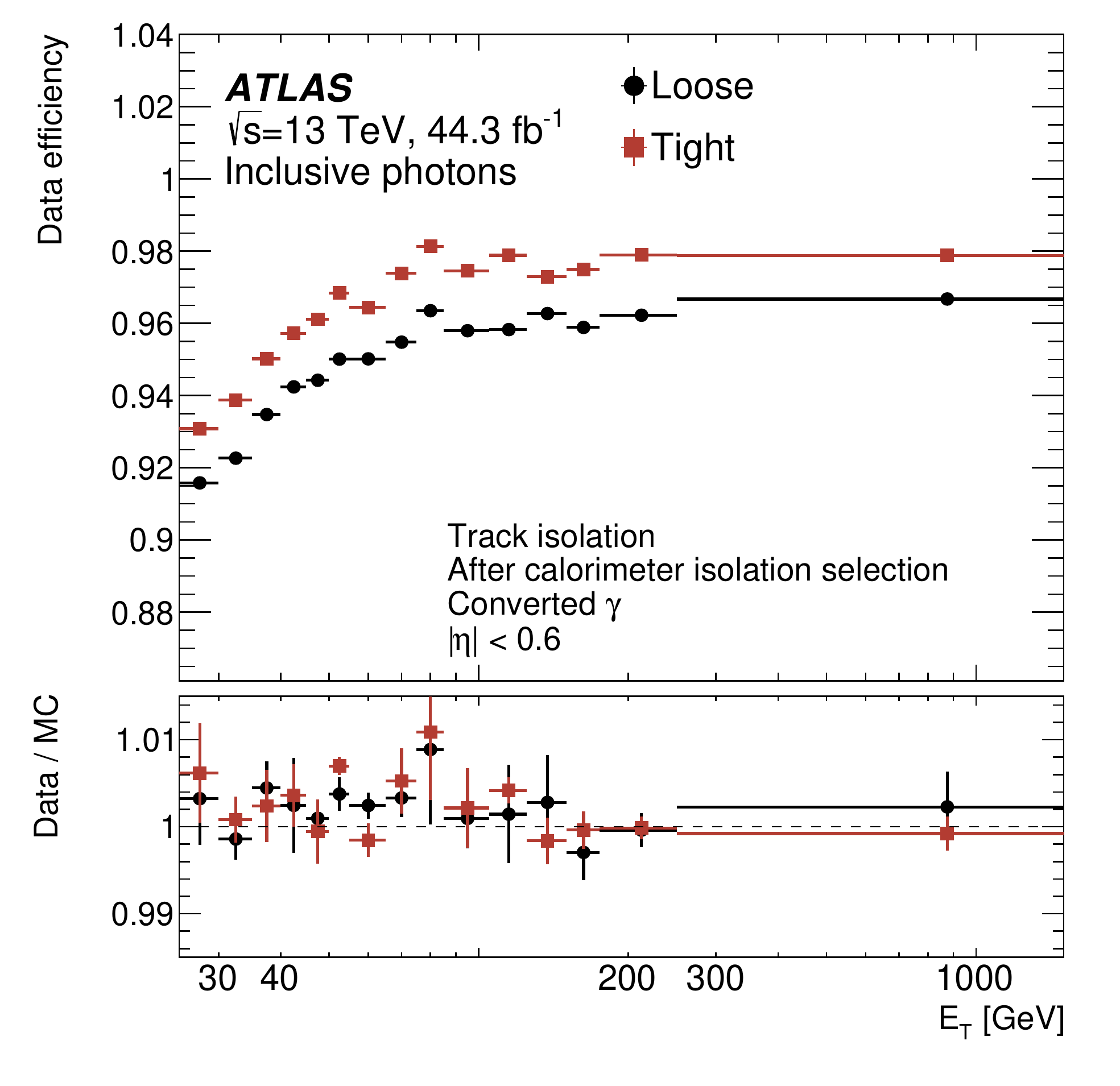}
\includegraphics[width=0.49\textwidth,angle=0]{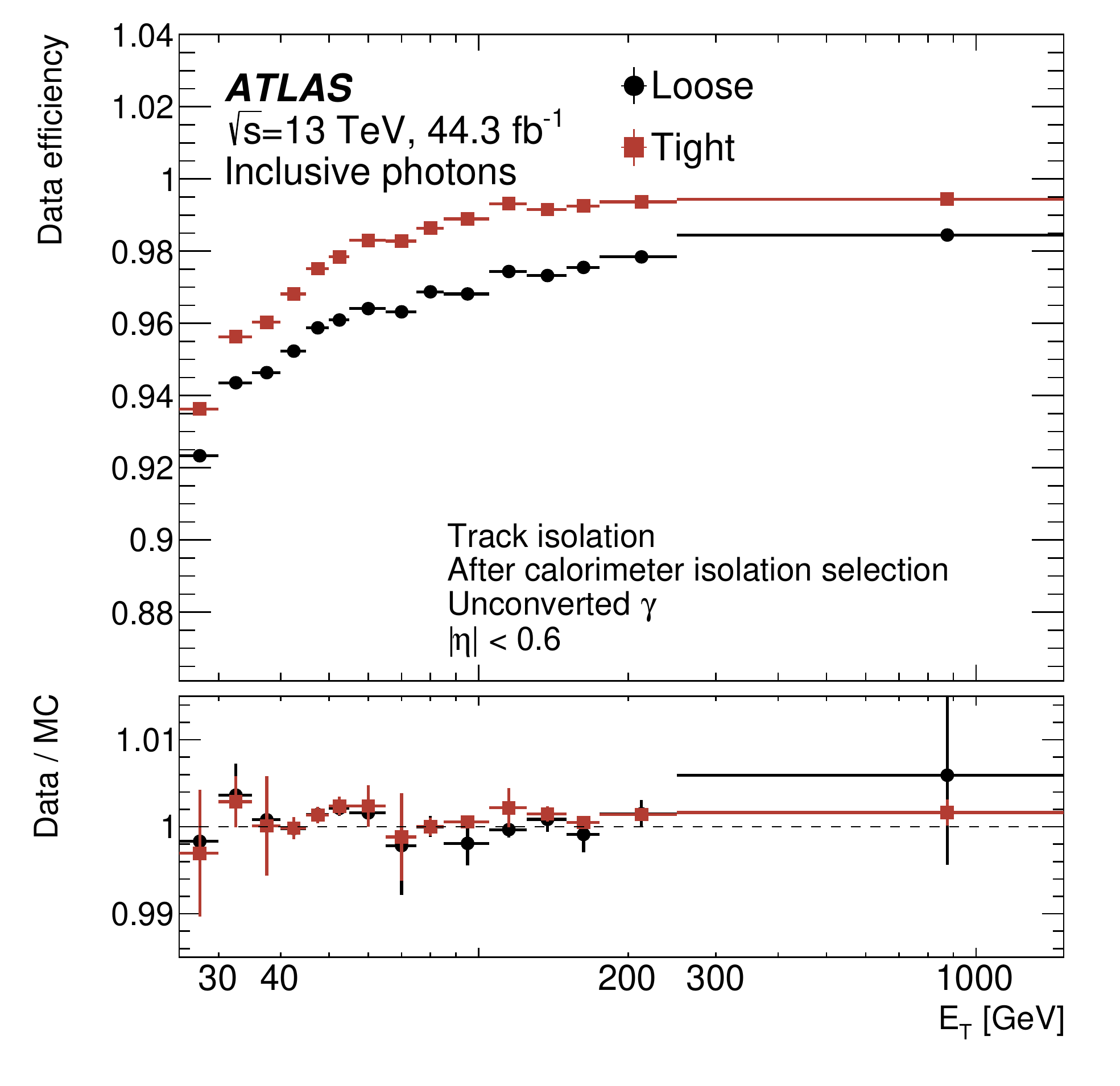}
\includegraphics[width=0.49\textwidth,angle=0]{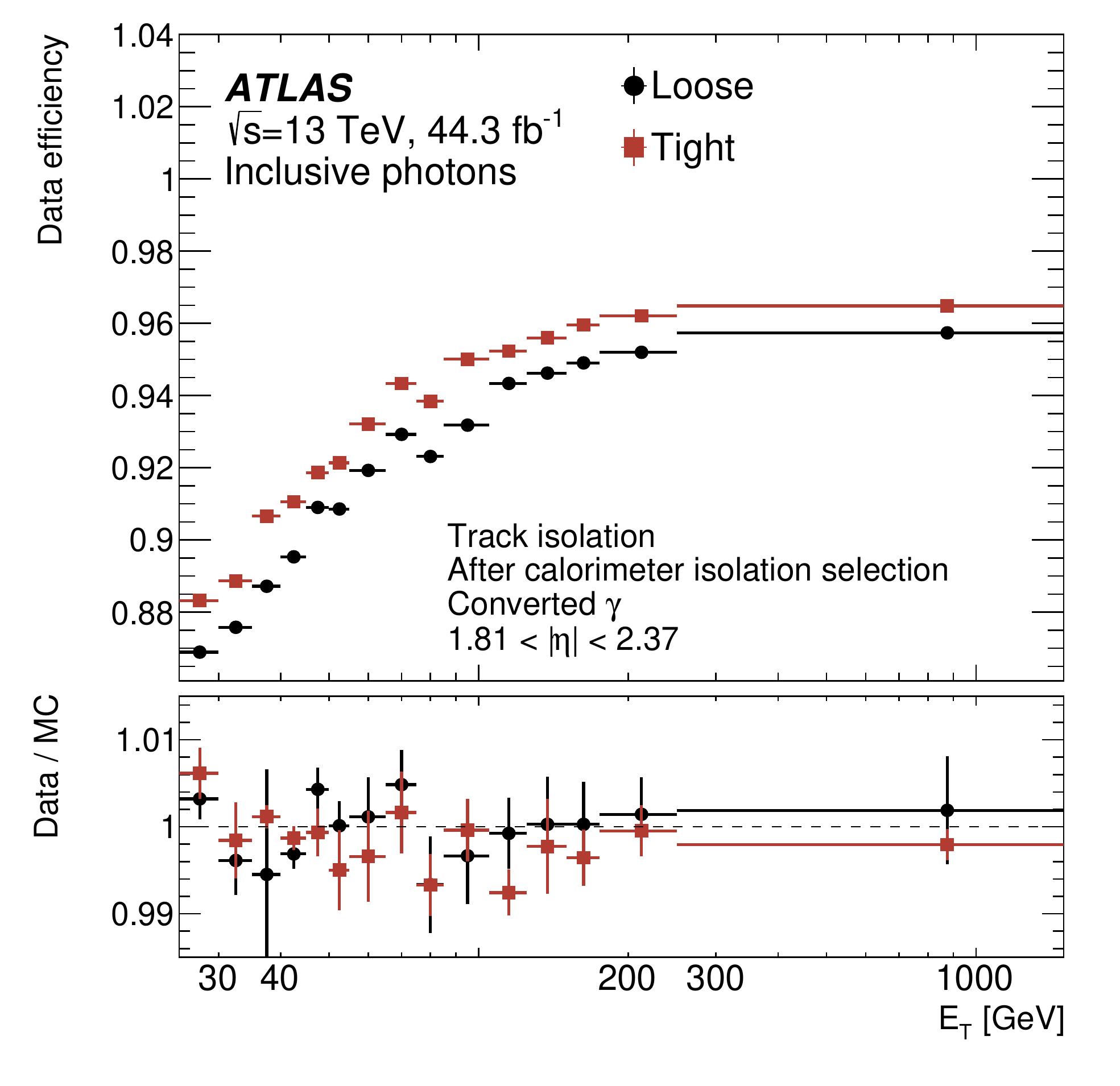}
\includegraphics[width=0.49\textwidth,angle=0]{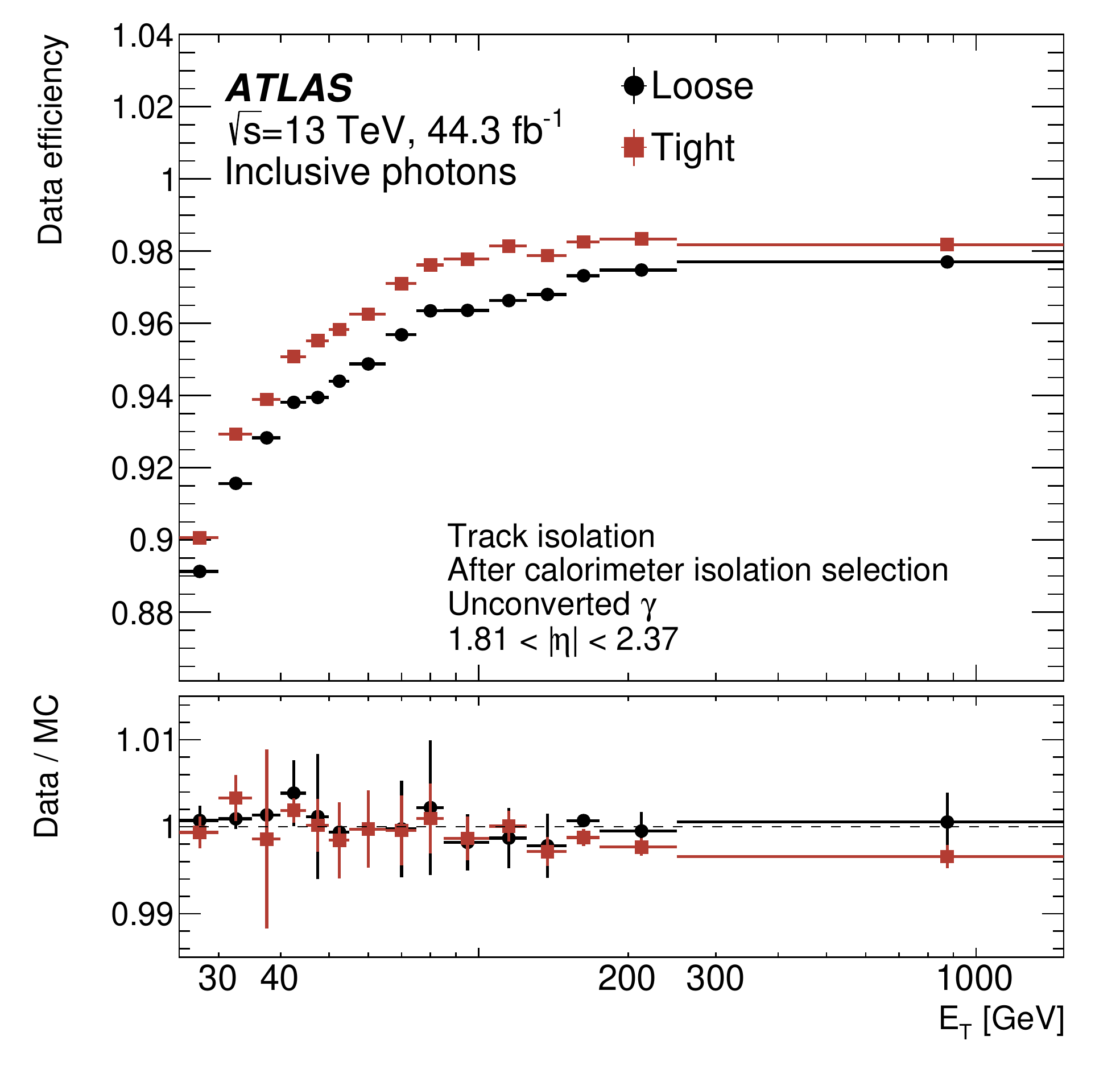}
\caption{Efficiency of the different \textit{track-only} isolation working points for photons from inclusive-photon events, as a function of photon \ET in two $\eta$ bins ($|\eta|<0.6$ top, and $|\eta|>1.81$ bottom). The results are shown for converted (left) and unconverted (right) photons.
The lower panel shows the ratio of the efficiencies measured in data and in simulation. The total uncertainties are shown, including the statistical and systematic components.
}
\label{fig:photon_trackOnly_SgPh}
\end{figure}

The \textit{track-only} isolation efficiencies measured with inclusive-photon events in 2017 data are shown in Figure~\ref{fig:photon_trackOnly_SgPh}.
The ratio of the data to MC simulation is close to unity.

\subsubsection{Combination of photon isolation scale factors}
The photon isolation scale factors are measured for the three isolation working points detailed in Table~\ref{tab:phWPs}\
using radiative $Z$ decays and inclusive-photon events. The different results are combined to obtain one set of scale factors
per working point, data-taking year, and photon conversion status.
The combination is performed in two steps.
First, the track-only and calo-only scale factors determined with inclusive-photon events are multiplied together to obtain a single set per configuration.
These inclusive-photon scale factors are further combined with those determined with radiative $Z$ decay events using a simple weighted average.
The uncertainties in the track-only and calo-only results obtained with inclusive-photon events
are treated as fully correlated in the combination, while the uncertainties in the radiative-$Z$ and
inclusive-photon measurement results are treated as uncorrelated.
The combination is performed for $25<\ET<100$~\gev; below 25~\gev, only results from radiative $Z$ decays are available, while above 100~\gev\ the results are obtained with inclusive-photon events only.
If, in a given ($|\eta|$, \ET) bin, the total uncertainty in the combined scale factor does not cover the difference between the values obtained from the two samples, it is scaled such that $\chi^2 = 1$. Above $1.5\ \tev$, the results obtained in the last bin used for the measurement are considered, with no change in the systematic uncertainty.
 
For \ET < 25~\GeV, the measurements achieve a typical uncertainty of about 2\%, and at worst 5--10\% for \ET < 15~\GeV. For \ET > 100~\GeV, uncertainties around 1--2\% are obtained. For $25<\ET<100$~\GeV, the
combination of the two channels reduces the scale factor uncertainties to about 1\% on average.
 
\section{Electron charge misidentification}
\label{sec:QMisID}
 
The reconstruction of the electric charge of an electron relies solely on the measurement of the curvature of its associated track in the inner detector.
Interactions of an electron with the detector material can create secondary particles: photons and electron--positron pairs.
The production of these secondary particles can lead to distortions of the primary electron track, e.g. hits from the secondary particle being included in the fit of the primary electron track, and the presence of additional tracks of secondary particles in the vicinity of the primary electron track.
Incorrect charge reconstruction can thus be caused either by an incorrect determination of the track curvature, or by the choice of an incorrect track.
 
For electrons at high transverse momentum, the first effect becomes dominant and leads to an almost linear increase with energy in the probability to determine the sign of the curvature incorrectly.
Final-state radiation emitted collinearly off the electron can also cause charge misidentification if the radiated photon subsequently converts to an electron--positron pair in the detector material. 
Here, the correct or incorrect charge is assigned with equal probability.
The electric charge is heavily used as a selection criterion in measurements with the ATLAS experiment, and hence understanding the effects of charge misidentification is important. Some specific signatures also require the suppression of electron charge misidentification in order to reduce background.
 
\subsection{Suppression of electron charge misidentification}
\label{sec:QMisID:optim}
 
The suppression of electron charge misidentification is based on the output discriminant of a boosted decision tree (BDT). A previous version, optimized for data recorded in 2015 and 2016, rejected 90\% of electrons with incorrectly reconstructed charge, removing only 3\% of electrons with correctly reconstructed charge~\cite{PERF-2017-01}. The optimization was based on simulated electrons and showed a higher rejection than observed in data. In the following, a re-optimization of the BDT is described. Data from \Zee\ decays are used to reduce efficiency losses due to mismodelling of the input variables in the BDT training. Furthermore, additional input variables have been studied.
 
To select a relatively clean sample of electrons with correctly and incorrectly reconstructed charge, one of the electrons is restricted to $\left|\eta\right|<0.6$, required to satisfy Tight identification and to pass the 97\% operating point of the previous BDT discriminant.
These requirements minimize charge misidentification for this electron.
Any additional reconstructed electron in the event is used to train the BDT, as a signal electron if it has an electric charge different from the first electron, and as a background electron if the electric charge is the same. To reduce background from converted photons from initial- or final-state radiation, the invariant mass of any pairs of electrons must lie within 5~\GeV\ of 90~\GeV\ in opposite-charge events and within 5~\GeV\ of 88~\GeV\ in same-charge events. The lower value used in same-charge events accounts for the fact that electrons with the incorrect charge have a higher probability for energy loss as discussed in \Sect{\ref{sec:QMisID:measurement}} and illustrated in \Fig{\ref{fig:eleID:chargeflipbkg}}.
 
Input quantities to the BDT are the electron \et\ and $\eta$, and a set of additional variables. In decreasing order of separation power, these are: the transverse impact parameter multiplied by the electron electric charge $q\times d_0$, the average charge of all tracks matched to the electron weighted by their number of hits in the
SCT detector
$\bar{q}_{\mathrm{SCT}}$, $E/p$ and $\Delta\phi_{\mathrm{res}}$. With $\bar{q}_{\mathrm{SCT}}$ the BDT includes for the first time the reconstructed properties of additional tracks in the vicinity of the electron,
which improves rejection in cases where the incorrect track is chosen as the primary electron track. 
 
The efficiency of the requirement on the BDT is 98\%  in $\Zee$ events for electrons satisfying Medium or Tight identification with the Tight isolation requirement, and that have the correct electric charge. Approximately 90\% of electrons with the same identification and isolation requirements but incorrect electric charge are removed.
This re-optimization of the BDT variables has improved the efficiency of the selection criterion, leaving the rejection of electrons with misidentified charge unchanged.
 
\subsection{Measurement of the probability for charge misidentification}
\label{sec:QMisID:measurement}
The probability for electron charge misidentification is measured in seven bins in $\eta$ and six \et bins in the range $20~\GeV<\et<95~\GeV$ in \Zee\ events.
The events were collected with the dielectron triggers discussed in \Sect{\ref{sec:data_set}} with transverse momentum thresholds of 17~\GeV\ or less and Loose trigger identification, allowing the measurement to be extended to lower values of \et\ and looser identification criteria than previous measurements. Both electrons in the event are selected with the same identification and isolation criteria and, respectively, fall into bins $i$ and $j$ in $\left(\eta,\et\right)$, yielding $N_{ij}$ \Zee\ events. Their invariant mass must lie within 10~\GeV\ of the nominal $Z$-boson mass.
The probabilities of the electron charge misidentification in bins $i$ and $j$, $\epsilon_i$ and $\epsilon_j$, maximize the Poisson probability $P\left(\lambda_{ij}|n_{ij}^{\mathrm{sc}}\right)$, where:
\begin{equation}
\lambda_{ij}=\left(\epsilon_i\left(1-\epsilon_j\right)+\left(1-\epsilon_i\right)\epsilon_j\right)N_{ij}+B_{ij}^{\mathrm{sc}},
\nonumber
\end{equation}
and $n_{ij}^{\mathrm{sc}}$ is the number of same-charge \Zee\ events. 
The number of background events in the sample where both electrons have the same electric charge, $B_{ij}^{\mathrm{sc}}$, consists of misidentified electrons from multijet production and electrons from converted photons from the aforementioned final-state radiation.
The two components are estimated in a sideband subtraction and from simulation, respectively. The selected data and the estimated background is shown in \Fig{\ref{fig:eleID:chargeflipbkg}} for an example bin. Sources of systematic uncertainties in the measurement are the estimation of the background from multijet production and final-state radiation, and the restriction of the dielectron invariant mass. Possible biases in the experimental method used to perform the measurement are evaluated by comparing the charge misidentification probability obtained in the likelihood maximization in simulation with those obtained using generator-level information.
 
The kinematic range of $\et>95$~\GeV\ is particularly relevant for searches for physics beyond the Standard Model with same-charge signatures.
For a measurement with high granularity, the double differential charge misidentification probabilities are factorized into an $\eta$- and an $\et$-dependent part.
This approach allows measurements in 5 bins in \et\ and 14 bins in $\eta$ with reasonable statistical precision from a sample of approximately 9000 electrons with the incorrect charge assignment (for Tight identification).
The systematic uncertainty in the parameterization is assessed by comparing, double differentially, the ratio of same-charge events and opposite-charge events, weighted with the charge misreconstruction probability, in data and simulation.
The systematic uncertainty is derived by incrementing the uncertainty in steps of 1\% until the $\chi^2$ value falls below 1, separately in each bin in \et.

\begin{figure}[t]
\centering
\subfloat[]{
\includegraphics[width=.49\textwidth]{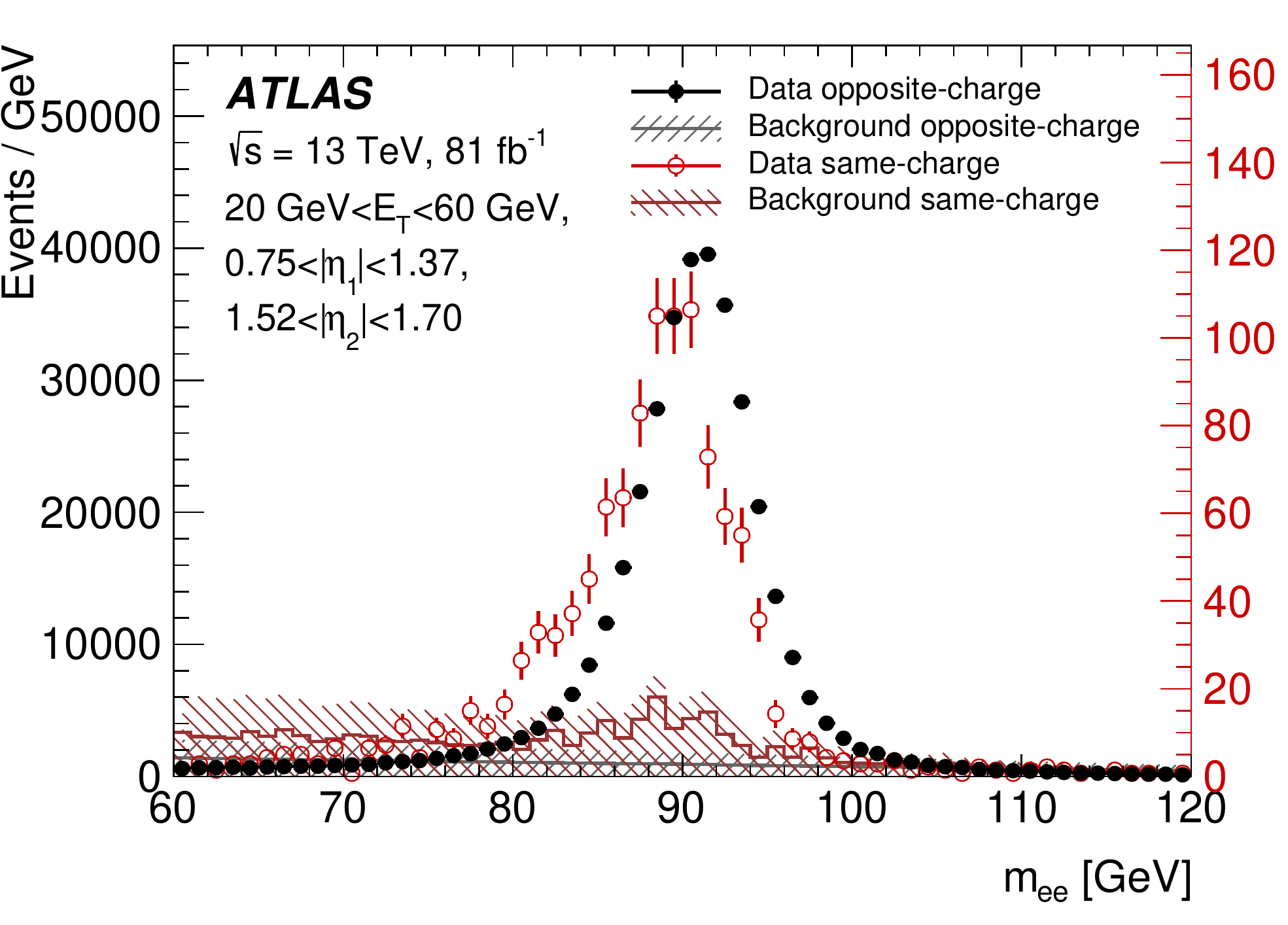}
\label{fig:eleID:chargeflipbkg}
}
\subfloat[]{
\includegraphics[width=.49\textwidth]{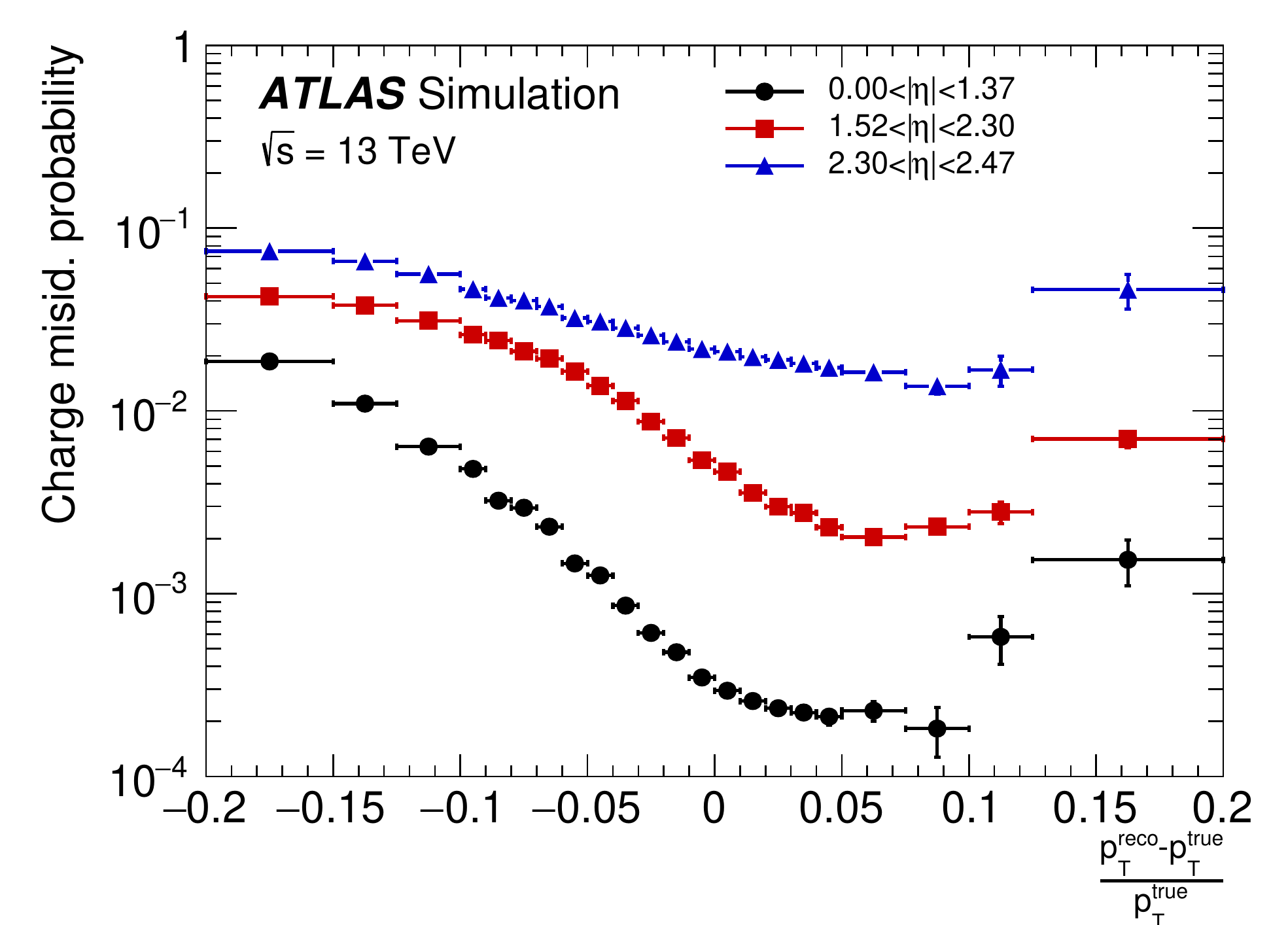}
\label{fig:eleID:chargeflipeloss}
}
\caption{(a) Dielectron invariant mass distribution of events from \Zee\ production used for the measurement of electron charge misidentification efficiencies.
The events are selected with a same-charge or an opposite-charge requirement where one electron falls into $0.75<\left|\eta\right|<1.37$ and the other into  $1.52<\left|\eta\right|<1.70$. Both electrons have $20~\GeV<\et<60~\GeV$. The estimated background from misidentified electrons and contributions from final state radiation are shown as a continuous line with its uncertainty as a shaded band.
(b) Charge misidentification probabilities as a function of the energy measurement residual, for electrons meeting the Tight identification and Tight isolation criteria, in simulated \Zee\ events. Only statistical uncertainties are shown.
}
\end{figure}
 
The interactions with material in the inner detector causing electron charge misidentification can also lead to significant energy loss and leakage of energy outside the EM cluster, introducing a correlation between the two effects.
In \Fig{\ref{fig:eleID:chargeflipeloss}}, the charge misidentification probability is shown as a function of the energy response, $(\pt^{\mathrm{reco}}-\pt^{\mathrm{true}})/\pt^{\mathrm{true}}$, in several bins of $\eta$. It increases with the difference between true and reconstructed electron energy. The same effect causes the differences in reconstructed invariant mass between opposite-charge and same-charge events shown in \Fig{\ref{fig:eleID:chargeflipbkg}}.
The correlation with the energy response complicates the measurement of charge misidentification probabilities in data.
The probability measurement is blind as to which of the two electrons has the incorrect charge assignment. Hence, the probabilities determined from the likelihood maximization are used to form data-to-simulation probability ratios.
No significant dependence of the data-to-simulation ratios on the dilepton invariant mass has been observed.
The charge misidentification probabilities in data are obtained by multiplying the data-to-simulation probability ratios by the charge misidentification probabilities computed in the simulation, where the electron with the incorrect charge assignment is unambiguous.
The probabilities in data are shown in \Fig{\ref{fig:eleID:chargeflipeff}} for several combinations of identification and isolation operating points. For Medium identification with Tight isolation, the electron charge misidentification probability in \Zee\ events is smallest in the central region of the detector at 0.05\%, 
and increases to 2.7\% 
at high $|\eta|$. As a function of \et\ it increases approximately linearly from 0.28\% 
at $\et=20$~\GeV\ to 1.7\% 
at $\et=120$~\GeV. With Tight instead of Medium identification, a reduction of charge misidentification by 25\%--50\%, depending on \et\ and $\eta$, is seen. The BDT presented in Section~\ref{sec:QMisID:optim} further reduces the misidentification probability by factor of about five, on average over the detector acceptance, and by up to a factor 10 at high pseudorapidity.
 
\begin{figure}[t]
\centering
\includegraphics[width=.49\textwidth]{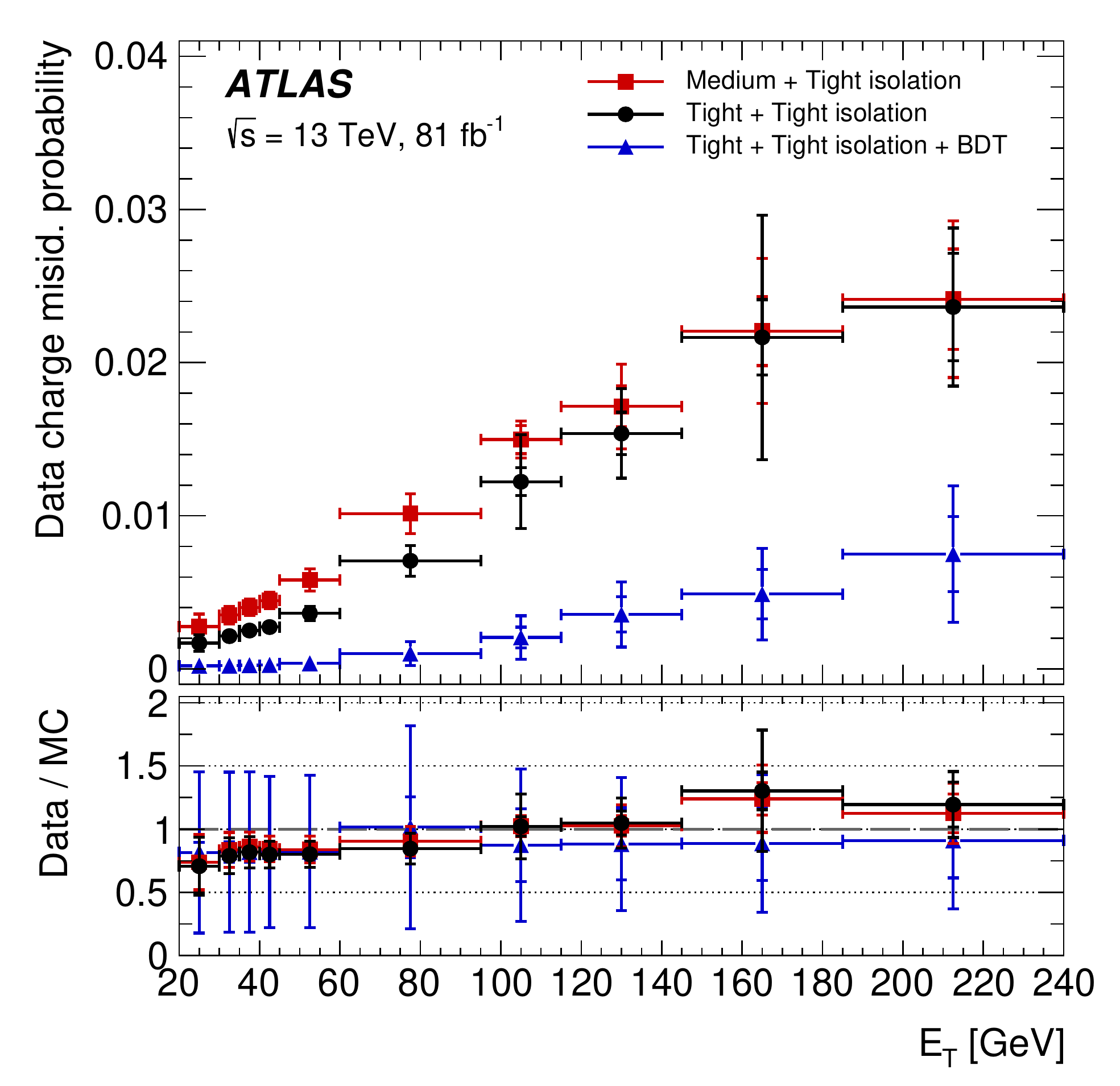}
\includegraphics[width=.49\textwidth]{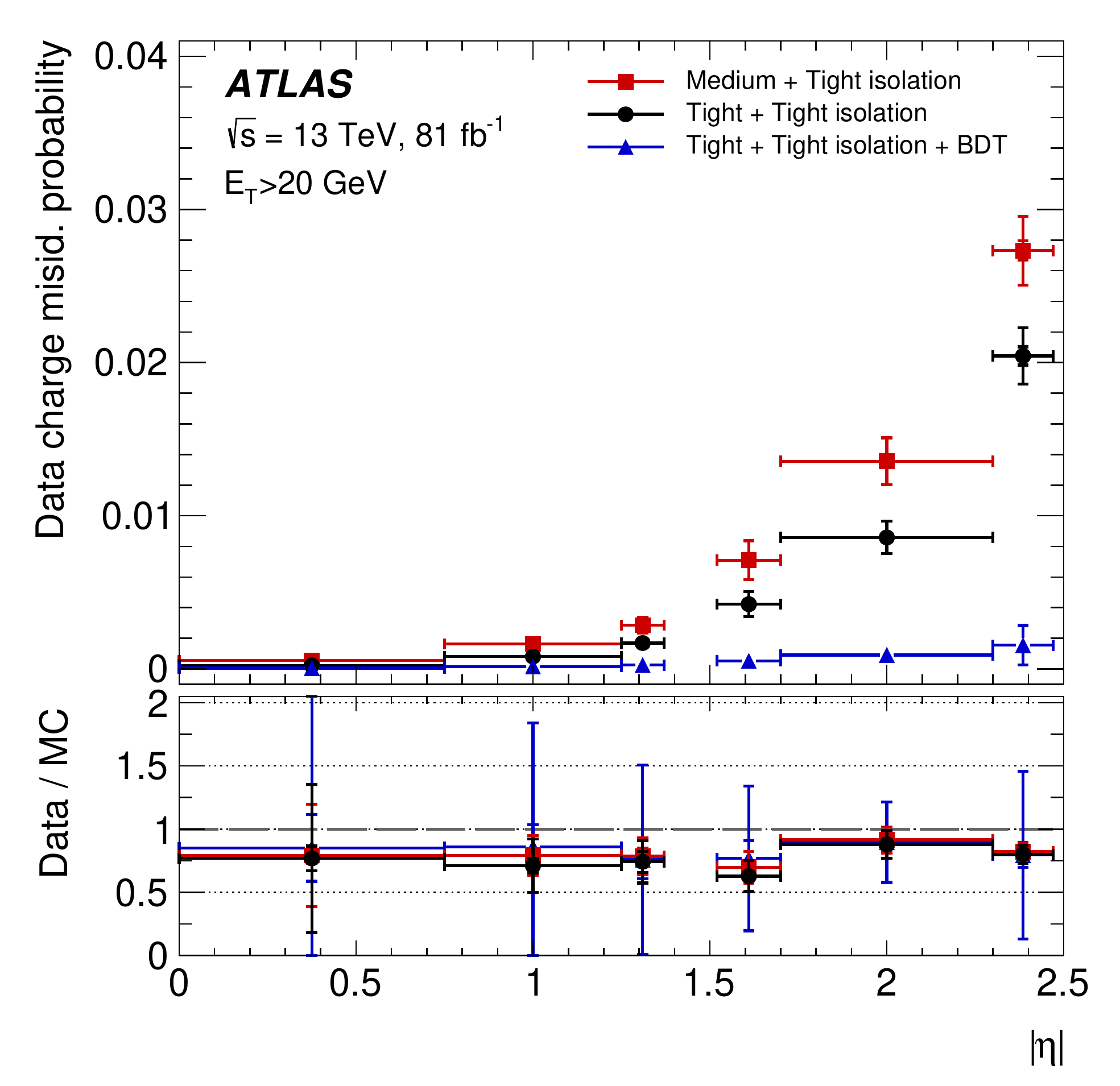}
\caption{Charge misidentification probabilities in data as a function of $\et$ (left) and $\left|\eta\right|$ (right). The energies of the electrons have been corrected for the energy loss in the interaction with the detector material, which is the primary source of charge misidentification. The inner uncertainties are statistical while the total uncertainties include both the statistical and systematic components.
}
\label{fig:eleID:chargeflipeff}
\end{figure}
 
\FloatBarrier
\section{Conclusions}
The reconstruction of electrons and photons based on a dynamical, topological cell clustering algorithm has been described, and the corresponding updates to the methods used for the
identification of the candidates and the estimation of their energy have been discussed. The rejection of non-isolated particles and of mismeasured electron candidates have been re-optimized accordingly.
 
The dynamical cell clustering algorithm provides an electron and photon reconstruction efficiency similar to that of the sliding-window reconstruction. A relative improvement of about 15\% is
obtained in the reconstruction efficiency for two-track photon conversions.
The misclassification of unconverted photons as single-track TRT conversions is reduced by a factor of two, while the single-track conversion reconstruction efficiency only decreases by 5 to 10\%.
The present algorithm also provides a better energy measurement, with a relative improvement in resolution by about 15\% in the
barrel, and about 20--25\% in the endcap, for electrons and converted photons. The resolution for unconverted photons is unchanged.
 
Energy scale and resolution corrections have been measured using electrons from \Zee\ decays. A significant dependence of the corrections on the amount of pile-up has been observed,
reflecting a mismodelling of the calorimeter activity in minimum-bias events. The uncertainty in the energy scale corrections ranges from $4\times 10^{-4}$ in the barrel to $2\times 10^{-3}$ in
the endcap. The uncertainty in the constant-term resolution corrections is typically 1--$2\times 10^{-3}$. The electron-based energy calibration has been verified for photons, using radiative
$Z$-boson decays, to a precision of 0.5\% at worst.
 
The identification of electrons and photons has been revisited to match the improved cell clustering procedure. For electrons, identification efficiencies vary from 93\% for the Loose
identification criterion, to 80\% for the Tight criterion, for electrons from $Z$-boson decays. The simulation models these efficiencies to a precision of 2\% for Loose electrons and 5\% for Tight
electrons, respectively. The efficiency correction factors are measured with a typical precision of 0.2\%. In the case of photons, the identification efficiency reaches 92\% for unconverted photons,
and 98\% for converted photons, for $\et\sim 70$~GeV and above. The precision of the efficiency correction factors ranges from 7\% at low \et to 0.5\% at high \et for unconverted photons, and
from 12\% to 1\% for converted photons.
 
Several electron and photon isolation selection criteria have been defined, targeting a range of processes with varying event activity. The efficiencies of the isolation selections vary from about
99\% for the loosest, to about 90\% for the tightest criterion, depending on the physics process. Tight isolation selections exhibit a steeply rising efficiency as a function of \et; for all isolation
criteria, the selection efficiency varies by about 10\% as a function of \muhat, for the range of \muhat\ spanned by the present dataset. Differences in efficiency between data and simulation range from 1\% to 5\%, depending on $|\eta|$ and \et.
 
A dedicated algorithm has been implemented to reject electrons with badly measured track parameters, with the main objective of reducing the fraction of electron candidates with wrongly measured
charge. This fraction, rising from less than 0.1\% in the barrel to about 3\% at high $|\eta|$ for all candidates, is reduced by a factor of three to five as a function of \et, and by up to a factor of ten at high $|\eta|$. The simulation is found to model the data within 20\% for the residual fraction of wrong-charge electron candidates, and the corresponding correction factors are measured with about 50\% precision.
 
The present results define the baseline performance of the ATLAS detector for searches and measurements using electrons and photons from LHC proton--proton collision data collected at $\sqrt{s}=13$~TeV.
 
\section*{Acknowledgements}
 
 
We thank CERN for the very successful operation of the LHC, as well as the
support staff from our institutions without whom ATLAS could not be
operated efficiently.
 
We acknowledge the support of ANPCyT, Argentina; YerPhI, Armenia; ARC, Australia; BMWFW and FWF, Austria; ANAS, Azerbaijan; SSTC, Belarus; CNPq and FAPESP, Brazil; NSERC, NRC and CFI, Canada; CERN; CONICYT, Chile; CAS, MOST and NSFC, China; COLCIENCIAS, Colombia; MSMT CR, MPO CR and VSC CR, Czech Republic; DNRF and DNSRC, Denmark; IN2P3-CNRS, CEA-DRF/IRFU, France; SRNSFG, Georgia; BMBF, HGF, and MPG, Germany; GSRT, Greece; RGC, Hong Kong SAR, China; ISF and Benoziyo Center, Israel; INFN, Italy; MEXT and JSPS, Japan; CNRST, Morocco; NWO, Netherlands; RCN, Norway; MNiSW and NCN, Poland; FCT, Portugal; MNE/IFA, Romania; MES of Russia and NRC KI, Russian Federation; JINR; MESTD, Serbia; MSSR, Slovakia; ARRS and MIZ\v{S}, Slovenia; DST/NRF, South Africa; MINECO, Spain; SRC and Wallenberg Foundation, Sweden; SERI, SNSF and Cantons of Bern and Geneva, Switzerland; MOST, Taiwan; TAEK, Turkey; STFC, United Kingdom; DOE and NSF, United States of America. In addition, individual groups and members have received support from BCKDF, CANARIE, CRC and Compute Canada, Canada; COST, ERC, ERDF, Horizon 2020, and Marie Sk{\l}odowska-Curie Actions, European Union; Investissements d' Avenir Labex and Idex, ANR, France; DFG and AvH Foundation, Germany; Herakleitos, Thales and Aristeia programmes co-financed by EU-ESF and the Greek NSRF, Greece; BSF-NSF and GIF, Israel; CERCA Programme Generalitat de Catalunya, Spain; The Royal Society and Leverhulme Trust, United Kingdom.
 
The crucial computing support from all WLCG partners is acknowledged gratefully, in particular from CERN, the ATLAS Tier-1 facilities at TRIUMF (Canada), NDGF (Denmark, Norway, Sweden), CC-IN2P3 (France), KIT/GridKA (Germany), INFN-CNAF (Italy), NL-T1 (Netherlands), PIC (Spain), ASGC (Taiwan), RAL (UK) and BNL (USA), the Tier-2 facilities worldwide and large non-WLCG resource providers. Major contributors of computing resources are listed in Ref.~\cite{ATL-GEN-PUB-2016-002}.
 

\clearpage

\printbibliography

\clearpage
 
\begin{flushleft}
{\Large The ATLAS Collaboration}

\bigskip

G.~Aad$^\textrm{\scriptsize 101}$,    
B.~Abbott$^\textrm{\scriptsize 128}$,    
D.C.~Abbott$^\textrm{\scriptsize 102}$,    
A.~Abed~Abud$^\textrm{\scriptsize 70a,70b}$,    
K.~Abeling$^\textrm{\scriptsize 53}$,    
D.K.~Abhayasinghe$^\textrm{\scriptsize 93}$,    
S.H.~Abidi$^\textrm{\scriptsize 167}$,    
O.S.~AbouZeid$^\textrm{\scriptsize 40}$,    
N.L.~Abraham$^\textrm{\scriptsize 156}$,    
H.~Abramowicz$^\textrm{\scriptsize 161}$
% ...
\bigskip
\\

% ...
$^{40}$Niels Bohr Institute, University of Copenhagen, Copenhagen; Denmark.\\
% ...
$^{53}$II. Physikalisches Institut, Georg-August-Universit\"{a}t G\"ottingen, G\"ottingen; Germany.\\
% ...
$^{70}$$^{(a)}$INFN Sezione di Pavia;$^{(b)}$Dipartimento di Fisica, Universit\`a di Pavia, Pavia; Italy.\\
% ...
$^{93}$Department of Physics, Royal Holloway University of London, Egham; United Kingdom.\\
% ...
$^{101}$CPPM, Aix-Marseille Universit\'e, CNRS/IN2P3, Marseille; France.\\
$^{102}$Department of Physics, University of Massachusetts, Amherst MA; United States of America.\\
% ...
$^{128}$Homer L. Dodge Department of Physics and Astronomy, University of Oklahoma, Norman OK; United States of America.\\
% ...
$^{156}$Department of Physics and Astronomy, University of Sussex, Brighton; United Kingdom.\\
% ...
$^{161}$Raymond and Beverly Sackler School of Physics and Astronomy, Tel Aviv University, Tel Aviv; Israel.\\
% ...
$^{167}$Department of Physics, University of Toronto, Toronto ON; Canada.\\
% ...
$^{a}$ Also at Borough of Manhattan Community College, City University of New York, New York NY; United States of America.\\
$^{b}$ Also at CERN, Geneva; Switzerland.\\
% ...

\end{flushleft}

 
\end{document}